\documentclass[aps,rmp,a4paper,twocolumn]{revtex4}
\usepackage{hyperref}
\usepackage{amsmath}
\usepackage{graphicx}
\usepackage{bm}
\newcommand{\bfvec}[1]{\bm{\mathrm{#1}}}
\begin{document}
%
\title{Tests of the standard electroweak model in beta decay\footnote{In print in Reviews of Modern Physics (2006).}\footnote{To our teacher and collaborator, J.P.~Deutsch.} }
\author{Nathal
Severijns\footnote{Electronic address:
nathal.severijns@fys.kuleuven.be} }
\author{Marcus
Beck\footnote{Electronic address: marcus.beck1@gmx.net} }
\affiliation{Instituut voor Kern- en Stralingsfysica, Katholieke
Universiteit Leuven, B--3001 Leuven, Belgium}
\author{Oscar
Naviliat-Cuncic\footnote{Electronic address:
naviliat@lpccaen.in2p3.fr} } \affiliation{Universit\'e de Caen
Basse-Normandie and Laboratoire de Physique Corpusculaire
CNRS-ENSI, F--14050 Caen, France}
%
%
%
%
\begin{abstract}
We review the current status of precision measurements in allowed
nuclear beta decay, including neutron decay, with emphasis on
their potential to look for new physics beyond the standard
electroweak model. The experimental results are interpreted in the
framework of phenomenological model-independent descriptions of
nuclear beta decay as well as in some specific extensions of the
standard model. The values of the standard couplings and the
constraints on the exotic couplings of the general beta decay
Hamiltonian are updated. For the ratio between the axial and the
vector couplings we obtain $C_A/C_V = -1.26992(69)$ under the
standard model assumptions. Particular attention is devoted to the
discussion of the sensitivity and complementarity of different
precision experiments in direct beta decay. The prospects and the
impact of recent developments of precision tools and of high
intensity low energy beams are also addressed.
\end{abstract}
%
%
%
\maketitle
\tableofcontents
\newpage
%
%
%
%
%
\section{INTRODUCTION}
\label{sec:introhist}

Nuclear beta decay has played a crucial role in the development of
the weak interaction theory. Three experimental foundations of the
standard electroweak model, {\em i.e.} {\it (i)\,} the assumption
of maximal parity violation; {\it (ii)\,} the assumption of
massless neutrinos; and {\it (iii)\,} the vector axial-vector
character of the weak interaction have their sources in the
detailed analysis of nuclear beta decay processes. The so-called
{\em universal V-A} theory has been established from the analogy
between nuclear beta decay and muon decay. The ensuing
confrontation of the weak interaction theory, constructed at low
energies, against the results obtained at higher energies,
motivated the development of a gauge theory and constituted a
significant step which led to the construction of the unified
electroweak model.

The beta decay theory, including some of its refinements like the
induced weak currents, has been firmly established and tested more
than three decades ago and has later been embedded into the wider
framework of the standard electroweak model. Since then the main
motivations of new experiments performed at low energies, with
ever increased statistical accuracy, have been to provide
precision tests of the discrete symmetries as well as to address
specific questions involving the light quarks, which are naturally
best studied in nuclear and neutron decays.

Nuclear beta decay is a semi-leptonic strangeness-conserving
process which, at the fundamental level and to lowest order,
involves the lightest leptons ($e$,$\nu_e$) and quarks ($u$,$d$)
interacting via the exchange of the charged vector bosons
$W_L^\pm$. The number of constraints on the standard model
provided by these low energy experiments is actually limited as is
also the number of relevant standard model parameters involved in
the description of the semi-leptonic beta decay. In this sense the
{\em tests of the standard model} considered at low energies
generally refer to the tests of the underlying fundamental
symmetries rather than to tests of the consistency of the theory
or the predictions of new phenomena. The main aim of precision low
energy experiments is to find deviations from the standard model
assumptions as possible indications of new physics.

Despite the great success of the standard model, many open
questions remain such as the hierarchy of fermion masses, the
number of generations, the origin of parity violation, the
mechanism behind CP violation, the number of parameters of the
theory, etc. These are expected to find explanations in extended
and unified theoretical frameworks involving new physics.

The production of intense sources and beams of $\beta$ emitters
(nuclei and neutrons), with high purity and possibly polarized,
enables a high statistical accuracy to be reached in the
determination of the parameters which describe the weak
interaction in nuclei and in the searches for deviations from
maximal parity violation or from time reversal invariance. In
addition, the rich spectra of nuclear states and the combination
with transitions involving other emitted particles ($\alpha$,
$\gamma$, $p$) following the beta decay transition, offer a large
diversity to the design of low energy experiments and to implement
different techniques. For well selected transitions the
uncertainties associated with hadronic effects can be well
controlled such that their impact remains below the experimental
accuracy and does not affect the extraction of reliable results.

The role of beta decay experiments to test the standard model
assumptions and to look for new physics has been discussed earlier
in several papers
\cite{deutsch95,herczeg95a,towner95,vanklinken96,
yerozolimsky00,herczeg01} with emphasis on specific aspects of
this sector.  We have heavily relied on these works to prepare the
present review and we invite the reader to consult them for more
details.


This review discusses nuclear beta decay within the framework of
the standard model and beyond, with emphasis on the sensitivity of
experiments looking for new physics. The article is organized as
follows. The formalisms used for the beta decay interaction are
presented in Sec.~\ref{sec:formalism}, where several
parameterizations are reviewed and the relations between them are
discussed. The expressions of the correlation coefficients as a
function of the relevant couplings, which are used and discussed
in the following sections, are given in the appendix.
Section~\ref{sec:fit} examines the present status of the standard
theory in terms of the values of the weak couplings and the
constraints derived from the most precise data available to date.
In particular, selected results obtained over the past decade have
been included. Several assumptions on the couplings are considered
and the updated values and constraints are discussed. In
Sec.~\ref{sec:experiments} the properties and correlation
parameters accessible to beta decay experiments are reviewed. As a
given correlation parameter provides information on several
questions concerning the tests of the standard model while a given
question can be addressed by considering different observables, a
two-way approach is needed. The potential sensitivity to new
physics and the current most precise results are presented for
each measured quantity. The current experimental difficulties and
future plans and developments are discussed. The conclusions on
the present achievements and some future perspectives in the field
are summarized in Sec.~\ref{sec:conclusion}.

Several fundamental questions addressed by other decay or capture
experiments, like the test of lepton number violation and the
determination of the nature of the neutrino in neutrino-less
double beta decay experiments, are not covered in this article.
The measurements which indicate that neutrinos are massive and
oscillate and their consequences have been the subject of a number
of dedicated reviews of this very important and rapidly changing
field \citep{kajita01,jung01,bemporad02}. The status and prospects
of double beta decay experiments \citep{zdesenko02,elliot02} and
muon decay experiments \citep{kuno01} have also recently been
reviewed.

%
%
%
%
\section{FORMALISMS OF ALLOWED BETA-DECAY}
\label{sec:formalism}

The elaboration of the weak interaction theory has been retraced
in many books and reviews. The most relevant original publications
have been summarized and analyzed in different contexts
\citep{kabir63,bertin84}. Several classical texts
\citep{konopinski66,schopper66,wu66} introduce the theory with
appropriate references to the early experiments and provide also
the basics of the phenomenological description of nuclear beta
decay. More recent texts \citep{commins83,Holstein89,greiner96}
place nuclear beta decay in the context of the unified electroweak
theory.

The beta transitions are traditionally divided into {\em allowed}
and {\em forbidden}. Allowed transitions correspond to processes
in which no orbital angular momentum is carried away by the pair
of leptons. Their selection rules are:

\begin{eqnarray}
\Delta J = J_i - J_f & = & 0, \pm 1 \\
\pi_i\pi_f & = & +1
\end{eqnarray}
where $J_i$ and $\pi_i$ ($J_f$ and $\pi_f$) designate the spin and
parity of the initial (final) state. The allowed transitions can
then be subdivided into singlet and triplet components depending
on whether the lepton spins are anti-parallel ($S = 0$) or
parallel ($S = 1$). In allowed transitions the singlet state can
only arise when $\Delta J = 0$ (Fermi selection rule) whereas the
triplet state corresponds to $\Delta J = 0,\pm 1$ (Gamow-Teller
selection rule). In this last case, transitions between states of
zero angular momentum ($0 \rightarrow 0$) are excluded since it is
impossible to generate a triplet state for $\bfvec{J}_i =
\bfvec{J}_f = \bfvec{0}$.


\subsection{The V-A Theory}
\label{sec:V-A}

\begin{figure}[!htb]
\includegraphics{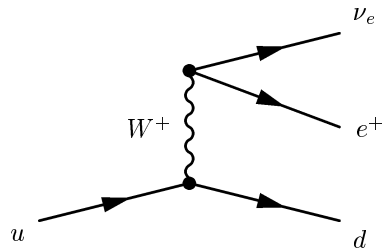}
\caption{The $\beta$-decay at the quark-lepton level, mediated by
the exchange of a weak boson.}
\label{fig:boson-exchange}
\end{figure}

The electroweak interaction is described by the standard model
\citep{salam64,weinberg67}. The symmetries of the underlying
$SU_L(2)\times U(1)$ gauge group determine the properties of the
interaction and generate the intermediate vector bosons.
Figure~\ref{fig:boson-exchange} shows a diagram of a $\beta$-decay
process described at the elementary quark-lepton level by the
exchange of a charged weak boson $W^+$.

In low-energy processes like $\beta$-decay, in which the typical
energies involved in the process are much smaller than the mass of
the weak bosons, the interaction can be described by a
four-fermion contact interaction. Such formulation was first
introduced with a vector interaction \citep{fermi34}, it was later
extended \citep{gamow36} to describe transitions which required
the introduction of other possible Lorentz invariants and it was
finally generalized \citep{feynman58,sudarshan58} as a universal
formulation of the weak interaction, incorporating the assumption
of maximal parity violation. The Hamiltonian of the {\em V-A
theory} resulting from such a four-fermion contact interaction has
the form of a current-current interaction

\begin{equation}
\label{eqn:H.V-A} {\cal H}_{V-A} = \frac{G_F}{\sqrt{2}} J_\mu^\dag
\cdot J_\mu + {\rm H.c.}
\end{equation}

\noindent where $G_F/(\hbar c)^3 = 1.16639(1)\times
10^{-5}$~GeV$^{-2}$ is the Fermi coupling and the current $J_\mu$
contains a hadronic and a leptonic contribution

\begin{equation}
\label{eqn:weak-current} J_\mu  =  J_\mu^{had} + J_\mu^{lep}
\end{equation}

\noindent The fact that the Fermi coupling has the dimension of
(mass)$^{-2}$ indicates that it cannot correspond to a fundamental
interaction strength whose value should not depend on a specific
system of units. If $g$ designates the coupling strength between
the weak boson and the fermions at each vertex of
Fig.~\ref{fig:boson-exchange} then, in the limit of low momentum
transfer, one has a simple relation between the Fermi coupling of
the {\em V-A} theory and the boson mass, $M_W$,

\begin{equation}
\label{eqn:GF} \frac{G_F}{\sqrt{2}} = \frac{g^2}{8M_W^2}
\end{equation}

\begin{figure}[!htb]
\includegraphics{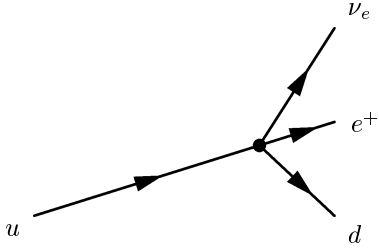}
\caption{Contact interaction of four fermions. The hadronic and
leptonic currents interact at a single point.}
\label{fig:contact-interaction}
\end{figure}

Figure~\ref{fig:contact-interaction} shows the same decay process
as in Fig.~\ref{fig:boson-exchange} but in which the four fermions
interact at a single point.

The $\beta$-decay of the neutron and of nuclei are described by a
diagram similar to Fig.~\ref{fig:contact-interaction}.

\subsection{Quark mixing}
\label{sec:qark-mixing}

The relative strength of the weak interaction in pure leptonic, in
semi-leptonic and in pure hadronic processes are not identical.
This has been incorporated into the electroweak theory by the
mechanism of quark mixing. The weak eigenstates of the quarks with
charge $-1/3$ are postulated to differ from the eigenstates of the
electromagnetic and strong interaction, which define the mass
eigenstates. In the case of the three quark families the mixing is
expressed by means of the Cabibbo-Kobayashi-Maskawa (CKM)
matrix~\citep{cabibbo63,kobayashi72}

\begin{displaymath}
\label{eqn:CKMmatrix} \left( \begin{array}{c}
d^\prime \\
s^\prime \\
b^\prime
\end{array} \right)
= \left( \begin{array}{ccc}
V_{ud} & V_{us} & V_{ub} \\
V_{cd} & V_{cs} & V_{cb} \\
V_{td} & V_{ts} & V_{tb}
\end{array} \right)
\left( \begin{array}{c}
d \\
s \\
b
\end{array} \right)
\end{displaymath}

\noindent Here the primes denote the weak eigenstates. The
normalization of the states requires the CKM-matrix to be unitary.
For the $u$- and $d$-quarks involved in nuclear $\beta$-decay, in
which the heavier quarks do not contribute to lowest order, we
have

\begin{equation}
\label{eqn:cabibbomixing} d' \simeq V_{ud} d = \cos \theta_C d
\end{equation}

\noindent where $\theta_C$ is the Cabibbo angle. The weak
interaction of the quark $d'$ introduces then the matrix element
$V_{ud}$ in the amplitude of the hadronic current.

\subsection{The general Hamiltonian}

The most general interaction Hamiltonian density describing
nuclear $\beta$-decay, including all possible interaction types
consistent with Lorentz-invariance, is given by
\citep*{lee56,jackson57a}

\begin{eqnarray}
\label{eqn:Hgeneral}
{\cal H}_{\beta} & = & \left( \bar{p} n \right) \left( \bar{e}
\left( C_S + C_S^\prime \gamma_5 \right) \nu \right)
\nonumber \\
& & + \left( \bar{p} \gamma _\mu n \right) \left( \bar{e} \gamma
_\mu \left( C_V + C_V^\prime \gamma_5 \right) \nu \right)
\nonumber \\
& & + \frac{1}{2} \left( \bar{p} \sigma _{\lambda \mu} n \right)
\left( \bar{e} \sigma _{\lambda \mu} \left( C_T + C_T^\prime
\gamma _5
\right) \nu \right) \nonumber \\
& & - \left( \bar{p} \gamma _\mu \gamma _5 n \right) \left(
\bar{e} \gamma _\mu \gamma _5 \left( C_A + C_A^\prime \gamma _5
\right) \nu \right) \nonumber \\
& & + \left( \bar{p} \gamma _5 n \right) \left( \bar{e} \gamma _5
\left( C_P + C_P^\prime \gamma _5 \right) \nu \right)
\nonumber \\
& & + ~{\rm H.c.}
\end{eqnarray}

\noindent with the tensor operator given by

\begin{eqnarray}
\sigma _{\lambda \mu} & = & - \frac{i}{2} \left( \gamma _\lambda
\gamma _\mu - \gamma _\mu \gamma _\lambda \right).
\end{eqnarray}

\noindent The interacting fields are here associated to the
nucleons and the leptons and the interactions are described by the
five operators: the scalar ${\cal O}_S = 1$, the vector ${\cal
O}_V = \gamma_\mu$, the tensor ${\cal O}_T = \sigma_{\lambda
\mu}/\sqrt{2}$, the axial-vector ${\cal O}_A = -i\gamma_\mu
\gamma_5$, and the pseudo-scalar ${\cal O}_P = \gamma_5$. The
coefficients $C_i$ and $C^\prime_i$ which appear in the leptonic
currents determine the relative amplitude of each interaction.
These amplitudes can be complex corresponding to a total of 20
real parameters\footnote{One of these coefficients can be absorbed
in the overall strength of the interaction provided it be the same
for all $\beta$ transitions. In the standard model and following
CVC (see below) one fixes $C_V = G_F V_{ud}/\sqrt{2}$.} which
determine the properties of the Hamiltonian with respect to space
inversion ($P$), charge conjugation ($C$) and time reversal ($T$)
symmetries.

The presence of both the $C_i$- and the $C_i^\prime$-coefficients
is related to the transformation properties under parity. Parity
invariance holds for either $C_i^\prime = 0$ or $C_i = 0$ and is
violated if both $C_i$ and $C_i^\prime$ are present. Maximum
parity violation corresponds to $\vert C_i \vert = \vert
C_i^\prime \vert$. Charge-conjugation invariance holds if
$Re(C_i/C_i^\prime)=0$ or $Re(C_i^\prime/C_i)=0$, {\em i.e.} if
the $C_i$ are real and the $C_i^\prime$ are purely imaginary, up
to an overall phase. When $C_i$ and $C_i^\prime$ have both a real
or both an imaginary part charge-conjugation is violated.
Time-reversal invariance holds if the $C_i$ and $C_i^\prime$ are
all real up to an overall common phase and is violated if at least
one of the couplings has an imaginary phase relative to the
others. The relations between the $C_i$-coefficients and the
symmetry properties of the interactions are summarized in
Table~\ref{tab:disc-sym}.

\begin{table}[!htb]
\begin{center}
\begin{tabular}{c@{\hspace{3mm}}@{\hspace{5mm}}l}
\hline \hline
Symmetry & Condition for violation \\
\hline
C & ($Re C_i \ne 0$ and $Re C_i^\prime \ne 0$) or\\
  & ($Im C_i \ne 0$ and $Im C_i^\prime \ne 0$)\\
P & $C_i \ne 0$ and $C_i^\prime \ne 0$ \\
T & $Im (C_i/C_j) \ne 0$ or $Im (C_i^\prime/C_j) \ne 0$\\
\hline \hline
\end{tabular}
\caption{The consequences on the couplings due to the violations
of the discrete symmetries.} \label{tab:disc-sym}
\end{center}
\end{table}

In the non-relativistic treatment of the nucleons it is easy to
show that the pseudo-scalar hadronic current, $\bar{p}\gamma_5 n$,
vanishes and therefore the pseudo-scalar term in
Eq.~(\ref{eqn:Hgeneral}) can be neglected in calculations of the
experimental observables. The scalar and vector interactions
contribute to the Fermi (F) transitions whereas the axial and
tensor interactions contribute to the Gamow-Teller (GT)
transitions.

The description of $\beta$-decay in the minimal electroweak model
involves only $V$- and $A$-interactions; parity is assumed to be
maximally violated along with charge conjugation and the effects
due to the standard $CP$ (or $T$) violation observed in the $K$
and $B$-meson systems are not expected to contribute at the
present level of experimental precision~\citep{herczeg97}. In
terms of the couplings this leads to $C_V/C_V^\prime = 1$,
$C_A/C_A^\prime = 1$, $C_S = C_S^\prime = C_T = C_T^\prime = C_P =
C_P^\prime = 0$, and $Im(C_i) = 0$ for all $i$.

In addition to the Lorentz invariants which are linear in the
fermion fields the hadronic current can involve terms which depend
on the field derivatives (``gradient'' type contributions)
associated with the hadronic structure. These are the so-called
induced weak currents~\citep{holstein74,holstein76,
grenacs85,mukhopadhyay98} and are discussed in
Sec.~\ref{sec:higherorder}. In Eq.~(\ref{eqn:Hgeneral}) it is
assumed that only one neutrino state is involved and that the
effects due to a possibly finite neutrino mass are negligible.

\subsection{Helicity projection formalism}
\label{sec:helicity-proj}

Alternative formulations of the local four-fermion interaction,
not including derivatives in the fermion fields, have been
proposed to make more explicit the helicity structure of the
interacting fermions~\citep{herczeg95a,herczeg01}. In such
formulations the most general interaction Hamiltonian, at the
quark-lepton level, involving a left-handed neutrino state
$\nu^{(L)}$ and a singlet right-handed neutrino state $\nu^{(R)}$,
can be written as\footnote{See Appendix~\ref{sec:appendix:conv}
for the conventions on the metric.}~\citep{herczeg01}

\begin{equation}
\label{eqn:H_beta} H_\beta = H_{V,A} + H_{S,P} + H_T
\end{equation}

\noindent where

\begin{eqnarray}
H_{V,A} & = & \bar{e} \gamma_\mu (1+\gamma_5) \nu^{(L)} \nonumber \\
& & \left[ a_{LL} \bar{u} \gamma_\mu (1+\gamma_5) d
+ a_{LR} \bar{u} \gamma_\mu (1-\gamma_5) d \right] \nonumber \\
& & + \bar{e} \gamma_\mu (1-\gamma_5) \nu^{(R)} \nonumber \\
& & \left[ a_{RR} \bar{u} \gamma_\mu (1-\gamma_5) d +
a_{RL} \bar{u} \gamma_\mu (1+\gamma_5)d \right] \nonumber \\
& & + {\rm H.c.} \label{eqn:Hhpf1}
\end{eqnarray}
\begin{eqnarray}
H_{S,P} & = & \bar{e} (1+\gamma_5) \nu^{(L)} \nonumber \\
& & \left[ A_{LL} \bar{u} (1+\gamma_5) d + A_{LR} \bar{u}
(1-\gamma_5) d \right]
\nonumber \\
& & + \bar{e} (1-\gamma_5) \nu^{(R)} \nonumber \\
& & \left[ A_{RR} \bar{u} (1-\gamma_5) d
+ A_{RL} \bar{u} (1+\gamma_5)d \right] \nonumber \\
& & + {\rm H.c.} \label{eqn:Hhpf2}
\end{eqnarray}
\begin{eqnarray}
H_T & = & \alpha_{LL} \bar{e} \frac{\sigma_{\lambda
\mu}}{\sqrt{2}} (1+\gamma_5)\nu^{(L)}
\bar{u} \frac{\sigma_{\lambda \mu}}{\sqrt{2}} (1+\gamma_5) d \nonumber \\
& & + \alpha_{RR} \bar{e} \frac{\sigma_{\lambda \mu}}{\sqrt{2}}
(1-\gamma_5) \nu^{(R)}
\bar{u} \frac{\sigma_{\lambda \mu}}{\sqrt{2}} (1-\gamma_5) d \nonumber \\
& & + {\rm H.c.} \label{eqn:Hhpf3}
\end{eqnarray}

The terms in Eq.~(\ref{eqn:Hhpf1}) have vector and axial-vector
interactions, those given in Eq.~(\ref{eqn:Hhpf2}) have scalar and
pseudo-scalar interactions and Eq.~(\ref{eqn:Hhpf3}) contains
tensor interactions. The first subscript of the couplings
$a_{ij}$, $A_{ij}$ and $\alpha_{ij}$ gives the chirality of the
neutrino and the second the chirality of the $d$-quark. The
neutrino states $\nu^{(L)}$ and $\nu^{(R)}$ are in general linear
combinations of the left-handed and right-handed components of the
neutrino mass-eigenstates~\citep{herczeg01}. In the standard model
all couplings are zero except $a_{LL}$ which becomes
$(a_{LL})_{SM} = G_F V_{ud} / \sqrt{2}$.

The $\beta$-decay of the nucleon due to the interaction given in
Eq.~(\ref{eqn:H_beta}) is given by~\citep{herczeg01}

\begin{equation}
\label{eqn:Hhsum} H_\beta^{(N)} \simeq H_{V,A}^{(N)} + H_S^{(N)} +
H_T^{(N)}
\end{equation}

\noindent where the pseudo-scalar contribution has been neglected
and were the three terms are

\begin{eqnarray}
H_{V,A}^{(N)} & = & \bar{e} \gamma_\mu (C_V+C_V'\gamma_5) \nu
\bar{p} \gamma_\mu n \nonumber \\
 & & - \bar{e} \gamma_\mu \gamma_5 (C_A+C_A'\gamma_5) \nu
\bar{p} \gamma_\mu \gamma_5 n
+ {\rm H.c.} \label{eqn:Hhnuc1} \\
H_{S,P}^{(N)} & = &  \bar{e} (C_S+C_S'\gamma_5) \nu
\bar{p}n + {\rm H.c.} \label{eqn:Hhnuc2} \\
H_T^{(N)} & = & \bar{e} \frac{\sigma_{\lambda \mu}}{\sqrt{2}}
(C_T+C_T'\gamma_5) \nu \bar{p} \frac{\sigma_{\lambda
 \mu}}{\sqrt{2}} n + {\rm H.c.} \label{eqn:Hhnuc3}
\end{eqnarray}

\noindent The terms of the Hamiltonian in Eq.~(\ref{eqn:Hhsum})
are hence identical to those in Eq.~(\ref{eqn:Hgeneral}). The
relations between the couplings $C_i$ and $C_i'$ which appear in
Eqs.~(\ref{eqn:Hhnuc1}-\ref{eqn:Hhnuc3}) and those in
Eqs.~(\ref{eqn:Hhpf1}-\ref{eqn:Hhpf3}) are\footnote{Here the
coefficients $C_i$ and $C_i'$ are the same as those in
Eq.~(\ref{eqn:Hgeneral}). The signs of $C_V'$, $C_A$, $C_S'$ and
$C_T'$ are here opposite to those in~\citet*{herczeg01}.}

\begin{eqnarray}
\label{eqn:cherczeg}
C_V & = & g_{V} (a_{LL} + a_{LR} + a_{RR} + a_{RL}) \label{eqn:ch1} \\
C_V^\prime & = & g_{V} (a_{LL} + a_{LR} - a_{RR} - a_{RL}) \label{eqn:ch2} \\
C_A & = & g_{A} (a_{LL} - a_{LR} + a_{RR} - a_{RL}) \label{eqn:ch3} \\
C_A^\prime & = & g_{A} (a_{LL} - a_{LR} - a_{RR} + a_{RL}) \label{eqn:ch4} \\
C_S & = & g_{S} (A_{LL} + A_{LR} + A_{RR} + A_{RL}) \label{eqn:ch5} \\
C_S^\prime & = & g_{S} (A_{LL} + A_{LR} - A_{RR} - A_{RL}) \label{eqn:ch6} \\
C_T & = & 2 g_{T} (\alpha_{LL} + \alpha_{RR}) \label{eqn:ch7} \\
C_T^\prime & = & 2 g_{T} (\alpha_{LL} - \alpha_{RR})
\label{eqn:ch8}
\end{eqnarray}

The constants $g_{i} \equiv g_{i}(0)$, $i=V,A,S,T$, are the values
of hadronic form factors in the limit of zero momentum transfer.
They are defined by~\citep{herczeg01}

\begin{eqnarray}
g_{V}(q^2) \bar{p} \gamma_\mu n & = & \langle p \vert \bar{u}
\gamma_\mu d
\vert n \rangle \label{eqn:gherczeg1}\\
g_{A}(q^2) \bar{p} \gamma_\mu \gamma_5 n & = & \langle p \vert
\bar{u}
\gamma_\mu \gamma_5 d \vert n \rangle \label{eqn:gherczeg2}\\
g_{S}(q^2) \bar{p} n & = & \langle p \vert \bar{u} d \vert n \rangle \label{eqn:gherczeg3}\\
g_{T}(q^2) \bar{p} \sigma_{\lambda \mu} n & = & \langle p \vert
\bar{u} \sigma_{\lambda \mu} d \vert n \rangle
\label{eqn:gherczeg4}
\end{eqnarray}

In the standard model $C_V=g_V\cdot (a_{LL})_{SM}$ and
$C_A=g_A\cdot (a_{LL})_{SM}$. The CVC hypothesis states that $g_V
= 1$ and in the absence of new interactions one has $g_A \approx
-1.27$ (Sec.\ref{sec:fit}). The determination of the couplings
$A_{ij}$ and $\alpha_{ij}$ from experiments requires the constants
$g_S$ and $g_T$ to be known, which can be calculated in various
quark models of the nucleons \citep{herczeg01}.

\subsection{Left-right symmetric models}
\label{sec:righthanded}

The observations of parity violation in the weak interaction is
embedded in the standard model by imposing left-handed fermions to
transform like $SU_L(2)$ doublets whereas right-handed fermions
transform as singlets. Extensions based on wider gauge symmetry
groups, have been proposed to provide a natural framework for the
understanding of the breaking of the left-right symmetry observed
in weak interactions
\citep{pati73,pati74,mohapatra75,senjanovic75,beg77}. The simplest
left-right symmetric models are based on the gauge group $SU(2)_L
\times SU(2)_R \times U(1)$, in which, in addition to the
transformations under $SU_L(2)$ above, the right-handed fermions
transform now as doublets under $SU_R(2)$ whereas the left-handed
ones transform as singlets.

The gauge symmetry of these models introduces additional bosons.
The mass eigenstates of the predominantly left-handed bosons are
denoted $W_1$ and $Z_1$ whereas those of the additional
predominantly right-handed bosons are denoted $W_2$ and $Z_2$. The
weak eigenstates, $W_L$ and $W_R$, are linear combinations of the
mass eigenstates

\begin{eqnarray}
\label{eqn:defzeta}
W_L & = & W_1 cos{\zeta} + W_2 sin{\zeta} \\
W_R & = & e^{i \omega } ( -W_1 sin{\zeta} + W_2 cos{\zeta} ).
\end{eqnarray}

\noindent where $\zeta$ is a mixing angle and $\omega$ is a
CP-violating phase. The couplings of the weak bosons to the quarks
and leptons of the first generation is given by \citep{herczeg01}

\begin{eqnarray}
\label{eqn:Lleftright} {\cal L}_{LR} = (g_L/\sqrt{2})
\left(\bar{u}_L \gamma_\mu V_{ud}^L d_L +
\bar{\nu}_{Li} \gamma_\mu U_{ie}^L e_L \right) W_L \nonumber \\
(g_R/\sqrt{2}) \left( \bar{u}_R \gamma_\mu V_{ud}^R d_R +
\bar{\nu}_{Rj} \gamma_\mu U_{je}^R e_R \right) W_R
\end{eqnarray}

\noindent where $g_L$ and $g_R$ are the gauge couplings associated
with $SU_L(2)$ and $SU_R(2)$ respectively and $V_{ud}^L$,
$V_{ud}^R$, $U_{ie}^L$ and $U_{je}^R$ are elements of mixing
matrices for quarks and leptons which are relevant to the first
generation. The interaction given in Eq.~(\ref{eqn:Lleftright})
contains only vector terms and it is seen to be invariant under
left-right symmetry.

At the level of nucleons, the Hamiltonian which describes nuclear
$\beta$-decay resulting from Eq.~(\ref{eqn:Lleftright}) contains
$V$- and $A$-interactions, as in Eq.~(\ref{eqn:Hhpf1}). The
relation between the fundamental parameters of
Eq.~(\ref{eqn:Lleftright}) and the effective couplings in
Eq.~(\ref{eqn:Hhpf1}) are \citep{herczeg01}

\begin{eqnarray}
a_{LL} & \simeq & {g_L^2 V_{ud}^L}/({8 m_1^2}) \\
a_{RR} & \simeq & a_{LL} (V_{ud}^R/V_{ud}^L) (g_R^2/g_L^2)~\delta \\
a_{LR} & \simeq & -a_{LL} e^{i\omega} (V_{ud}^R/V_{ud}^L)
(g_R/g_L)~\zeta \\
a_{RL} & \simeq & -a_{LL} e^{i\omega} (g_R/g_L)~\zeta
\end{eqnarray}

\noindent where $\delta = (m_1/m_2)^2$, with $m_1$ ($m_2$) the
mass of the $W_1$ ($W_2$) boson.

In the simple limit of so-called manifest left-right symmetry, in
which $g_R = g_L$, $V_{ud}^R = V_{ud}^L$ and $\omega = 0$ one has
$a_{RR} = \delta\cdot a_{LL}$ and $a_{LR} = a_{RL} = -\zeta\cdot
a_{LL}$. Substituting these in Eqs.~(\ref{eqn:ch1}-\ref{eqn:ch4})
results in

\begin{eqnarray}
\label{eqn:Cs-vs-rhc} C_V & = & g_{V} a_{LL} (1 -2\zeta + \delta)
\label{eqn:CvsRHC1} \\
C_V^\prime & = & g_{V} a_{LL} (1 - \delta)
\label{eqn:CvsRHC2} \\
C_A & = & g_{A} a_{LL} (1 + 2\zeta + \delta)
\label{eqn:CvsRHC3} \\
C_A^\prime & = & g_{A} a_{LL} (1 - \delta) \label{eqn:CvsRHC4}
\end{eqnarray}

Comparing $C_i$ with $C_i'$ it appears that, in the limit of no
mixing ($\zeta \rightarrow 0$), parity violation arises solely
from the difference between the masses of $W_1$ and $W_2$.

\subsection{Leptoquark exchange}
\label{sec:leptoquark}

Leptoquarks are bosons which couple to quark-lepton pairs. As such
they carry lepton numbers, baryon numbers and fractional charges.
Only spin-zero (scalar) and spin-one (vector) leptoquarks
occur~\citep{herczeg01}.

\begin{figure}[!htb]
\includegraphics{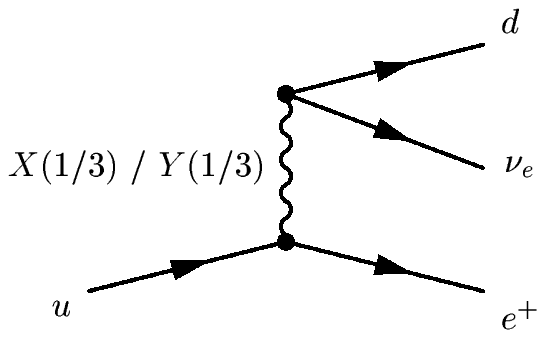}
\includegraphics{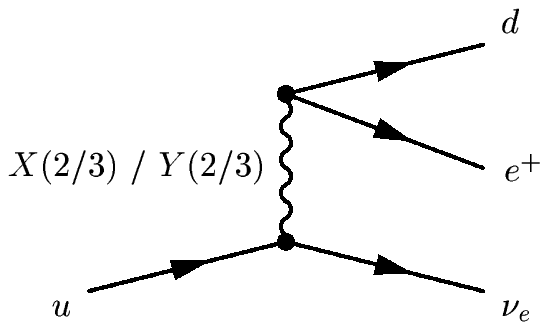}
\caption[Leptoquark interactions]{Leptoquark exchanges
contributing to nuclear $\beta$-decay. Top: $|Q| = 1/3$
leptoquarks; bottom: $|Q| = 2/3$ leptoquarks.}
\label{fig:leptoquark}
\end{figure}

Transitions in nuclear $\beta$-decay can be mediated by
leptoquarks with charges $|Q| = 2/3$ and $|Q| = 1/3$. Figure
\ref{fig:leptoquark} illustrates possible decay channels mediated
by vector and scalar leptoquarks, denoted $X$ and $Y$
respectively.

The four fermion interaction generated by the exchange of
$X_{|Q|}$ and $Y_{|Q|}$ leptoquarks has the form
\citep{herczeg95a}\footnote{See Appendix~\ref{sec:appendix:conv}
for the conventions on the metric.}

\begin{eqnarray}
H_{X(2/3)} & = & \bar{u}\gamma_{\mu}(1+\gamma_5)\nu_e^L \nonumber \\
 & & \times [f_{LL}\bar{e} \gamma_{\mu}(1+\gamma_5) d
     + f_{LR}\bar{e}\gamma_{\mu}(1-\gamma_5)d] \nonumber \\
 & & + \bar{u}\gamma_{\mu}(1-\gamma_5)\nu_e^R \nonumber \\
 & & \times [f_{RL}\bar{e} \gamma_{\mu}(1+\gamma_5) d
     + f_{RR}\bar{e}\gamma_{\mu}(1-\gamma_5) d] \nonumber \\
 & & + {\rm H.c.} \label{eqn:lepto1}
\end{eqnarray}
\begin{eqnarray}
H_{X(1/3)} & = & \bar{d}^c\gamma_{\mu}(1+\gamma_5)\nu_e^L \nonumber \\
 & & \times [h_{LL}\bar{e} \gamma_{\mu}(1+\gamma_5) u^c
     +  h_{LR}\bar{e}\gamma_{\mu}(1-\gamma_5)u^c] \nonumber \\
 & & +\bar{d}^c\gamma_{\mu}(1-\gamma_5)\nu_e^R \nonumber \\
 & & \times [h_{RL}\bar{e} \gamma_{\mu}(1+\gamma_5) u^c
     + h_{RR}\bar{e}\gamma_{\mu}(1-\gamma_5)u^c] \nonumber \\
 & & + {\rm H.c.} \label{eqn:lepto2}
\end{eqnarray}
\begin{eqnarray}
H_{Y(2/3)} & = & \bar{u}(1+\gamma_5)\nu_e^L \nonumber \\
 & & \times [F_{LL}\bar{e} (1+\gamma_5) d
     + F_{LR}\bar{e} (1-\gamma_5)d] \nonumber \\
 & & +\bar{u}(1-\gamma_5)\nu_e^R \nonumber \\
 & & \times[F_{RL}\bar{e}(1+\gamma_5) d
     + F_{RR}\bar{e}(1-\gamma_5)d] \nonumber \\
 & & + {\rm H.c.} \label{eqn:lepto3}
\end{eqnarray}
\begin{eqnarray}
H_{Y(1/3)} & = &
  \bar{d}^c(1+\gamma_5)\nu_e^L \nonumber \\
 & & \times[H_{LL}\bar{e} (1+\gamma_5) u^c
     + H_{LR}\bar{e}(1-\gamma_5)u^c] \nonumber \\
 & & + \bar{d}^c(1-\gamma_5)\nu_e^R \nonumber \\
 & & \times [H_{RL}\bar{e} (1+\gamma_5) u^c
     + H_{RR}\bar{e}(1-\gamma_5)u^c] \nonumber \\
 & & + {\rm H.c.} \label{eqn:lepto4}
\end{eqnarray}

The first subscript of the couplings indicates the neutrino
chirality and the second indicates the chirality of the fourth
fermion in the coupling. This Hamiltonian can be transformed to
the four fermion interaction of the form given in
Eq.~(\ref{eqn:Hgeneral}) by a Fierz transformation~\citep[see {\em
e.g.}][]{greiner96}. The relation between the coefficients $C_i$
and $C_i'$ in Eq.~(\ref{eqn:Hgeneral}) and the couplings in
Eqs.~(\ref{eqn:lepto1}-\ref{eqn:lepto4}) resulting from the
exchange of $X_{|Q|}$ and $Y_{|Q|}$ leptoquarks are summarized in
Tables \ref{tab:CforXlq} and \ref{tab:CforYlq} \citep{herczeg95a}.

\begin{table}[!htb]
\begin{center}
\begin{tabular}{l@{\hspace{7mm}}l@{\hspace{5mm}}l}
\hline \hline
 & X(2/3) & X(1/3)\\
\hline
 $C_V$ & $g_{V}(f_{LL}+f_{RR})$ & $g_{V}(-h_{LL}-h_{RR})$ \\
 $C_V'$ & $g_{V}(f_{LL}-f_{RR})$ & $g_{V}(-h_{LL}+h_{RR})$ \\
 $C_A$ & $g_{A}(f_{LL}+f_{RR})$ & $g_{A}(h_{LL}+h_{RR})$ \\
 $C_A'$ & $g_{A}(f_{LL}-f_{RR})$ & $g_{A}(h_{LL}-h_{RR})$ \\
 $C_S$ & $2g_{S}(-f_{LR}-f_{RL})$ & $2g_{S}(-h_{LR}+h_{RL})$ \\
 $C_S'$ & $2g_{S}(-f_{LR}+f_{RL})$ & $2g_{S}(-h_{LR}-h_{RL})$ \\
 $C_T$ & $0$ & $0$ \\
 $C_T'$ & $0$ & $0$ \\
\hline \hline
\end{tabular}
\caption{\label{tab:CforXlq} Coefficients resulting from vector
leptoquark exchange.}
\end{center}
\end{table}

\begin{table}[!htb]
\begin{center}
\begin{tabular}{l@{\hspace{7mm}}l@{\hspace{5mm}}l}
\hline \hline
    & Y(2/3) & Y(1/3)\\
\hline
 $2C_V$ & ${g_{V}}(-F_{LR}-F_{RL})$ & ${g_{V}}(- H_{LR} - H_{RL})$ \\
 $2C_V'$ & ${g_{V}}(-F_{LR}+F_{RL})$ & ${g_{V}}( -H_{LR} + H_{RL})$ \\
 $2C_A$ & ${g_{A}}(F_{LR}+F_{RL})$ & ${g_{A}}( -H_{LR} -H_{RL})$ \\
 $2C_A'$ & ${g_{A}}(F_{LR}-F_{RL})$ & ${g_{A}}( -H_{LR} + H_{RL})$ \\
 $2C_S$ & ${g_{S}}(-F_{LL}-F_{RR})$ & ${g_{S}}( -H_{LL} - H_{RR})$ \\
 $2C_S'$ & ${g_{S}}(-F_{LL}+F_{RR})$ & ${g_{S}}( -H_{LL}+H_{RR})$ \\
 $2C_T$ & ${g_{T}}(-F_{LL}-F_{RR})$ & ${g_{T}}( H_{LL} + H_{RR})$ \\
 $2C_T'$ & ${g_{T}}(-F_{LL}+F_{RR})$ & ${g_{T}}( H_{LL} - H_{RR})$ \\
\hline \hline
\end{tabular}
\caption{\label{tab:CforYlq} Coefficients resulting from scalar
leptoquark exchange.}
\end{center}
\end{table}

Notably, vector leptoquarks can generate $V$, $A$ and $S$
interactions whereas scalar leptoquarks can in addition generate
$T$ interactions.

\subsection{Higher-order corrections}
\label{sec:higherorder}

Additional effects may become important when the precision of the
measurements reaches a level below $10^{-2}$ to $10^{-3}$. Such
effects are called generically higher-order corrections and can
have various sources like the possible presence of forbidden
matrix elements due to the breakdown of the allowed approximation,
the induced weak currents due to the hadronic structure of the
nucleons and the radiative corrections of higher order. Other
effects like the finite mass of the recoiling nucleus can affect
some observables like the shape of the energy spectrum of
electrons \citep[]{holstein74,holstein76} and can be of importance
for specific experiments. Among the higher-order effects we focus
briefly here on the induced weak currents due to their role to
establish and test some of the symmetries of the weak interaction.

The fact that the strength of the weak interaction between quarks
is not the same as in muon decay is further complicated in hadrons
due to the presence of the strong interaction. In semi-leptonic
processes this gives rise to the induced weak currents which can
be observed by the departures of the experimental properties from
their leading order description.

The structure of the vector and axial vector hadronic currents,
consistent with Lorentz invariance and including recoil terms, has
the general form \citep{weinberg58,goldberger58,fujii59}

\begin{eqnarray}
\label{eqn:inducedV}
 V^h_\mu & = & \bar{p} \biggl[ g_V(q^2) \gamma _\mu \nonumber \\
 & & \left. + f_M(q^2) \sigma _{\mu \nu } \frac{q_\nu }{2M} + if_S(q^2)
 \frac{q_\mu }{m_e} \right] n
\end{eqnarray}
\begin{eqnarray}
\label{eqn:inducedA}
 A^h_\mu & = & \bar{p} \biggl[ g_A(q^2) \gamma _\mu \gamma _5 \nonumber \\
 & & \left. + f_T(q^2) \sigma _{\mu \nu} \gamma _5 \frac{q_\nu }{2M}
+ i f_P(q^2) \frac{q_\mu}{m_e} \gamma _5 \right] n
\end{eqnarray}

\noindent where $q_\mu = (p_i - p_f)_\mu$ is the four-momentum
transfer and $M$ and $m_e$ are respectively the nucleon and the
electron mass. The form factors $g_V$, $g_A$, $f_i$ ($i = M, S, T,
P$) are arbitrary functions of the Lorentz scalar $q^2$. The
values of these form factors in the limit of zero momentum
transfer, $q^2 \rightarrow 0$, are called the vector,
axial-vector, weak magnetism, induced scalar, induced tensor, and
induced pseudo-scalar couplings respectively. In particular $g_V =
g_V(0)$ and $g_A = g_A(0)$ are the leading-order $V$ and $A$
couplings whereas the other terms are the induced weak currents.

The study of the symmetries of the induced currents introduced the
concept of $G$-parity. A $G$-parity transformation is defined by a
charge conjugation operation followed by a rotation by $\pi$
around the $y$-axis in isospin space

\begin{equation}
G =  C e^{i\pi T_2}
\end{equation}

This transformation is a symmetry of the strong interaction and it
is interesting to study the properties of the terms in the
currents above under $G$ such as to determine, at least at the
phenomenological level, whether all the terms allowed by Lorentz
invariance are dynamically possible. By definition, vector
currents with $G$-parity $+1$ and axial currents with $G$-parity
$-1$ are called {\em first class} currents whereas those with the
opposite parities are called {\em second class} currents (SCC)
\citep{weinberg58}.

The dominant vector and axial currents, the weak magnetism and the
induced pseudo-scalar belong to the first class whereas the
induced scalar and the induced tensor are second class. The
requirement that the hadronic $V$ and $A$ currents have a definite
$G$-parity implies that SCC cannot exist, hence $f_S(q^2)=0$ and
$f_T(q^2)=0$. Such requirement appears in the elaboration of a
unified electroweak theory \citep{weinberg58} and was a very
strong motivation for the search of SCC. The tests performed so
far indicate that the strengths of SCC are consistent with zero
\citep{grenacs85,towner95}.

The fact that the electromagnetic current between nucleons
exhibits a similar isospin structure as the strangeness-conserving
weak vector current led \citet*{feynman58} to postulate that these
currents formed a multiplet of vector current operators. This was
the first significant step toward a formal unification of the weak
and electromagnetic interactions and a precursor of the
$SU_L(2)\times U(1)$ gauge theory of electroweak interactions. One
of the consequences of this hypothesis is the conservation of the
vector current (CVC) with the result that $g_V(q^2)=1$,
independently of the nucleus. In other words, the vector coupling
constant is not renormalized in the nuclear medium leading to the
``universality'' of the weak vector current. Another consequence
of CVC is that, for $\beta$-transitions between analogue states,
the weak magnetism form factor is related to the difference
between the anomalous isovector magnetic moments of the respective
nuclear states, while for non-analogue transitions it is related
to the corresponding isovector M1 $\gamma$-decay rate.

The axial-vector current has no electromagnetic analogue and is
not a conserved current. However, the hypothesis of a partially
conserved axial current (PCAC) has been introduced as a valid
symmetry in the limit of the pion mass tending to zero. One of the
consequences of PCAC relates the induced pseudo-scalar form factor
to the axial-vector form factor. In semi-leptonic processes, the
pseudo-scalar coupling is multiplied by the mass of the charged
lepton in the expression of the transition rate. For processes
such as nuclear $\beta$-decay the pseudo-scalar term gives a
negligible contribution whereas in muon capture the contribution
is enhanced due to the larger muon mass. Muon capture processes
have then served for the determination of the pseudo-scalar
couplings \citep{gorringe04} and the tests of PCAC.

In summary, the study of the induced weak currents in nuclei has
contributed to the establishment of relations between the
symmetries of the electromagnetic currents and those of the weak
currents and as such has played a crucial role in identifying and
testing the symmetry structure of the electroweak theory. The
three main consequences of the symmetry properties are {\em i)}
the conserved vector current (CVC); {\em ii)} the partially
conserved axial current (PCAC); {\em iii)} the absence of second
class currents (SCC). They provide additional tests of the
standard electroweak model in nuclear $\beta$-decay. The
experimental constraints on the possible existence of SCC as well
as the status of the tests of CVC and PCAC have been reviewed by
\citet{grenacs85}, \citet{towner95} and \citet{hardy05b}, and are
discussed in Sec.~\ref{sec:ftvalues} and Sec.~\ref{sec:CVC_SCC}.

\subsection{Correlation coefficients}
\label{sec:CorrCoeff}

The coupling constants $C_i$ and $C_i^\prime$, which determine the
dynamics of $\beta$-decay, have to be determined from experiments.
\citet*{jackson57b} calculated several decay rate distributions
from the general Hamiltonian, Eq.~(\ref{eqn:Hgeneral}), for
allowed transitions including Coulomb corrections. The
distribution in the electron and neutrino directions and in the
electron energy from oriented nuclei is given by

\begin{eqnarray}
\label{eqn:jtw1} \lefteqn{ \omega(\langle \bfvec{J} \rangle \vert
E_e, \Omega _e, \Omega_\nu)
dE_e d\Omega _e d\Omega _\nu = } & & \nonumber \\
& & \frac{F(\pm Z, E_e)}{(2\pi )^5} p_e E_e (E_0 - E_e)^2 dE_e
d\Omega _e
d\Omega _\nu \times \nonumber \\
& & \frac{1}{2} \xi \left\{ 1 + a \frac{\bfvec{p}_e \cdot
\bfvec{p}_\nu}{E_e E_\nu} +
b \frac{m}{E_e} \nonumber \right. \\
 & & \left. + c \left[ \frac{\bfvec{p}_e \cdot \bfvec{p}_\nu}{3 E_e E_\nu} -
\frac{(\bfvec{p}_e \cdot \bfvec{j}) (\bfvec{p}_\nu \cdot
\bfvec{j})}{E_e E_\nu} \right] \left[ \frac{J(J+1) - 3 \langle
\bfvec{J}
\cdot \bfvec{j} \rangle ^2}{J (2J-1)} \right] \right. \nonumber \\
& & \left. + \frac{\bfvec{J}}{J} \cdot \left[ A
\frac{\bfvec{p}_e}{E_e} + B \frac{\bfvec{p}_\nu}{E_\nu} + D
\frac{\bfvec{p}_e \times \bfvec{p}_\nu}{E_e E_\nu} \right]
\right\}
\end{eqnarray}

The distribution in electron and neutrino direction and electron
polarization from non-oriented nuclei is given by

\begin{eqnarray}
\label{eqn:jtw2} \lefteqn{ \omega(\bfvec{\sigma} \vert E_e, \Omega
_e, \Omega_\nu)
dE_e d\Omega _e d\Omega _\nu = } & & \nonumber \\
& & \frac{F(\pm Z, E_e)}{(2\pi )^5} p_e E_e (E_0 - E_e)^2 dE_e
d\Omega _e
 d\Omega _\nu \times \nonumber \\
& & \frac{1}{2}\xi \left\{ 1 + a \frac{\bfvec{p}_e \cdot
\bfvec{p}_\nu}{E_e E_\nu} + b \frac{m}{E_e} + \bfvec{\sigma} \cdot
\left[ G \frac{\bfvec{p}_e}{E_e} +
H \frac{\bfvec{p_\nu}}{E_\nu} \right. \right. \nonumber \\
& & \left. \left. + K \frac{\bfvec{p}_e}{E_e + m}
\frac{\bfvec{p}_e \cdot \bfvec{p}_\nu}{E_e E_\nu} + L
\frac{\bfvec{p}_e \times \bfvec{p}_\nu}{E_e E_\nu} \right]
\right\}
\end{eqnarray}

The distribution in the electron energy and angle and in the
electron polarization from oriented nuclei is given by

\begin{eqnarray}
\label{eqn:jtw3} \lefteqn{ \omega(\langle \bfvec{J} \rangle ,
\bfvec{\sigma} \vert E_e, \Omega _e)
dE_e d\Omega _e = } & & \nonumber \\
& & \frac{F(\pm Z, E_e)}{(2\pi )^4} p_e E_e (E_0 - E_e)^2 dE_e
d\Omega _e
 \times \nonumber \\
& & \xi \left\{ 1 + b \frac{m}{E_e} + \frac{\bfvec{p}_e}{E_e}
\cdot \left( A \frac{\langle \bfvec{J} \rangle}{J} + G
\bfvec{\sigma } \right) +
 \bfvec{\sigma} \cdot \left[ N \frac{\langle \bfvec{J} \rangle}{J} \right.
 \right. \nonumber \\
& & \left. \left. + Q \frac{\bfvec{p}_e}{E_e + m} \left(
\frac{\langle \bfvec{J} \rangle}{J} \cdot \frac{\bfvec{p}_e}{E_e}
\right) + R \frac{\langle \bfvec{J} \rangle}{J} \times
\frac{\bfvec{p}_e}{E_e} \right] \right\}
\end{eqnarray}

In Eqs.~(\ref{eqn:jtw1}-\ref{eqn:jtw3}), $E_e$, $p_e$, and
$\Omega_e$ denote the total energy, momentum, and angular
coordinates of the $\beta$-particle and similarly for the
neutrino; $\langle \bfvec{J} \rangle$ is the nuclear polarization
of the initial nuclear state with spin $\bfvec{J}$; \bfvec{j} is a
unit vector in the direction of \bfvec{J}; $E_0$ is the total
energy available in the transition; $m$ is the electron rest mass;
$F(\pm Z, E_e)$ is the Fermi-function and $\bfvec{\sigma}$ is the
spin vector of the $\beta$-particle.  The upper (lower) sign
refers to $\beta^-$ ($\beta^+$)-decay. The $a$, $b$, $c$, $A$,
$B$, etc., are the correlation coefficients, the most relevant
ones being listed in Appendix~\ref{sec:appendix:coeffs}. The
coefficients $c$, $H$, $K$ and $L$ are mentioned here for
completeness but are of no practical importance since there are no
precise measurements of them relevant to test the weak
interaction.

In decays leading to an intermediate unstable state in the
daughter nucleus followed by $\gamma$, $\alpha$ or proton
emission, the delayed particle carries part of the information of
the decay according to Eqs.~(\ref{eqn:jtw1}-\ref{eqn:jtw3}). These
decays can also serve to determine the correlation coefficients
\cite{holstein74,holstein76}.

For a given correlation coefficient complementary information can
be extracted from both the leading term and from its Coulomb
correction (terms of order $\alpha Z$). For pure Fermi or pure
Gamow-Teller transitions the correlation coefficients become
independent of the nuclear matrix elements avoiding the need to
accurately know the details of the nuclear structure.

%
%
%
%
\section{STATUS OF THE V-A THEORY}
\label{sec:fit}

The determination of the couplings which enter the general
$\beta$-decay Hamiltonian can be performed by considering accurate
experimental results from correlation measurements in allowed
transitions. The two most general analyses performed so far
\citep{paul70,boothroyd84} have determined to which extent the
presence of non-standard couplings were excluded by the
experimental data. Both analyses were consistent with the {\it
V-A} theory but allowed substantial deviations from it. Other
adjustments realized later were either less general
\citep{deutsch95}, excluded explicitly scalar and tensor
contributions \citep{carnoy92}, or were limited to some specific
decays
\citep{abele,towner03}.

In this section we present a new least-squares adjustment with the
aim to update the status of the phenomenological {\it V-A}
description of nuclear $\beta$-decay. The analysis is similar to
the one performed by \citet{boothroyd84} with some modifications
explained below. The inclusion of new experimental data with high
precision improves significantly the determination of the standard
couplings and the constraints on the exotic couplings.

\subsection{General Assumptions}

It is assumed that all transitions considered in this analysis can
be described in the allowed approximation. The most general
Hamiltonian describing nuclear $\beta$-decay is given in
Eq.~(\ref{eqn:Hgeneral}), which includes all possible Lorentz
invariant operators (scalar, vector, axial-vector, tensor and
pseudo-scalar). In this description, neutrinos are assumed to be
massless, the interaction is considered to be local and to involve
the fermion fields linearly. As indicated above, the pseudo-scalar
contribution cancels in the non-relativistic description of the
nucleons so that it is neglected from the expressions of the
correlation parameters in allowed transitions \citep{jackson57b}.

We will not restrict here the couplings $C_i$ and $C'_i$ to be all
real, as has been the case for previous general analyses
\citep{paul70,boothroyd84} although we consider such case as a
particular framework.

The expressions of the correlation coefficients which are
accessible to experiments can be calculated from the $\beta$-decay
Hamiltonian, Eq.~(\ref{eqn:Hgeneral}), and can be expressed as
functions of the coupling constants and the nuclear matrix
elements \citep{jackson57b}. Those of the parameters considered
here are presented in the Appendix~\ref{sec:appendix:coeffs}. In
the fits discussed below, the expressions of the correlation
parameters $a$, $A$, $B$, $G$, $D$ and $R$ have been divided by
the term $(1 + b \langle W^{-1} \rangle)$, where $W$ is the total
energy of the beta particle. In particular this has also been
applied to the angular correlation coefficient $a$ and not only to
parameters resulting from measurements of asymmetries.

We have considered the following experimental inputs: the Fierz
interference term $b_F$ from the ${\cal F}t$-values of the
super-allowed $0^+\rightarrow 0^+$ transitions; the lifetime of the
neutron, $\tau_n$; the electron-neutrino angular correlation, $a$;
the Fierz interference term, $b$; the angular distribution of
electrons from polarized neutrons or from polarized nuclei, $A$;
the angular distribution of neutrinos from polarized neutrons,
$B$; the electron longitudinal polarization, $G$ in units of
$v/c$; the ratio between the longitudinal polarizations of
electrons emitted from pure Fermi and pure Gamow-Teller
transitions, $P_F/P_{GT}$; the ratio between the polarizations of
electrons emitted from polarized nuclei along two directions
relative to the nuclear spin, $P^-/P^+$; the ratio between the
polarizations of electrons emitted from polarized and unpolarized
nuclei, $P^-/P^0$; the time reversal violating triple correlation
coefficients, $D$ and $R$.

The ${\cal F}t$-values of super-allowed transitions
and the neutron lifetime depend both on the weak interaction
coupling $G_F V_{ud}/\sqrt{2}$ but not so their ratio. The neutron
lifetime was then expressed in a ratio relative to the ${\cal F}t$
of super-allowed transitions. The average ${\cal F}t$-values used
in the calculations are: ${\cal F}t = (3073.5\pm 1.2)$~s for the
fits in which the assumptions result in $b_F = 0$~\cite{hardy05b}
or ${\cal F}t = (3072.5\pm 2.2)$~s for the fits where $b_F$ can be
non zero~\cite{towner05}.

\subsection{Least-squares method}

The expressions of the correlation parameters depend non-linearly
on a set of $M$ parameters $a_k, k=1,\dots ,M$. Given the model
functions $y(x,$\boldmath$a$\unboldmath$)$ one defines the merit
function $\chi^2$ which is minimized to determine the best-fit
parameters. The $\chi^2$ merit function is defined by

\begin{equation}
\chi^2 = \sum_{i=1}^{N} \left[ \frac{y_i - y(x_i, a)}{\sigma_i}
\right]^2
\end{equation}

In the present case $x_i$ is just a label for the input
measurement $i$; $y_i$ is the measured value and $\sigma_i$ is the
corresponding (1$\sigma$) experimental error. The functions
$y(x,$\boldmath$a$\unboldmath$)$ correspond to the theoretical
expressions of the correlation parameters given in the Appendix
\ref{sec:appendix:coeffs}. The parameters
\boldmath$a$\unboldmath\, are defined below as ratios of the
different couplings $C_i$ and $C'_i$. The principle of non-linear
$\chi^2$ minimization can be found elsewhere \citep{eadie71}. For
the present adjustment we have used the Levenberg-Marquardt method
which has become a standard for non-linear least-squares
algorithms \citep{numrecCpp}.

\subsection{Selection of data}

The experimental data used in the least-squares adjustment is
given in Tables~\ref{tab:neutron} and \ref{tab:nuclear}.

Table~\ref{tab:neutron} contains only data from neutron decay and
the columns give respectively: the measured parameter,
the experimental value, the experimental error (1$\sigma$) on the
value, an estimate of the average $\langle W^{-1} \rangle$,
where $W$ the $\beta$ particle total energy
in units of $m_e c^2$, and the reference for the quoted
value. The parameter $\lambda$ in this table is defined as
$\lambda = (A - B)/(A + B)$.

Table~\ref{tab:nuclear} contains data from pure
Fermi and pure Gamow-Teller transitions. The columns list
respectively: the parent nucleus in the transition, the atomic
number $Z$ of the daughter nucleus, the initial ($J$) and final
($J'$) spins of the transition, the transition type, the measured
parameter, the experimental value, the experimental
error on the value, an estimate of the average $\langle W^{-1} \rangle$
and the reference for the quoted value. For
the relative beta longitudinal polarization measurements, the
values listed in Table~\ref{tab:nuclear} are not directly the
values of the measured polarization ratios but the ratio between
the experimental result (which is a ratio between longitudinal
polarizations) and its corresponding value expected within the
{\em V-A} theory. This is of no concern for the ratio $P_F/P_{GT}$
obtained with unpolarized nuclei, in which case the expected ratio
within the {\em V-A} theory is unity, but applies to the other two
cases where the measured ratios depend on the experimental
conditions.

\begin{table*}
\caption{Experimental data from neutron decay used in the
least-squares fits}
\begin{tabular}{
c@{\hspace{3mm}} r@{\hspace{8mm}} r@{\hspace{5mm}}
c@{\hspace{5mm}} l@{\hspace{3mm}} }
%
\hline\hline
parameter & value~~ & error~~ & $\langle W^{-1} \rangle$ & ~~Reference
\\
\hline
%
%
%
\\
$a$ & $-0.0910$ & 0.0390 &  0.604 &   \citet{grigoriev68}\\
    & $-0.1017$ & 0.0051 &  0.655 &   \citet{stratowa78}\\
    & $-0.1054$ & 0.0055 &  0.655 &   \citet{byrne02}\\
%
\\
$A$ &  $-0.1040$ & 0.0110 &  0.716 &   \citet{krohn75}\\
    &  $-0.1160$ & 0.0110 &  0.537 &   \citet{krohn75}\\
    &  $-0.1200$ & 0.0100 &  0.594 &   \citet{erozolimsky79}\\
    &  $-0.1140$ & 0.0120 &  0.724 &   \citet{erozolimsky79}\\
    &  $-0.1120$ & 0.0062 &  0.561 &   \citet{erozolimsky79}\\
    &  $-0.1146$ & 0.0019 &  0.581 &   \citet{bopp86}\\
    &  $-0.1189$ & 0.0012 &  0.534 &   \citet{abele97}\\
    &  $-0.1160$ & 0.0015 &  0.582 &   \citet{liaud97}\\
    &  $-0.1135$ & 0.0014 &  0.558 &   \citet{yerozolimsky97}\\
    &  $-0.1189$ & 0.0008 &  0.534 &   \citet{abele02}\\
%
\\
$B$ &   0.9950 & 0.0340 &  0.655 &   \citet{erozolimsky70}\\
    &   0.9894 & 0.0083 &  0.554 &   \citet{kuznetsov95}\\
    &   0.9801 & 0.0046 &  0.594 &   \citet{serebrov98}\\
%
\\
$\lambda$ &  $-1.2686$ & 0.0047 &  0.581 &   \citet{mostovoi01}\\
%
\\
$\tau_n$ & 891.00 &  9.00 &  0.655  &  \citet{spivak88}\\
         & 877.00 & 10.00 &  0.655  &  \citet{paul89}\\
         & 887.60 &  3.00 &  0.655  &  \citet{mampe89}\\
         & 888.40 &  3.30 &  0.655  &  \citet{nesvizhevsky92}\\
         & 882.60 &  2.70 &  0.655  &  \citet{mampe93}\\
         & 889.20 &  4.80 &  0.655  &  \citet{byrne96}\\
         & 885.40 &  1.00 &  0.655  &  \citet{arzumanov00}\\
         & 886.80 &  3.40 &  0.655  &  \citet{dewey03}\\
         & 878.50 &  0.76 &  0.655  &  \citet{serebrov05a}\\
%
\\
$D$ &  $-0.00270$ & 0.00500 &  0.655  &      \citet{erozolimsky74}\\
    &  $-0.00110$ & 0.00170 &  0.650  &      \citet{steinberg76}\\
    &  $ 0.00220$ &  0.00300 &  0.619  &    \citet{erozolimsky78}\\
    &  $-0.00060$ &  0.00130 &  0.655  &    \citet{lising00}\\
    &  $-0.00024$ &  0.00071 &  0.602  &    \citet{soldner04}\\
\\
\hline\hline
\end{tabular}
\label{tab:neutron}
\end{table*}
\begin{table*}
\caption{Data from measurements in nuclear decays used in the
least-squares fits}
\begin{tabular}{
c@{\hspace{5mm}} r@{\hspace{8mm}} c@{\hspace{8mm}}
c@{\hspace{8mm}} c@{\hspace{3mm}} c@{\hspace{3mm}}
r@{\hspace{8mm}} r@{\hspace{8mm}} c@{\hspace{5mm}}
l@{\hspace{3mm}} }
%
\hline\hline
isotope & $Z$ & $J$ & $J'$ & type~~ & parameter & value~~ &
error~~ & $\langle W^{-1} \rangle$ & ~~Reference
\\
\hline
%
%
%
\\
$^6$He & 3 & 0 & 1 & GT/$\beta^-$  &  $a$ &
                  $-0.33000$ &  0.01000 &  0.286  & \citet{johnson61} \\
& & & & & &       $-0.33080$\footnote{ Value quoted by F.~Gluck
(1998) after including radiative corrections.}
               &  0.00300 &  0.286  & \citet{johnson63} \\
& & & & & &       $-0.31900$ &  0.02800 &  0.199  & \citet{vise63} \\
%
\\
  $^8$Li  & 4 &  2 & 2 & GT/$\beta^-$ &   $R$  &
                 0.00090 & 0.00220 &  0.062  & \citet{huber03} \\
%
\\
$^{12}$B & 6 & 1 & 0 & GT/$\beta^-$ &  $G$ &
                 $-0.98000$ &  0.06000 &  0.055  &  \citet{lipnik62} \\
%
\\
$^{12}$N &  6 & 1 & 0 & GT/$\beta^+$ & $P^-/P^+$ &
                 1.00060 &  0.00340 &  0.079    & \citet{thomas01} \\
%
\\
$^{14}$O & 7  & 0 & 0 & F/$\beta^+$ & $G$ &
                 0.97000 &  0.19000 &  0.338  & \citet{hopkins61} \\
%
\\
$^{14}$O/$^{10}$C &  7/5 &  &  & F-GT/$\beta^+$ &  $P_F/P_{GT}$ &
                0.99960 & 0.00370 &  0.292  &  \citet{carnoy91}\\
%
\\
$^{18}$Ne & 9 & 0 & 0 & F/$\beta^+$ & $a$ &
               1.06000 & 0.09500 &  0.289  & \citet{egorov97}\\
%
\\
$^{23}$Ne &   11 & 2.5 & 1.5 & GT/$\beta^-$ &   $a$ &
              $-0.37000$&  0.04000 &  0.243  & \citet{allen59}\\
& & & & & &      $-0.33000$ &  0.03000  &  0.243 &       \citet{carlson63}\\
%
\\
$^{26}$Al/$^{30}$P & 12/14 &  &  & F-GT/$\beta^+$ & $P_F/P_{GT}$ &
              1.00300 & 0.00400 &  0.189  &     \citet{wichers87}\\
%
\\
$^{32}$Ar &  17 & 0 & 0 & F/$\beta^+$ &   $a$ &
            0.99890 & 0.00650 &  0.210 & \citet{adelberger99}\\
%
\\
$^{38m}$K &  18 & 0 & 0 & F/$\beta^+$ &   $a$ &
           0.99810 & 0.00480 &  0.161  & \citet{gorelov05}\\
%
\\
$^{68}$Ga &  30 & 1 & 0 & GT/$\beta^+$ &   $G$ &
           0.99000 & 0.09000 &  0.307  &  \citet{ullman61}\\
%
\\
$^{107}$In & 48 & 4.5 & 3.5 & GT/$\beta^+$ & $P^-/P^+$ &
          0.92600 & 0.04100 &  0.311  & \citet{severijns93}\\
& & & & & $P^-/P^0$ & 0.98980 & 0.00820 &  0.311 & \citet{camps97}\\
%
\\
$^{114}$In &  50 & 1 & 0 & GT/$\beta^-$ &    $b$ &
          0.05000 &  0.02000&  0.399  & \citet{daniel61}\\
& & & & & &      0.00500 &  0.02200 &  0.399  &       \citet{daniel64}\\
%
\\
& & & & &     $A$ &   $-1.01300$ &  0.02400 & 0.662 &  \citet{severijns89b}\\
%
\\
& & & & &     $G$ &   $-0.96900$ &  0.03700 &  0.449 &  \citet{vanklinken66}\\
%
\\
$^{127}$Te & 53 & 1.5 & 2.5 & GT/$\beta^-$ & A &
             0.56900 & 0.05100 &  0.721  & \citet{vanneste86}\\
%
\\
$^{129}$Te & 53 & 1.5 & 2.5 & GT/$\beta^-$ & A &
            0.64500 & 0.05900 &  0.528  & \citet{vanneste86}\\
%
\\
$^{133}$Xe & 55 & 1.5 & 2.5 & GT/$\beta^-$ & A &
           0.59800 & 0.07300 &  0.818  & \citet{vanneste86}\\
%
\\
several & & 0 & 0 & F/$\beta^+$ & $b_F$ & 0.0001 & 0.0026 &  & \citet{hardy05b}\\
\\
\hline\hline
\end{tabular}
\label{tab:nuclear}
\end{table*}

The following selection criteria have been adopted: {\em i)}\,
except for the neutron decay, only pure Fermi and pure
Gamow-Teller transitions have been considered. The inclusion of
data from other mixed transitions like $^{19}$Ne, $^{21}$Na or
$^{35}$Ar would require to review the relevant spectroscopic data
of those transitions. Such data is necessary to calculate the
expected values of the correlation coefficients within the {\em
V-A} theory, with sufficient accuracy; {\em ii)}\, when in a given
transition several values for a correlation coefficient were
available, all inputs having an error which was at least ten times
larger than the error of the most precise measurement have been
eliminated. Exceptions to this rule concern some cases where the
quoted values for a given parameter have been published for
different energies of the $\beta$ particles; {\em iii)}\, all
experimental data from transitions having a ``large'' $\log{(ft)}$
value have been eliminated. Such slow transitions require a closer
look to the validity of the description within the allowed
approximation and to the effects of recoil order corrections. The
phenomenological classification of transitions based on the
$\log{(ft)}$ values considers as {\em allowed} those having values
in the range $\log{(ft)} = 5.7\pm1.1$ \citep{deShalit74}. We
therefore eliminated all inputs from transitions having
$\log{(ft)} > 6.8$. This concerns $^{22}$Na ($\log{(ft)} = 7.4$),
$^{32}$P ($\log{(ft)} = 7.9$) and $^{60}$Co ($\log{(ft)} = 7.5$),
and excludes 61 inputs which were previously used in the analysis
by \citet{boothroyd84}. These authors concluded that the question
on the validity of the allowed approximation had to be
reconsidered when including more accurate experimental data. In
this context, a recent measurement of the $\beta-\gamma$
directional correlation \citep{bowers99} addresses the competition
between the suppressed allowed matrix elements and the relevant
forbidden ones in the decay of $^{22}$Na.

\subsection{Results}

\subsubsection{Real couplings fit}
\label{sec:fit-realcouplings}

The first framework involves a maximum of seven parameters $a_k$,
assumed all to be real. Expressed as a function of the couplings
these parameters are defined as

\begin{alignat}{3}
\label{eq:pars-defs}
a_1 &= C_A/C_V, &\qquad a_2 &= C_S/C_V, &\qquad a_3 &= C_T/C_A, \\
a_4 &= C'_V/C_V, &\qquad a_5 &= C'_A/C_A, &\qquad a_6 &= C'_S/C_V, \notag \\
a_7 &= C'_T/C_A \notag
\end{alignat}

Several subsets of these parameters can be considered as free
parameters, corresponding to different assumptions in terms of the
presence of exotic couplings and of maximal parity violation.

For comparison with previous work~\cite{boothroyd84} the inputs on
the $D$ and $R$ coefficients, which are mainly sensitive to
imaginary couplings, have been excluded from the data subset in
this first framework.

{\bf Case 1: Standard one-parameter fit.} The simplest model to be
considered corresponds to the {\em V-A} limit. Here it is assumed
that $C'_V/C_V = C'_A/C_A = 1$, and that the scalar and tensor
couplings are zero. The only free parameter is $C_A/C_V$. This
ratio is determined by the neutron data as the coefficients from
the considered nuclear transitions do not depend on it. The fit of
the corresponding 26 experimental inputs with that single free
parameter gives $C_A/C_V = -1.27293(46)$, where the error is only
statistical at $1\sigma$. For this fit one obtains $\chi^2 =
74.08$ for $\nu = 25$ degrees of freedom. It is however
interesting to observe the effect of excluding the single recent
measurement of the neutron lifetime~\cite{serebrov05a} from the
input data set. The fit of the remaining 25 data gives $C_A/C_V =
-1.26992(63)$ with $\chi^2 = 25.86$ for $\nu = 24$ degrees of
freedom. Accounting for the $\pm 1.2$~s error on the ${\cal
F}t$-value results in

\begin{equation}
C_A/C_V = -1.26992(69)
\end{equation}

\noindent This value has an error which is more than an order of
magnitude smaller than that obtained by \citet{boothroyd84} and a
factor of 4.2 smaller than the presently recommended value
$\lambda = -1.2695(29)$~\cite{eidelman04}.

{\bf Case 2: Left-handed three-parameter fit.} This model allows
the presence of scalar and tensor couplings with the constraints
$C'_V/C_V = 1$, $C'_A/C_A = 1$, $C'_S/C_V = C_S/C_V$, and
$C'_T/C_A = C_T/C_A$. The three free parameters are $C_A/C_V$,
$C_S/C_V$ and $C_T/C_A$. The minimization of the $\chi^2$
converges to a single minimum with the following values and
$1\sigma$ statistical errors: $C_A/C_V = -1.27296(69)$, $C_S/C_V =
0.00045(127)$ and $C_T/C_A = 0.0086(31)$, implying a non-zero
tensor component. At this minimum $\chi^2 = 82.45$ for $\nu = 47$
degrees of freedom. Excluding again the recent neutron lifetime
measurement leads however to a significantly different minimum,
with $\chi^2 = 40.91$ for $\nu = 46$ degrees of freedom. The
values of the parameters at this minimum are then

\begin{alignat}{2}
C_A/C_V &= &-1.26994(82) \\
C_S/C_V &= &0.0013(13) \\
\label{eq:ReCTCA} C_T/C_A &= &0.0036(33)
\end{alignat}

\noindent where the errors include the effect of the $\pm 1.2$~s
error on the ${\cal F}t$-value.

For $C_A/C_V$ the error is again more than an order of magnitude
smaller than that obtained by \citet{boothroyd84}. The error on
$C_S/C_V$ is a factor of about 2 smaller. However, the error on
$C_T/C_A$ is larger by a factor of 4. This is attributed to the
exclusion of the 61 data points from $^{22}$Na, $^{32}$P and
$^{60}$Co, which are all three pure Gamow-Teller transitions. The
95.5\% confidence level (CL) limits are obtained by taking
$2\sigma$ of the quoted values as the correlations between the
parameters are small.

{\bf Case 3: Vector Axial-vector three-parameter fit.} In this
model the scalar and tensor couplings are excluded. The three
remaining parameters are $C_A/C_V$, $C'_V/C_V$ and $C'_A/C_A$
which all contain standard couplings. Allowing these parameters to
be free provides a test of the degree of maximal parity violation
with vector and axial couplings. The minimization results in two
equivalent minima. Excluding the recent neutron lifetime
measurement from the data set leads to $\chi^2 = 40.93$ for $\nu =
46$ degrees of freedom. The central values of the parameters at
these minima are

\begin{center}
\begin{tabular}{l@{\hspace{5mm}}r@{\hspace{5mm}}r}
\hline\hline
& min.A & min.B\\
\hline
$C_A/C_V$ & $-1.2702$ & $-1.2701$ \\
$C'_V/C_V$ & $0.920$ & $1.087$ \\
$C'_A/C_A$ & $0.920$ & $1.087$\\
\hline\hline
\end{tabular}
\end{center}

\begin{figure}[!htb]
\includegraphics[height=75mm,width=82mm]{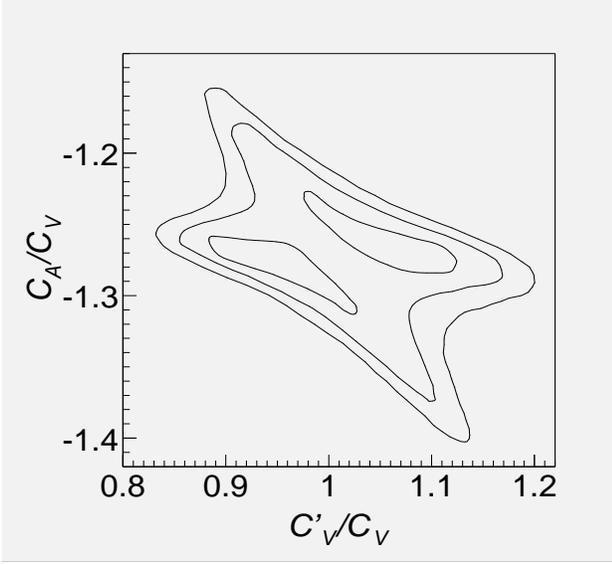}
\caption{Contours of constant $\chi^2$ around the two minima
obtained for the fit case 3. The lines correspond to $1\sigma$,
$2\sigma$ and $3\sigma$ confidence levels.}
\label{fig:3-CVpCV-CACV}
\end{figure}

\noindent The situation is similar to that encountered earlier by
Boothroyd {\em et al.} although the relative distance between the
minima is here significantly smaller. Due to the correlations
between the parameters the quotation of independent confidence
intervals requires the $\chi^2$ hyper-surface to be scanned.
Figure~\ref{fig:3-CVpCV-CACV} shows the projection of the
iso-$\chi^2$ contours onto the plane of the parameters $C'_V/C_V$
and $C_A/C_V$. The lines around the minima correspond to the
$1\sigma$, $2\sigma$ and $3\sigma$ contours of the confidence
regions, obtained by varying the values of all three parameters
near the minima. The corresponding global limits of the confidence
regions, at the $2\sigma$ level, are

\begin{alignat}{3}
-1.372 &< C_A/C_V  &< & -1.180 \\
0.857 &< C'_V/C_V  &< & 1.169 \\
0.868 &< C'_A/C_A  &< & 1.153
\end{alignat}

\noindent The limits on $C'_A/C_A$ are similar to those obtained
earlier~\cite{boothroyd84} whereas the intervals for the other two
parameters have significantly been reduced.

{\bf Case 4: Right-handed Scalar and Tensor three-parameter fit.}
In this model the three free parameters are $C_A/C_V$, $C_S/C_V$
and $C_T/C_A$ with the conditions $C'_V/C_V = 1$, $C'_A/C_A = 1$,
$C'_S/C_V = -C_S/C_V$, and $C'_T/C_A = -C_T/C_A$. This corresponds
to the assumption of left-handed couplings in the standard sector
and right-handed couplings for the scalar and tensor. The
minimization of the $\chi^2$ converges to two minima. Again,
excluding the recent neutron lifetime measurement leads to $\chi^2
= 39.96$ for $\nu = 46$ degrees of freedom. The central values of
the parameters at the minima are

\begin{center}
\begin{tabular}{l@{\hspace{5mm}}r@{\hspace{5mm}}r}
\hline\hline
& min.A & min.B\\
\hline
$C_A/C_V$ & $-1.2689$ & $-1.2689$ \\
$C_S/C_V$ & $0.033$ & $-0.033$ \\
$C_T/C_A$ & $0.052$ & $-0.052$\\
\hline\hline
\end{tabular}
\end{center}

\noindent The two minima differ only by the signs of the scalar
and tensor couplings which change simultaneously.
Such a scenario has recently been considered in the analysis of
selected data from neutron decay~\cite{mostovoi00}. Although the
technique used there for the determination of the couplings
differs from the one presented here, the conclusions regarding the
scalar and tensor couplings are similar, namely that for each
minimum the two couplings have the same sign. However, the
independent quotation of CL intervals for each parameter requires
both minima to be considered simultaneously and to account for the
correlations with $C_A/C_V$. The $2\sigma$ confidence regions
obtained by scanning the $\chi^2$ hyper-surfaces are

\begin{alignat}{3}
-1.272 &< C_A/C_V &< &-1.265 \\
-0.067 &< C_S/C_V &< &0.067 \\
-0.081 &< C_T/C_A &< &0.081
\end{alignat}

\begin{figure}[!htb]
\includegraphics[height=76mm,width=74mm]{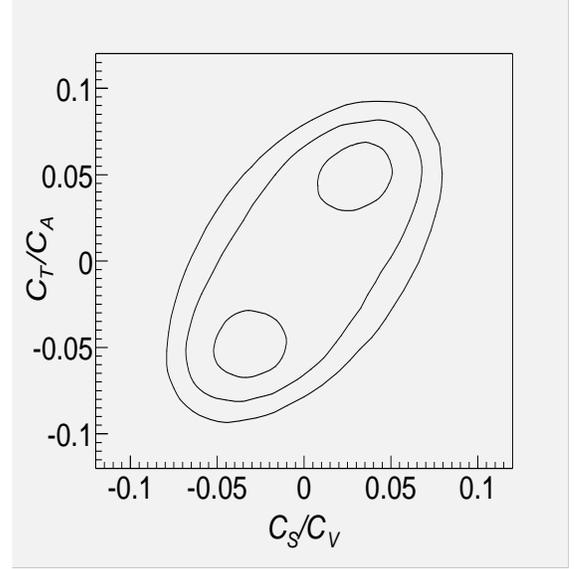}
\caption{Projections of contours of constant $\chi^2$ on the plane
of parameters $C_S/C_V$ and $C_T/C_A$ for the fit case 4.}
\label{fig:4-CSCV-CTCA}
\end{figure}

\begin{figure}[!htb]
\includegraphics[height=80mm,width=80mm]{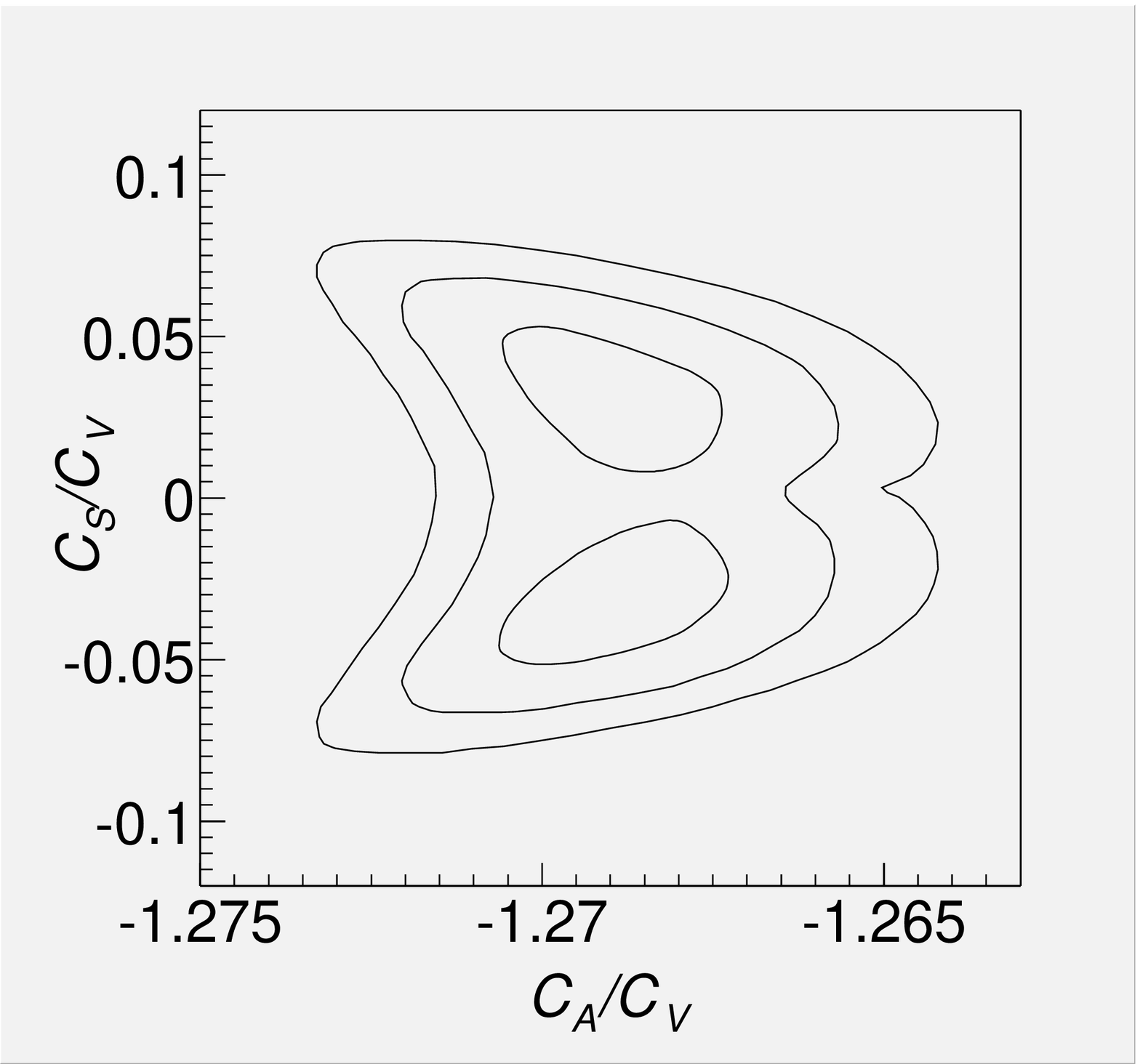}
\caption{Projections of contours of constant $\chi^2$ on the plane
of parameters $C_A/C_V$ and $C_S/C_V$ for the fit case 4.}
\label{fig:4-CACV-CSCV}
\end{figure}

\begin{figure}[!htb]
\includegraphics[height=85mm,width=85mm]{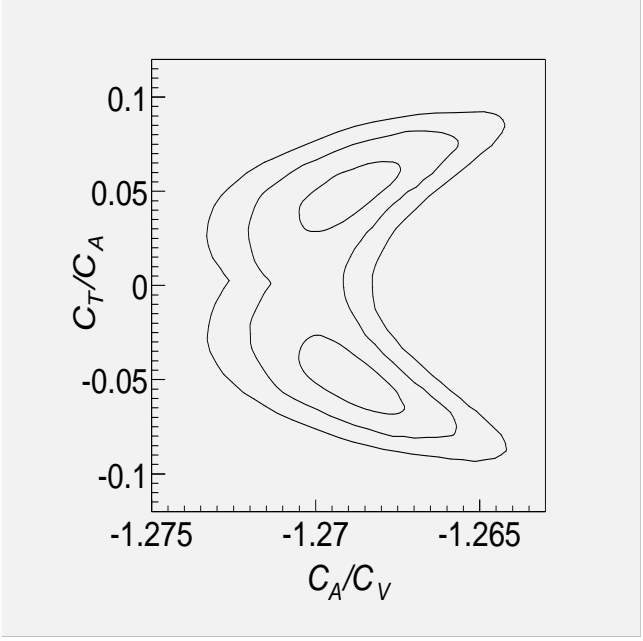}
\caption{Projections of contours of constant $\chi^2$ on the plane
of parameters $C_A/C_V$ and $C_T/C_A$ for the fit case 4.}
\label{fig:4-CACV-CTCA}
\end{figure}

The contours of the confidence regions for the three pairs of
parameters are shown in Figs.~\ref{fig:4-CSCV-CTCA},
\ref{fig:4-CACV-CSCV} and \ref{fig:4-CACV-CTCA}. Again, the lines
around each minimum correspond to the three levels of constant
$\chi^2$ : $\chi^2_0 + 1$, $\chi^2_0 + 2^2$ and $\chi^2_0 + 3^2$,
where $\chi^2_0$ is the value of the $\chi^2$ at the minimum. The
global confidence region for each parameter is obtained by varying
the values of the other two parameters around the minima.

{\bf Case 5: Five-parameter fit.} A first generalization of the
case 2 above consists in relaxing the constraint on the exotic
couplings, allowing $C'_S$ be different from $C_S$ and $C'_T$ from
$C_T$. In this model the five free parameters are then $C_A/C_V$,
$C_S/C_V$, $C'_S/C_V$, $C_T/C_A$ and $C'_T/C_A$ keeping the
condition $C'_V/C_V = C'_A/C_A = 1$. When the recent neutron
lifetime is excluded from the data, two equivalent minima are
found with $\chi^2 = 38.67$ for $\nu = 44$ degrees of freedom. At
each minimum the signs of $C_S/C_V$ and $C_T/C_A$ are the same but
opposite to those of $C'_S/C_V$ and $C'_T/C_A$. The $2\sigma$
intervals obtained from the projections of the $\chi^2$
hyper-surface are

\begin{alignat}{3}
-1.272 &< C_A/C_V &< &-1.265 \\
-0.064 &< C_S/C_V &< &0.066 \\
-0.064 &< C'_S/C_V &< &0.065\\
-0.077 &< C_T/C_A &< &0.086 \\
-0.077 &< C'_T/C_A &< &0.087
\end{alignat}

\begin{figure}[!htb]
\includegraphics[height=82mm,width=82mm]{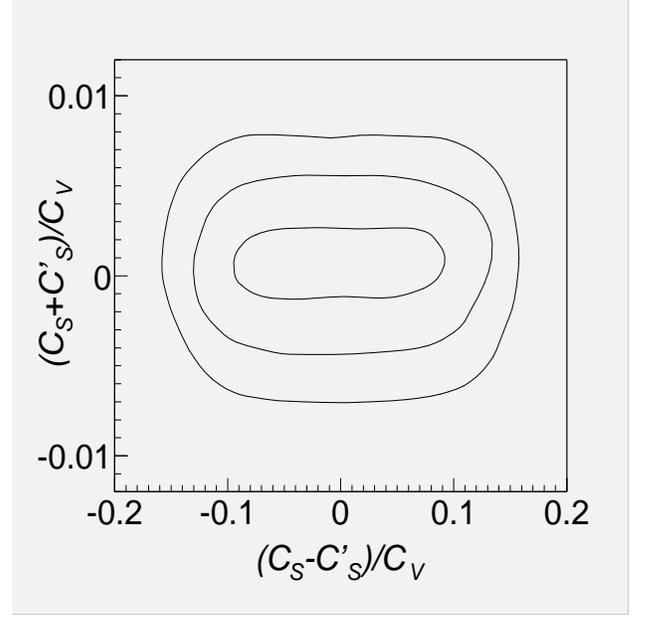}
\caption{Projections of contours of constant $\chi^2$ for the
combination of parameters $(C_S-C'_S)/C_V$ and $(C_S+C'_S)/C_V$.}
\label{fig:5-diff-sum-CS}
\end{figure}

\begin{figure}[!htb]
\includegraphics[height=87mm,width=85mm]{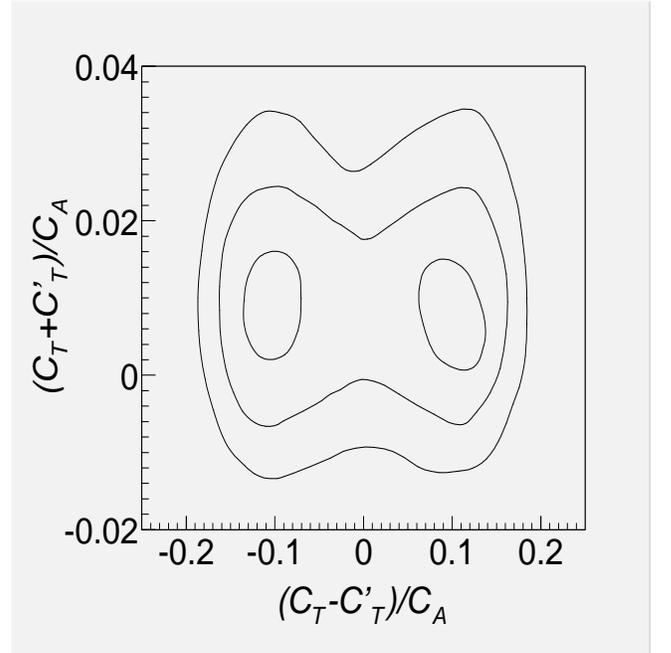}
\caption{Projections of contours of constant $\chi^2$ for the
combination of parameters $(C_T-C'_T)/C_A$ and $(C_T+C'_T)/C_A$.}
\label{fig:5-diff-sum-CT}
\end{figure}

Because of the correlations between $C_S$ and $C'_S$ and between
$C_T$ and $C'_T$, it is interesting to consider here the exclusion
plots associated with the differences and the sums of these
coefficients. This is also useful for a direct comparison with
some experiments because the differences and sums of scalar and
tensor couplings enter several correlation coefficients. The
exclusion plots associated with this case are shown in
Figs.~\ref{fig:5-diff-sum-CS} and \ref{fig:5-diff-sum-CT}.

{\bf Case 6: Seven-parameter fit.} With the definition of
parameters indicated above, Eq.~(\ref{eq:pars-defs}), the most
general fit is obtained by allowing all seven parameters to be
free. However, when the number of parameters increases the search
for equivalent minima is more difficult, the convergence is less
robust and several local minima with similar $\chi^2$ can be
found. When considering the correlation between the parameters,
the present case combines actually the cases 3 and 5 considered
above. It is nevertheless possible to scan the $\chi^2$
hyper-surface by varying all parameters to obtain the $2\sigma$
confidence level for each parameter. The outcome of such scan
results in the following limits

\begin{alignat}{3}
-1.40 &< C_A/C_V &< &-1.17 \\
0.87  &< C'_V/C_V &< &1.17 \\
0.86 &< C'_A/C_A &< &1.16 \\
-0.065 &< C_S/C_V &< &0.070 \\
-0.067 &< C'_S/C_V &< &0.066\\
-0.076 &< C_T/C_A &< &0.090 \\
-0.078 &< C'_T/C_A &< &0.089
\end{alignat}

\noindent It is seen that the surface $\chi^2 = \chi^2_0 + 2^2$
corresponding to the 95.5\% CL region encloses the {\em V-A}
assumptions, $C_S = C'_S = C_T = C'_T = 0$ and $C'_V/C_V =
C'_A/C_A = 1$.
%
\subsubsection{Imaginary couplings fit}

When allowing for the presence of imaginary phases in the
couplings, the total number of real parameters doubles with
respect to the case in which all couplings are assumed to be real.
In this second framework there are then 14 real parameters.

The most sensitive input data to imaginary parts in the couplings
are the $D$ and $R$ triple correlation coefficients and there are
actually only two such coefficients in the input data as far as
the dependence on the couplings is concerned. It is then necessary
to make additional assumptions in order to determine possible
imaginary parts while achieving also a robust convergence to a
minimum.

{\bf Case 7: Two-parameter fit with axial and vector couplings.}
The triple correlation coefficient $D$ in a mixed transition is
particularly sensitive to a possible phase between the two
standard couplings $C_V$ and $C_A$. The experimental results of
such measurements are generally interpreted assuming $C_S = C'_S =
C_T = C'_T = 0$, $C'_V/C_V = 1$ and $C'_A/C_A = 1$. Under these
assumptions there remains only two parameters which are the real
and imaginary parts of the ratio $C_A/C_V$. Here again, only the
neutron data contributes to the determination of these parameters.
Including all the 31 neutron data the minimization converges to a
minimum with $\chi^2 = 75.25$ for $\nu = 29$ degrees of freedom
whereas excluding the recent measurement of the neutron lifetime
results in $\chi^2 = 27.03$ for $\nu = 28$. In both cases the
values of the real part of $C_A/C_V$ are identical to those
obtained for the case 1 and both fits give for the imaginary part

\begin{equation}
{\rm Im}(C_A/C_V) = -0.0012(19)
\end{equation}

{\bf Case 8: Single-parameter fit with imaginary tensor
couplings.} The other parameter sensitive to the possible presence
of imaginary couplings is the $R$ triple correlation. The only
input data considered here arises from the decay in $^8$Li. This
decay is known to proceed by a predominantly Gamow-Teller
transition which is then driven by the axial and eventually tensor
couplings. It can here be assumed that all the couplings are
left-handed ({\em i.e.} $C'_i = C_i$) and real except for
$C_T/C_A$. As the determination of the imaginary part of $C_T/C_A$
is solely determined by the $R$ triple correlation, the result is
independent of assumptions on the other parameters leading to the
following value and $1\sigma$ error

\begin{equation}
{\rm Im}(C_T/C_A) = 0.0014(33)
\end{equation}

\noindent It is interesting to notice that the uncertainty on the
imaginary part of the tensor coupling is similar to that obtained
on the real part in the most constrained fit,
Eq.~(\ref{eq:ReCTCA}).

\subsection{Conclusions}
\label{sec:fit-conclusion}

This section provided a quantitative summary of the experimental
progress achieved over the past two decades. The values on the
standard couplings and the constraints on exotic couplings have
significantly been improved due to precision data from neutron
decay, from measurements of relative longitudinal polarization of
beta particles and of beta-neutrino correlations in nuclear
decays. However, the recent measurement of the neutron
lifetime~\cite{serebrov05a} strongly affects the consistency of
the fits and the values or ranges of the parameters. This
obviously calls for an urgent confirmation or clarification of
that experimental result.

The general fit with seven free real parameters (case 6) results
in the following 95.5\% CL limits for the exotic couplings,

\begin{alignat}{2}
\left|{C_S}/{C_V}\right| &<  &0.070 \\
\left|{C'_S}/{C_V}\right| &<  &0.067 \\
\left|{C_T}/{C_A}\right| &<  &0.090 \\
\left|{C'_T}/{C_A}\right| &<  &0.089
\end{alignat}

Considering that $|C_A/C_V| \approx 1.27$ it appears from the
results above that, in absolute terms, the limits on the
amplitudes of tensor contributions are a factor of about two
larger than those on the scalar contributions.

The fit with the three real parameters $C_A/C_V$, $C'_V/C_V$ and
$C'_A/C_A$ (case 3) results in the following 95.5\% CL limits for
the standard couplings,

\begin{alignat}{2}
-1.372 &< {C_A}/{C_V} &< -1.180 \\
0.857 &< {C'_V}/{C_V} &< 1.169 \\
0.868 &< {C'_A}/{C_A} &< 1.153
\end{alignat}

The limits on the imaginary parts of $C_A/C_V$ and of $C_T/C_A$
are almost independent of the constraints on other parameters
under the assumptions considered above.

%
%
%
%
\section{EXPERIMENTAL TESTS}
\label{sec:experiments}

This section gives an overview of recent as well as ongoing and
planned experiments in beta decay to test symmetries of the
standard electroweak model and to search for new physics. We will
concentrate here on experiments and projects that were ongoing or
started after previous reviews of this field
\cite{abele,deutsch95,towner95,vanklinken96,yerozolimsky00,
herczeg01}. Other recent reviews can be found in \citet{erler05}
and \citet{nico05b}.

First, the status and prospects for testing the unitarity of the
quark mixing matrix will be discussed. This will be followed by an
overview of searches for possible scalar and/or tensor type
contributions to the weak interaction. Thereafter, the present
situation with respect to the discrete symmetries of parity and
time reversal will be reviewed. Finally, the direct searches for
the electron neutrino mass will be discussed and the status of CVC
and second class currents will briefly be presented.

\subsection{Unitarity of the CKM quark mixing matrix}
\label{sec:unitarity}

The CKM matrix (see Sec.~\ref{sec:qark-mixing}) relates the quark
weak interaction eigenstates to the quark mass eigenstates and, as
such, is a unitary matrix, {\em i.e.}

\begin{equation}
\label{unitarity_general} \sum_k V^*_{ki} V_{kj} = \delta_{ij} .
\end{equation}

As up to now only the matrix elements $V_{ud}$ and $V_{us}$ have
been determined with sub-percent precision, the most precise test
of unitarity to date is obtained from the first row of the matrix,
{\em i.e.}
\begin{equation}
\label{eqn:unitarity} \sum_i V_{ui}^2 = V_{ud}^2 + V_{us}^2 +
V_{ub}^2 \; \; \; \; \; \; (i = d, s, b)
\end{equation}
which should be equal to unity. The leading element, $V_{ud}$,
depends only on the quarks in the first generation and can
therefore be determined most precisely. The $V_{us}$ matrix
element is obtained from $K$ decays. The third matrix element,
$V_{ub}$, is obtained from $B$ meson decays \citep{battaglia04}.
If the CKM-matrix would turn out to be non-unitary this could
point either to the existence of a fourth generation of fermions
or to other new physics beyond the standard model, such as
right-handed currents or non-{\em V, A} contributions to the weak
interaction \citep[see][]{hardy05a,hardy05b}.

The $V_{ud}$ element can be deduced from the $\mathcal{F}$t values
of superallowed $0{^+}\rightarrow 0{^+}$ pure Fermi $\beta$
transitions, from neutron decay and from pion beta decay.
Combining each of these three values for $V_{ud}$ with the two
other matrix elements just mentioned then yields three almost
independent tests of the unitarity of this matrix. We will review
here the current status of these.

\subsubsection{Superallowed Fermi transitions}
\label{sec:ftvalues}

Currently, the $\mathcal{F}$t value of eight superallowed
$0{^+}\rightarrow 0{^+}$ pure Fermi transitions, $^{14}$O,
$^{26}$Al$^m$, $^{34}$Cl, $^{38}$K$^m$, $^{42}$Sc, $^{46}$V,
$^{50}$Mn and $^{54}$Co, has been determined with a precision
better than $1 \times 10^{-3}$ and of four others, $^{10}$C,
$^{22}$Mg, $^{34}$Ar and $^{74}$Rb with a precision better than $4
\times 10^{-3}$ \citep{hardy90,towner03,hardy05a}.

The relation between the $\mathcal{F}$t value and $V_{ud}$ is
\begin{equation}
\label{eqn:Ft} \mathcal{F} t \equiv ft\, (1+\delta_R)\,
(1-\delta_C) = \frac{K}{2\, G^2_F\, V^2_{ud}\, (1+\Delta_R^V)}
\end{equation}
\noindent
where {\it f} is the statistical rate function (see e.g. Appendix
A in \citet{hardy05a}) and
\begin{equation}
t = \frac{t_{1/2}} {BR} \left(1 + \frac{\varepsilon}{\beta^+}
\right)
\end{equation}
is the partial half-life for the transition that is obtained from
the half-life, $t_{1/2}$, of the parent nucleus corrected for the
branching ratio, $BR$, of the transition and for electron capture,
$\varepsilon / \beta^+$. Note that the right hand side of
Eq.~(\ref{eqn:Ft}) contains only fundamental constants and
parameters determined by the weak interaction, while the left hand
side contains the experimentally determined quantities and
calculated nuclear corrections. The determination of the
$ft$-value for a specific transition requires advanced
spectroscopic methods as the half-life, the branching ratio as
well as the transition energy, $Q_{EC}$, which is required to
calculate $f$, have to be known with good precision.

Further, $\delta_R$ and $\Delta_R^V$ are the transition-dependent
and nucleus-independent radiative corrections, while $\delta_C$ is
the isospin symmetry-breaking correction. These must be
calculated. The transition-dependent radiative correction
$\delta_R$ can be split into a nuclear structure independent part,
$\delta_R^{'}$, and a nuclear structure dependent part,
$\delta_{NS}$, with $\delta_R = \delta_R^{'} + \delta_{NS}$. The
first is calculated from QED
\citep{sirlin67,sirlin86,sirlin87,jaus87,towner02} and is
currently evaluated up to order $Z^2 \alpha^3$, assigning an
uncertainty equal to the magnitude of this order $Z^2 \alpha^3$
contribution as an estimate of the error made by stopping the
calculation there. For the twelve above mentioned transitions the
values of $\delta_R^\prime$ range from 1.39\% to 1.65\%
\citep*{towner02}. The nuclear structure dependent part,
$\delta_{NS}$, was calculated in the nuclear shell model with
effective interactions and ranges from +0.03\% to $-0.36$\%
\citep{towner02}. For the nucleus-independent correction the
currently adopted value is $\Delta_{R}^{V}$ = 0.0240(8)
\citep{towner02,sirlin95}. Several independent calculations were
performed for the isospin symmetry-breaking correction $\delta_C$
\citep*{towner77,barker92,ormand95,sagawa96,towner02,wilkinson02,wilkinson04}.
As only the calculations by \citet{ormand95} and by
\citet{towner02} are constrained by experiments, thus offering an
independent means to access their reliability, only these are
usually retained. They are in reasonably good agreement, yielding
values from about 0.2\% to 0.6\% depending on the nucleus
involved, although there is some (small) scatter between the two
calculations. A detailed discussion of all corrections can be
found in \citet{towner02}. Further, on the right hand side of
Eq.~(\ref{eqn:Ft}) one has the constants

\begin{equation}
\frac{K}{(\hbar c)^6} = \frac{2 \pi^3 \hbar \ln{2}}{(m_e c^2)^5} =
8120.271(12) \times 10^{-10}\, \text{GeV}^{-4}\, \text{s}
\end{equation}
and

\begin{equation}
\frac{G_F}{(\hbar c)^3} = 1.16639(1) \times 10^{-5}\,
\text{GeV}^{-2}\: .
\end{equation}

\noindent The value for the Fermi coupling constant $G_F$ is known
from the purely leptonic decay of the muon \citep{eidelman04}. It
is related by CVC (Sec.~\ref{sec:higherorder}) to the vector
coupling constant $G_V$ in nuclear beta-decay, $G_V = V_{ud}\:
G_F\: g_V(q^2 \rightarrow 0)$, with $g_V$ the vector form factor
and $g_V(q^2 \rightarrow 0) = 1$ the vector coupling constant with
$q$ the momentum transfer to the leptons in the decay.

\begin{figure}[!htb]
\begin{center}
\includegraphics[height = 6 cm, width=\columnwidth]{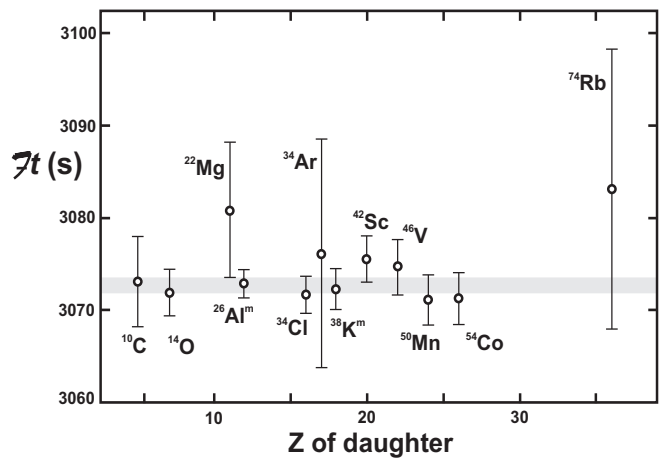}
\caption[$\mathcal{F}$t values]{$\mathcal{F}$t values for the
superallowed $0{^+}\rightarrow 0{^+}$ transitions. The shaded band
is the 1$\sigma$ result from the best least-squares one-parameter
fit. From Hardy and Towner \citeyearpar{hardy05b}. \label{fig:Ft}}
\end{center}
\end{figure}

According to the CVC hypothesis \citep{feynman58} the
$\mathcal{F}t$ value should be the same for all superallowed
$0{^+}\rightarrow 0{^+}$ transitions. The fit to a constant of the
corrected $\mathcal{F}t$ values for the twelve transitions yields
$\mathcal{F} t = 3072.7(8)$ s \citep{hardy05a}
(Fig.~\ref{fig:Ft}), confirming the CVC hypothesis at the $3
\times 10^{-4}$ precision level. Taking into account an additional
error related to the above mentioned systematic difference between
the two calculations of $\delta_c$ by \citet*{towner02} and
\citet*{ormand95} one gets $\mathcal{F} t = 3073.5(12)$ s
\citep{hardy05a} which leads to
\begin{equation}
|V_{ud}| = 0.9738(4)  ~~~  {\rm (superallowed ~ transitions)}
\end{equation}

\subsubsection{Neutron decay}
\label{sec:neutron-unitarity}

The matrix element $|V_{ud}|$ can also be determined from the
decay of the free neutron. The $ft$-value for the neutron is given
by
\begin{equation}
\label{eqn:ft-neutron} f_n\: \tau_n \: (1+\delta_R)\: = \frac{K\,
/\, \ln{2}}{G_F^2\: V^{2}_{ud}\: (1+\Delta_R^V)\: (1 + 3
\lambda^2)}
\end{equation}
with $\tau_n$ the lifetime of the free neutron and $f_n
(1+\delta_R) = 1.71489(2)$ the phase space factor
\citep{wilkinson82,towner95}. The factor $\lambda$ is the ratio of
the effective vector and axial vector weak coupling constants,
$\lambda = G_A^{\prime} / G_V^{\prime}$, with $G_A^{\prime 2} =
G_A^2 (1+\Delta_R^A)$ and $G_V^{\prime 2} = G_V^2 (1+\Delta_R^V)$.
Here, $G_A = V_{ud}\: G_F\: g_A(q^2 \rightarrow 0)$, with $g_A$
the axial vector form factor and $g_A(q^2 \rightarrow 0) \approx
-1.27$ the axial vector coupling constant. The factors
$\Delta_R^A$ and $\Delta_R^V$ are the nucleus-independent
radiative corrections. Since the neutron is a single nucleon, no
nuclear structure correction $\delta_{NS}$ or isospin
symmetry-breaking correction $\delta_C$ have to be applied (see
also \citealp{garcia01}). However, one now has to determine the
ratio $\lambda$ which enters because the decay of the neutron
proceeds through a mixed Fermi/Gamow-Teller transition. This is
usually obtained in measurements of the beta asymmetry parameter
$A$.
%
%
%
%
\noindent Eq.~(\ref{eqn:ft-neutron}) can be rewritten to obtain
$|V_{ud}|$ as
\begin{equation}
\label{eq:Vud-neutron} |V_{ud}|^2 = \frac{ 4903.7(38) }{ \tau_n (1
+ 3 \lambda^2) } .
\end{equation}

\begin{figure}[!htb]
\begin{center}
\includegraphics{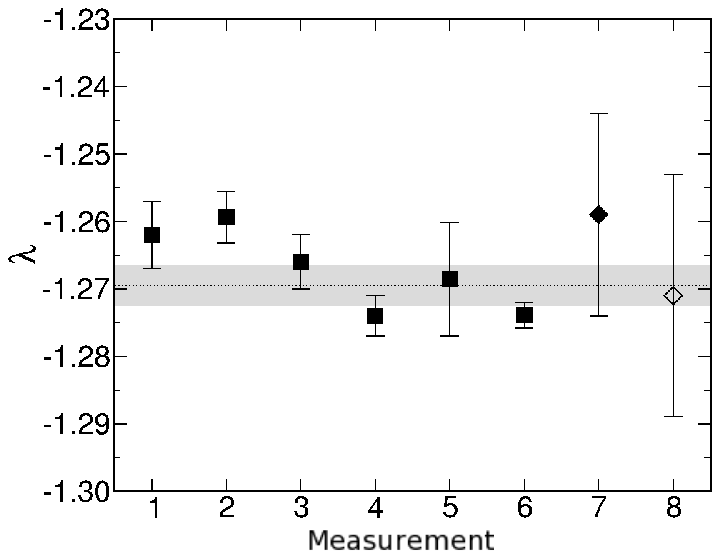}
\caption[lambda-n]{Input data for the world average value of
$\lambda$ from measurements in neutron decay (See also
Table~\ref{tab:neutron}). The most precise results are from
measurements of the $\beta$ asymmetry parameter $A$
(1:~\citealp{bopp86}; 2:~\citealp{yerozolimsky97};
3:~\citealp{liaud97}; 4:~\citealp{abele97}; 6:~\citealp{abele02}).
Data points 7~\citep{stratowa78} and 8~\citep{byrne02} refer to
measurements of the $\beta-\nu$ correlation coefficient $a$ and
data point 5~\citep{mostovoi01} is from a simultaneous measurement
of the $\beta$ asymmetry and the $\nu$ asymmetry parameters $A$
and $B$. The band indicates the weighted average adopted by the
Particle Data Group~\cite{eidelman04}. \label{fig:lambda-n}}
\end{center}
\end{figure}

\begin{figure}[!htb]
\begin{center}
\includegraphics{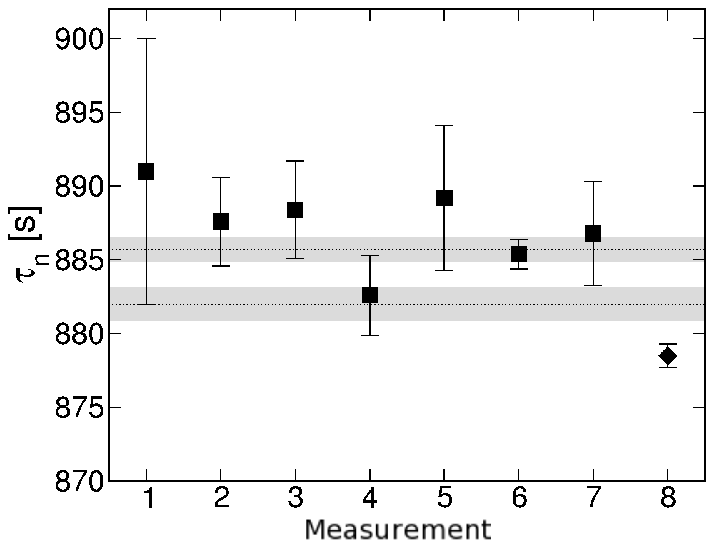}
\caption[tau-n]{Input data for the world average value of $\tau_n$
(See also Table~\ref{tab:neutron}). 1:~\citealp{spivak88};
2:~\citealp{mampe89}; 3:~\citealp{nesvizhevsky92};
4:~\citealp{mampe93}; 5:~\citealp{byrne96};
6:~\citealp{arzumanov00}, 7:~\citealp{dewey03} and
8:~\citealp{serebrov05a}. The upper band shows the weighted
average of the first seven values and the lower band the weighted
average of all values. \label{fig:tau-n}}
\end{center}
\end{figure}

The world average value for $\lambda$ recommended by the Particle
Data Group \citep*{eidelman04} is
\begin{equation}
\lambda = -1.2695(29) ,
\end{equation}
\noindent which is extracted mainly from measurements of the
$\beta$ asymmetry parameter (Fig.~\ref{fig:lambda-n}). The value
for the neutron lifetime recommended by the Particle Data Group is
\begin{equation}
\tau_n = 885.7\pm0.8~s  ,
\end{equation}
%
\noindent which is the weighted average (with $\chi^2/\nu = 0.76$)
of seven independent results (Fig.~\ref{fig:tau-n}). It is
dominated, however, by the value reported by \citet{arzumanov00}.
Recently, a new measurement of the neutron lifetime was reported
\citep{serebrov05a}. The result, $\tau_n =
(878.5\pm0.7_{stat}\pm0.3_{syst})$~s, although being obtained with
a method rather similar to the one used by \citet{arzumanov00},
differs by 6.5 standard deviations from the former world average.
As a consequence, the data set for $\tau_n$ (Fig.~\ref{fig:tau-n})
is now dominated by two very precise but conflicting results (see
Fig.~\ref{fig:tau-n}). In Sec.\ref{sec:fit} it was shown that
including the result of \citet{serebrov05a} worsens the $\chi^2 /
\nu$ for the multi-parameter fits to the various couplings by a
factor of about 2 to 3. The same is true here, i.e. combining all
neutron lifetime data leads to a world average $\tau_n =
(882.0\pm1.1)$~s with $\chi^2/\nu$~=~1.90 (the uncertainty was
increased accordingly). We therefore adopt the same approach with
respect to this result as in Sec.\ref{sec:fit}. Using then the
world average neutron lifetime $\tau_n$ not including the most
recent measurement, and the adopted value for $\lambda$,
Eq.~(\ref{eq:Vud-neutron}) yields
\begin{equation}
|V_{ud}| = 0.9741(20)  ~~~  {\rm (neutron ~ decay)}
\end{equation}
\noindent which agrees with the value obtained from the
superallowed Fermi transitions but is a factor five less precise.

%
%


\subsubsection{Pion beta decay}
\label{sec:pion-unitarity}

The value of $|V_{ud}|$ can also be obtained from pion beta decay,
$\pi^+\: \rightarrow\: \pi^0\: e^+\: \nu_e$ \citep[see
e.g][]{towner99}. As this is a $0^-\: \rightarrow\: 0^-$ pure
vector transition, no separation of vector and axial-vector
components is required. In addition, like neutron decay, it has
the advantage that no nuclear structure-dependent corrections have
to be applied. A major disadvantage, however, is that pion beta
decay has a very weak branch, of the order of $10^{-8}$, leading
to severe experimental difficulties. The values for the lifetime
and branching given by the Particle Data Group \citep{eidelman04},
are $\tau_\pi = (2.6033 \pm 0.0005) \times 10^{-8}$~s and $BR =
1.025(34) \times 10^{-8}$. Since the precision of this branching
ratio is about an order of magnitude worse than the theoretical
uncertainties, a new experiment was performed at the Paul Scherrer
Institute. The analysis has yielded $BR = (1.036 \pm 0.004_{stat}
\pm 0.005_{syst}) \times 10^{-8}$ \citep{pocanic04}, corresponding
to
\begin{equation}
|V_{ud}| = 0.9728(30)  ~~~  {\rm (pion ~ \beta ~ decay).}
\end{equation}
%
This is in agreement with but still much less precise than the
value obtained from the superallowed  Fermi $\beta$ decays.

\subsubsection{Status of unitarity}
\label{sec:status-unitarity}

\begin{figure}[!htb]
\begin{center}
\includegraphics[width=\columnwidth]{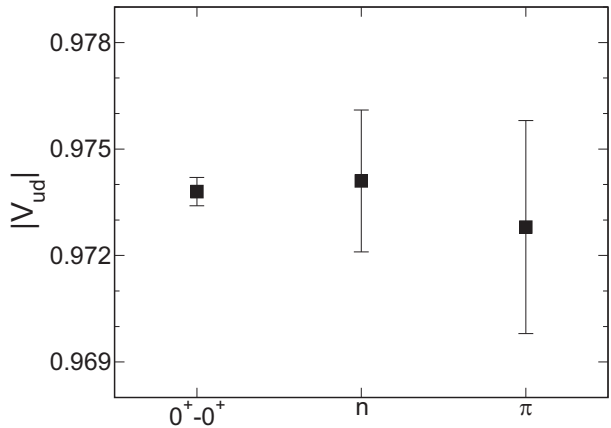}
\caption[V_ud]{Values for $|V_{ud}|$ obtained from the average
$\mathcal{F}$t value for the $0{^+}\rightarrow 0{^+}$ superallowed
Fermi $\beta$ decays~\cite{hardy05a}, from neutron decay (see
text) and from pion $\beta$ decay~\cite{pocanic04}.
\label{fig:Vud}}
\end{center}
\end{figure}

The values obtained for $V_{ud}$ from the superallowed Fermi
decays, from neutron decay and from pion $\beta$ decay are
compared to each other in Fig.~\ref{fig:Vud}. The values for the
two other matrix elements in the first row that are recommended by
the Particle Data Group are $|V_{us}|$ = 0.2200(26) and $|V_{ub}|$
= 0.00367(47) \citep*{eidelman04}. Note that the $V_{ub}$ matrix
element is so small that it does not contribute to the unitarity
test at the present level of precision. As a consequence, since
this test of unitarity is not even sensitive to the third quark
generation, it will not be sensitive to a possible fourth
generation either, except in some non-hierarchical scenarios where
the couplings of fourth generation quarks would be larger than
those of the third generation.

The third column of Table~\ref{tab:unitarity} lists the results of
the unitarity test when the values for $|V_{ud}|$ obtained from
the three different types of $\beta$ decay are combined with the
current Particle Data Group value for $|V_{us}|$.
For the superallowed Fermi decays a 2.5$\sigma$ deviation from the
standard model is observed. The current data for the decay of the
neutron and for pion $\beta$ decay are in agreement with unitarity
but the error bars are a factor of three to five larger than is
the case for the superallowed Fermi transitions.

\begin{table}[!htb]
\begin{center}
\begin{ruledtabular}
\begin{tabular}{ll|ll}
   decay & $|V_{ud}|$ & $|V_{us}|$=0.2200(23) & $|V_{us}|$=0.2254(21)$^1$\\
\hline
 $0{^+}\rightarrow 0{^+}$ &  0.9738(4) & 0.9967(13) & 0.9991(12) \\
\hline
 neutron & 0.9741(20)$^2$ & 0.9973(40) & 0.9997(40) \\
\hline
 pion & 0.9728(30) & 0.9946(59) & 0.9971(59) \\
\end{tabular}
\caption{Results of the unitarity test for the first row of the
CKM matrix when combining the values of $|V_{ud}|$ obtained from
$0{^+}\rightarrow 0{^+}$ superallowed nuclear decays, from neutron
decay and from pion beta decay (second column) with the value for
$|V_{us}|$ adopted by the Particle Data Group \citep{eidelman04}
(third column). The results of the unitarity test when the
weighted average for $|V_{us}|$ from the recent results obtained
in kaon decays is used is shown in the fourth column (see also
Table~\ref{tab:Vus}). In all cases $|V_{ub}| = 0.00367(47)$
\citep{eidelman04} was used. ($^{1}$: see Table~\ref{tab:Vus};
$^{2}$: see Sec.~\ref{sec:neutron-unitarity})}
\label{tab:unitarity}
\end{ruledtabular}
\end{center}
\end{table}

In the past years several new determinations of $|V_{us}|$ were
reported, namely by the E865 experiment at Brookhaven National
Laboratory \citep{sher03}, the KTeV Collaboration at Fermilab
\citep{alexopoulos04}, the NA48 Collaboration at CERN
\citep{lai04} and the KLOE Collaboration at Frascati
\citep{ambrosino06a,franzini04}. All experiments have determined
$|V_{us}| f_{+}(0)$ from charged kaon and/or neutral kaon decays,
with the form factor $f_{+}(0)$ taking into account $SU(3)$
breaking and isospin breaking effects. \citet{leutwyler84}
calculated
\begin{equation}
f_{+}(0) = f_{+}^{K^{0}\pi^{-}} = 0.961(8),
\end{equation}
a value that was confirmed by lattice calculation
\citep{becirevic05}, while chiral perturbation theory yields
values that are slightly larger, i.e. 0.974(11) to 0.981(10)
\citep{bijnens03,cirigliano04,jamin04}. In the case of charged
kaons (E865 experiment) the correction factor is
\begin{eqnarray}
f_{+}(0) & = & f_{+}^{K^{+}\pi^{0}} \simeq 1.022 f_{+}^{K^{0}\pi^{-}} \nonumber \\
& & = 0.982 \pm 0.008 \pm 0.002,
\end{eqnarray}
\noindent due to $\pi-\eta$ mixing induced by $m_d - m_u$ mass
splittings \citep{czarnecki04}. An overview of the results for
$|V_{us}|$ obtained from these new measurements is given in
Table~\ref{tab:Vus}.

\begin{table}[!htb]
\begin{center}
\begin{ruledtabular}
\begin{tabular}{l|lll}
  Experiment & Decay & $|V_{us}|f_{+}(0)$ {\it $^{1)}$\,} & $|V_{us}|$ {\it $^{2)}$\,}\\
\hline
 E865 & $K^{+} , e3$ &  0.2243(22)(7)$^3$ & 0.2284(23)(20) \\
\hline
 KTeV & $K_{L} , e3,\mu3$ & 0.2165(12)$^4$ & 0.2253(13)(20) \\
\hline
 NA48 & $K_{L} , e3$ & 0.2146(16)$^5$ & 0.2233(17)(20) \\
 \hline
 KLOE$^6$ & $K_{L} , e3,\mu3$ & 0.21673(59) & 0.2255(6)(20)$^7$ \\
 \hline
 \hline
 weighted average &  &  & 0.2254(21) \\
\end{tabular}
\caption{Results for $|V_{us}|$ obtained from the recent
measurements of $|V_{us}| f_{+}(0)$ in neutral and charged kaon
decays. ($^{1}$: for K$^+$ decay $f_{+}(0)$=0.982(8), while for
K$_L$ decay $f_{+}(0)$=0.961(8) (see text); $^{2}$: the first
error is due to experimental uncertainties; the common error of
0.0020 is related to the uncertainty of $f_{+}(0)$; $^{3}$:
Sher~\emph{et al.}~\citeyearpar{sher03}; $^{4}$:
Alexopoulos~\emph{et al.}~\citeyearpar{alexopoulos04}; $^{5}$:
Lai~\emph{et al.}~\citeyearpar{lai04}; $^{6}$: a result obtained
at KLOE for the $K_S,e3$ decay is not included here as only a
preliminary value, i.e. $|V_{us}|$ = 0.2254(17)~\cite{franzini04},
is available to date; $^{7}$: Ambrosino~\emph{et
al.}~\citeyearpar{ambrosino06a}).} \label{tab:Vus}
\end{ruledtabular}
\end{center}
\end{table}


\noindent All values are in good agreement with each other leading
to the weighted average
\begin{equation}
|V_{us}| = 0.2254(21)  ~~~  {\rm (kaon ~ decays).}
\label{eq:Vus}
\end{equation}
The central value is obtained as the weighted average using only
the first error in Table~\ref{tab:Vus}. The error is the
quadrature of 0.0020, from $f_+(0)$, and 0.0005 from experiment,
and is thus dominated by the $f_+(0)$ calculations. This new value
for $V_{us}$ is 2.6$\sigma$ larger than the value recommended by
the Particle Data Group \citep{eidelman04}. Combining this with
the values of $V_{ud}$ leads to perfect agreement with unitarity
as can be seen in the last column of Table~\ref{tab:unitarity}.

\begin{figure}[!htb]
\begin{center}
\includegraphics{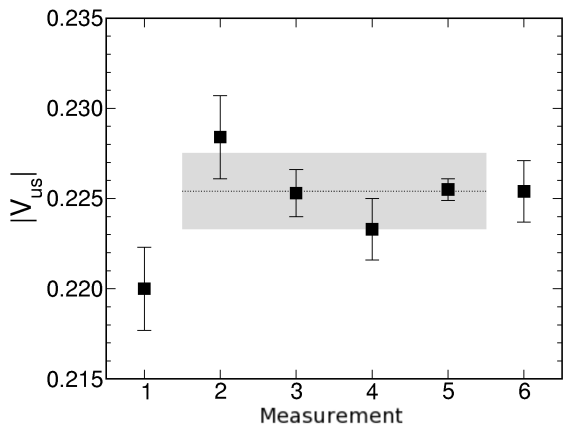}
\caption[V_us]{Values for $|V_{us}|$ from the Particle Data Group
analysis (1:~\citealp{eidelman04}) and from recent results in $K$
decays (2:~\citealp{sher03}, 3:~\citealp{alexopoulos04},
4:~\citealp{lai04}, 5:~\citealp{ambrosino06a}, 6:preliminary
result from KLOE~\citealp{franzini04}). The shaded band indicates
the weighted average of the published new results from $K$-decays
(refs. 2 - 5). See also Table~\ref{tab:Vus}. \label{fig:Vus}}
\end{center}
\end{figure}

\noindent $|V_{us}|$ can also be extracted from hyperon $\beta$
decay data. It is interesting to note that the new values for
$|V_{us}|$ from kaon decays are in good agreement with the value
$|V_{us}|$ = 0.2258(27) that was previously obtained from the
analysis of semi-leptonic hyperon decays \citep*{garcia92}.
However, the analysis leading to this result has theoretical
uncertainties because of first-order $SU(3)$ symmetry-breaking
effects in the axial-vector couplings. \citet{cabibbo03} have
therefore reanalyzed the hyperon $\beta$ decay data using a
technique that is not subject to these effects by focusing the
analysis on the vector from factors. They obtained $|V_{us}|$ =
0.2250(27), again in good agreement with the new values from kaon
decays (Table~\ref{tab:Vus}, Fig.~\ref{fig:Vus}).

Recent experimental results on hadronic $\tau$ decays into strange
particles obtained by the OPAL Collaboration \cite{gamiz05},
yielded $V_{us}$= 0.2208(34). This is somewhat lower than the
values from kaon decays but, within the error bar, still in
agreement with these. The error is dominated by experiment and
should be improvable in the future. The main complications in this
type of analysis are discussed by \citet{maltman05}.

Further, \citet{marciano04} showed that combining the ratio of the
experimental kaon and pion decay widths
\begin{equation}
\frac{\Gamma( K \rightarrow \mu \nu (\gamma))} {\Gamma( \pi
\rightarrow \mu \nu (\gamma))} \label{branching}
\end{equation}
\noindent with lattice gauge theory calculations of the ratio $f_K
/ f_\pi$ of the kaon and pion decay constants and the value
$|V_{ud}|$ from superallowed $\beta$-decays, provides a precise
value for $|V_{us}|$. Using
\begin{equation}
f_K / f_{\pi} = 1.120(4)(13)
\end{equation}
\noindent from \citet{aubin04}, \citet{marciano04} found
\begin{equation}
|V_{us}| = 0.2219(25)  ~~~  {\rm (lattice ~ 1),}
\end{equation}
\noindent while the value
\begin{equation}
f_K / f_{\pi} = 1.198(3)(^{16}_{-5})
\end{equation}
\noindent of \citet{bernard05} leads to
\begin{equation}
|V_{us}| = 0.2241(25)  ~~~  {\rm (lattice ~ 2).}
\end{equation}
\noindent Although both values agree with Eq.~\ref{eq:Vus} they
still differ by about 1$\sigma$. Since, in addition, the accuracy
on $|V_{us}|$ is in both cases dominated by the error on $f_K /
f_\pi$, further improvements in the lattice determination of this
ratio would be desirable. A reduction of the combined error on
$f_K / f_\pi$ by a factor of 2 to 4 may indeed be possible
\cite{marciano04}. It is finally to be noted that the absolute
branching ratio for the $K^+ \rightarrow \mu^+ \nu (\gamma)$ decay
was recently remeasured with the KLOE detector
\cite{ambrosino06b}. The result is in agreement but slightly more
precise than the value adopted by the Particle Data Group
\cite{eidelman04} that was used by \citet{marciano04} to calculate
$\Gamma( K \rightarrow \mu \nu (\gamma)) / \Gamma( \pi \rightarrow
\mu \nu (\gamma))$ (Eq.~\ref{branching}).

Combining now the weighted average value from the three types of
$\beta$ decay, {\em i.e.} $|V_{ud}|$ = 0.9738(4) (note that this
is identical to the value from the superallowed Fermi decays) with
the weighted average $|V_{us}|$ = 0.2254(21) from the recent
measurements in kaon decays, yields for the current test of
unitarity
\begin{equation}
\sum_i V_{ui}^2 = V_{ud}^2 + V_{us}^2 + V_{ub}^2 = 0.9991(12) \: ,
\end{equation}
showing no sign for physics beyond the standard model at the
present level of precision.

Finally, it is to be noted that the values of $f_{+}(0) =
f_{+}^{K^{0}\pi^{-}} \simeq$ 0.974 - 0.981 obtained from chiral
perturbation theory \cite{jamin04,bijnens03,cirigliano04} result
in a significantly lower weighted average in the last column of
Table~\ref{tab:Vus}, i.e. $|V_{us}|$= 0.2208(21) - 0.2224(21),
leading to
\begin{equation}
\sum_i V_{ui}^2 = V_{ud}^2 + V_{us}^2 + V_{ub}^2 = 0.9970(12)  -
0.9977(12) \: ,
\end{equation}
\noindent which again deviates by 1.9$\sigma$ to 2.5$\sigma$ from
unitarity. Resolving this ambiguity in the value of $f_{+}(0)$
should therefore be vigorously pursued. \citet{marciano04} has
pointed out that improvements in this respect may be expected from
lattice gauge theory calculations.



Depending on the outcome of new and more precise calculations of
the factor $f_{+}(0)$ in kaon decay the long standing so-called
``unitarity problem''\citep*{towner03} (see below) may finally be
solved. However, because of its impact on the result of the
unitarity test, the issue of the value of $V_{us}$ should be
definitely settled.

Since the precision on $|V_{ud}|$ from neutron decay is still well
below what is presently obtained for the superallowed $0^+
\rightarrow 0^+$ transitions, new experiments in neutron decay are
important too. The value of $|V_{ud}|$ from neutron decay is
statistics limited and there is still room for improvement with
the present experimental techniques. The same holds for pion beta
decay.

Over the last decades significant progress was made in improving
the precision and reliability of the experimental input data for
the $0{^+}\rightarrow 0{^+}$ transitions and for neutron decay,
but also in calculating the corrections that have been described
above. In a recent critical analysis \citep*{towner03}, prior to
the new results for $V_{us}$, it was pointed out that if only the
data for the $0{^+}\rightarrow 0{^+}$ transitions are at variance
with unitarity this could be due to a non-perfect understanding of
the nuclear structure dependent corrections $\delta_{NS}$ and
$\delta_C$ since these are absent in neutron decay. If the data
for both the $0{^+}\rightarrow 0{^+}$ transitions and neutron
decay are at variance with unitarity this might be due to a
non-perfect knowledge of the nucleus independent but model
dependent radiative correction $\Delta_R^V$. As long as the new
value for $V_{us}$ is not firmly established it would be useful to
address both types of corrections in detail again. Whereas both
the neutron and the pion results are still statistics limited, the
dominant contribution to the precision of $|V_{ud}|$ obtained from
the $0{^+}\rightarrow 0{^+}$ nuclear decays comes from
$\Delta_R^V$, which is responsible for most of the uncertainty of
the result $|V_{ud}|$ = 0.9738(4) \citep{hardy05a,hardy05b}. It is
interesting to note that a new calculation of $\Delta_R^V$ by
\citet{marciano06} reduces the error on this radiative correction
by a factor of 2, leading to
\begin{equation}
|V_{ud}| = 0.97377(27)   .
\end{equation}
\noindent for the data listed by \citet{hardy05a}. Finally, since
$\Delta_R^V$ also contributes to neutron decay experiments,
neutron decay would be able to test whether there are important
systematic problems with the nucleus-dependent corrections
($\delta_C$ and $\delta_{NS}$) but cannot test unitarity with a
significantly better precision than the nuclear decays.

If the new value $|V_{us}|$ = 0.2251(21) is confirmed, unitarity
is validated at the 10$^{-3}$ precision level for the
$0{^+}\rightarrow 0{^+}$ transitions and this then permits to set
stringent limits on different types of new physics (see
Sec.~\ref{sec:exotic} and Sec.~\ref{sec:parity}).


\subsubsection{Prospects for superallowed Fermi transitions}
\label{sec:prospects Fermi}

A number of precision nuclear spectroscopy experiments are ongoing
or planned to check the nuclear structure dependent corrections
for the $0{^+}\rightarrow 0{^+}$ superallowed Fermi transitions.
The total nuclear structure dependent correction ($\delta_C -
\delta_{NS}$) is the second largest contribution to the error
budget on $|V_{ud}|$ \citep*{towner03}. These corrections have
been validated only to about 10\% of their values, which range
from 0.25 to 0.77\% for the eight transitions that are currently
best known. In order to improve on this, the available data set
for the $0{^+}\rightarrow 0{^+}$ transitions is now being
significantly extended. New technical developments such as
improved detection techniques at isotope separators and recoil
separators, Penning traps for precision mass and $Q$-value
measurements and improved production techniques for exotic
isotopes permit precision measurements on several new
$0{^+}\rightarrow 0{^+}$ transitions of $T_z = -1$ nuclei with $18
< A < 42$ like $^{18}$Ne, $^{22}$Mg, $^{26}$Si, $^{30}$S,
$^{34}$Ar, $^{38}$Ca and $^{42}$Ti, as well as on a number of
$0{^+}\rightarrow 0{^+}$ transitions in the decay of $T_z = 0$
nuclei with $A > 54$ like $^{62}$Ga, $^{66}$As, $^{70}$Br and
$^{74}$Rb \citep*{hardy05b}. With the first group of transitions
the present range of values for the total nuclear structure
dependent correction ($\delta_C - \delta_{NS}$) will be extended
from 0.77\% to 1.12\%. For the second group this correction has
values between 1.4 and 1.5\% \citep*{hardy02}. The aim is to
determine the $ft$-values for these transitions, which cover a
wide range of calculated values for ($\delta_C - \delta_{NS}$),
with a precision that is comparable to the present set of the
eight best-known transitions. If the $\mathcal{F}t$-values that
will be obtained after applying the calculated corrections are in
agreement with CVC, this will verify the calculated corrections
and act to reduce the uncertainty attributed to them, which are
currently based only on theoretical estimates. If not, this will
point to some other problem to be investigated in detail. First
data are already available for most of the nuclei mentioned above
\cite{hardy05b}. For $^{22}$Mg, $^{34}$Ar and $^{74}$Rb results
are sufficiently precise that they were included in the latest
analysis of the $0{^+}\rightarrow 0{^+}$ transitions
\citep{hardy05a}. Further, in the case of $^{74}$Rb a first
experimental result has been obtained for the
isospin-symmetry-breaking correction $\delta_C$ = 1.81(29)\%
\citep{kellerbauer04} which is in good agreement with the
theoretical value of $1.50(40)\%$ \citep*{towner03}.

Finally, a new determination of the $Q$-value of the superallowed
decay of $^{46}$V obtained from the masses of both $^{46}$V and
its decay daughter $^{46}$Ti, together with an investigation of an
earlier $Q$-value measurement of $^{46}$V has uncovered a set of 7
measurements that cannot be reconciled with modern data
\cite{savard05}. An analysis of the data used by \citet{hardy05a}
taking into account the new $Q$-value for $^{46}$V and neglecting
those 7 measurements leads to a shift of the average
$\mathcal{F}t$ value for the superallowed $0{^+}\rightarrow 0{^+}$
transitions of about 1$\sigma$ \cite{savard05}. Given the high
precision that is now routinely available in Penning trap based
mass measurements it would thus be desirable that the $Q$-values
for all these transitions be determined again.



\medskip
\subsubsection {Experiments in neutron decay}
\label{sec:unitarity_neutron}

In neutron decay new measurements of the lifetime and of several
correlation coefficients ({\em viz.} the $\beta$ asymmetry
parameter $A$ and the $\beta-\nu$ correlation coefficient $a$) are
ongoing and planned, which should lead to a reduction of the error
on $V_{ud}$. All experiments use cold, very cold and even
ultra-cold neutrons (UCN). Cold neutrons have energies in the
range $0.1 - 5$~meV, corresponding to wavelengths of $4 - 29$~\AA
and velocities of $140 - 1000$~m/s. UCN have energies of only
$\sim$10$^{-7}$~eV, corresponding to wavelengths of
$\sim900~$\AA and velocities of $\sim$~5~m/s so that they move
extremely slowly. Neutrons as slow as possible are required for
these measurements since in many experiments the neutron decay is
measured during its motion through an experimental set-up. It is
then desirable that the neutron spends as much time as possible in
the set-up as the slower it moves, the greater the probability
that it will decay inside the set-up.

Most neutron decay experiments obtain their neutrons from nuclear
reactors and spallation sources containing a moderator. The energy
spectrum of the neutrons produced at the different facilities
contains very few cold to UCN, the fraction of UCN amounting
typically to $\sim$10$^{-11}$ only. Cold neutrons are formed in
the rare process in which a thermal neutron looses almost all of
its energy in a single inelastic collision. The number of cold
neutrons can be increased by passing the beam of thermal neutrons
through an extra moderator, {\em e.g.} a container with liquid
deuterium ($T \approx 23-25$~K). The neutrons then reach a new
thermal equilibrium at the temperature of liquid deuterium, so
that the maximum of the Maxwellian spectrum is shifted to the
energy range of cold neutrons. The increase of the fraction of
very cold and UCN in this way requires very low temperatures for
the extra moderator ($\sim$10$^{-3}$~K for UCN) which causes
practical difficulties. However, with current techniques (see
below) neutrons can now be stopped completely and be stored for a
time typically as long as their lifetime in a certain volume such
that one can simply wait for their decay. This resulted in an
enormous gain in measurement efficiency because the neutron loss
rate (which is the main limitation in lifetime measurements with
cold neutrons) as well as other sources of systematic errors are
significantly reduced.

At the current level of precision most techniques using cold
neutrons for lifetime experiments have reached their systematic
limits, such that significant progress in precision can only be
made when UCN are used. Lifetime experiments at present-day UCN
sources have provided values for the neutron lifetime which are a
few times more precise than those from beam experiments (see {\em
e.g.} \citealp{arzumanov00,serebrov05a} and Fig.~\ref{fig:tau-n}).
The UCN have too low energy to penetrate the surface of a material
and therefore undergo total external reflection at all angles. The
probability to be absorbed on each bounce has been measured to be
of the order of less than one in ten thousand in several
materials. UCN can therefore be stored for several hundreds of
seconds \citep{huffman00b}, and can also be guided through pipes
with sharp bends. All this enables experiments with UCN to be
shielded from the production source of the neutrons, both by
physical shielding (since the neutrons can be guided around the
shielding material) and by time (as one can store the neutrons
until the background caused by their production has died away). If
UCN are also used for measurements of correlations between the
neutron spin and the momenta of the leptons emitted in free
neutron decay a further increase in precision can be expected here
too. Several such experiments are under preparation (see {\em
e.g.}~\citealp{carr00}).


\medskip
{\it a. Neutron lifetime}
\label{sec:neutron-lifetime-experiments}
\medskip

Since the neutron decays via a mixed transition, any correlation
experiment in neutron decay has to be combined with the neutron
lifetime in order to fix the mixing ratio $\lambda$
(Sec.~\ref{sec:neutron-unitarity}). To determine the neutron
lifetime both beam and storage experiments are used. In the first
case decays from a neutron beam passing through an apparatus are
observed, while in the second neutrons are stored for a while in a
volume inside the apparatus and the remaining neutrons are
counted.

\begin{figure}[!htb]
\begin{center}
\includegraphics[width=\columnwidth]{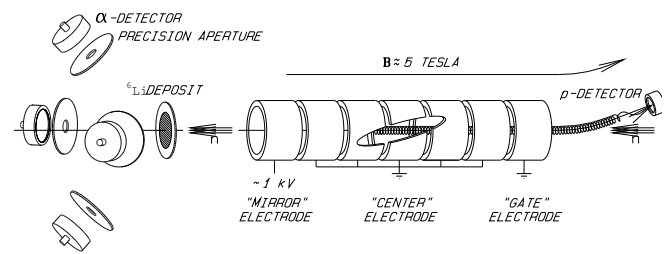}
\caption[Neutron lifetime at NIST]{Schematic drawing of the NIST
Penning trap neutron lifetime experiment. Details are given in the
text. From Dewey \citeyearpar{dewey03}.
\label{fig:n-lifetime-NIST}}
\end{center}
\end{figure}

At NIST a measurement was recently performed with the set-up shown
in Fig.~\ref{fig:n-lifetime-NIST}. In this type of beam experiment
\citep{dewey03} one measures simultaneously both the number $N$ of
neutrons in a well-defined volume of a neutron beam and the number
of neutron decays $dN / dt$ in the same volume. The lifetime is
then determined from the ratio $\tau_n$ = $N / (dN/dt)$. The
number of decays is obtained by trapping the protons from neutron
decay in a cylindrical Penning trap and sending them at regular
intervals onto a detector for counting \cite{byrne90,byrne96}. The
number of neutrons that are present in the decay volume is
determined by counting the number of $\alpha$-particles or tritons
emitted from the prompt decay of $^7$Li after neutron capture on a
well characterized isotopic target of $^6$LiF. The resulting
value, $\tau_n$ = $(886.6 \pm 1.2_{stat} \pm 3.2_{syst})$~s
\citep{dewey03,nico05a}, is the most precise measurement of the
neutron lifetime to date using an in-beam method. The error is
dominated by systematics which is mainly caused by uncertainties
in the mass of the (LiF)-$^6$Li deposit and the neutron capture
$^6$Li($n$,$t$) cross section. Continuing efforts to measure the
neutron count rate are underway by both calorimetric and
coincidence techniques, which should reduce the present
uncertainty by about a factor of two.


The second strategy for measuring the neutron decay rate is based
on the storage of UCN. The storage volume is defined either by
material surfaces, by gravity or by the interaction of the neutron
magnetic moment with a magnetic field gradient. Conceptually these
experiments are rather simple. UCN are first injected and trapped
in a storage volume with suitable walls, the ``bottle''. After a
certain storage period the bottle is emptied and the number of
surviving neutrons, $N(t)$, is measured. Repeating this experiment
with different storage times yields the decay curve of the
neutrons $N(t) = N(0) \exp{(-t/\tau_n)}$ which is then fitted to
extract the lifetime $\tau_n$. Special care has to be taken to
correct for leakage, mainly due to absorption and inelastic
scattering on the walls of the bottle. The total probability of
neutron losses in the storage volume, $P_s = 1/\tau_s$, is
determined by the sum of the probability of the neutron decay,
$P_n = 1/\tau_n$, and the probability of leakage, $P_l =
1/\tau_l$, {\em viz.} $1/\tau_s = 1/\tau_n + 1/\tau_l$. Several
methods are used to separate the different loss mechanisms
\citep{mampe89,nesvizhevsky92}. Two experiments were recently
performed at the ILL with the walls of the storage bottle being
coated with a film of hydrogen-free Fomblin oil. In the first the
neutrons were trapped in a material bottle with variable volume
and Fomblin oil at $\sim$250~K, yielding $\tau_n = (885.4 \pm
0.9_{stat} \pm 0.4_{syst})$~s \citep{arzumanov00}. The second
experiment used low temperature Fomblin oil (at $\sim$110 K) so as
to further reduce systematic errors \cite{serebrov05a}. The
result, $\tau_n$ = (878.5$\pm$0.7$_{stat}\pm$0.3$_{syst})$~s,
surprisingly differs by 5.6$\sigma$ from the previous result and
by 6.5$\sigma$ from the former world average value. It is
interesting to note in this respect that a new experiment using a
gravitational storage system with a wall coating of low
temperature Fomblin oil (105 K to 150 K) is planned at ILL
\cite{yerozolimsky05}.

Recently, progress was made at NIST towards magnetic trapping of
UCN \citep{huffman00a, huffman00b}. Due to the magnetic moment of
the neutron a magnetic field gradient will, depending on its
orientation, either accelerate neutrons and let them pass, or
retard them by creating a potential barrier without material
substance. By using a magnetic field as a boundary to reflect
neutrons, the problem of losses due to interactions with material
walls can be avoided. Together with the reduction of several other
systematic errors and a high yield this is expected to lead to
significantly improved precision. The UCN are produced by
inelastic scattering of cold ($8.9$ \AA) neutrons with phonons in
superfluid $^4$He (at T$<$250~mK) and are confined in a
three-dimensional magnetic trap using superconducting magnets
. The electrons emitted by the trapped neutrons ionize helium
atoms in the superfluid resulting in scintillation light pulses
that are recorded with nearly 100\% efficiency. The neutron
lifetime can be directly determined from the scintillation rate as
a function of time. A proof-of-principle of this technique has
been demonstrated \citep{huffman00b}. The apparatus is equipped
with a larger magnet for a measurement of the neutron lifetime at
the $10^{-3}$ level \citep{dewey01}. A further gain in precision
by at least another order of magnitude is anticipated
\citep*{alonso99,gabriel03} when combining this apparatus with a
higher-flux cold neutron source, such as the Spallation Neutron
Source (SNS) at the Oak Ridge National Laboratory.


The storage of UCN in a small magnetic trap made of permanent
magnets was also demonstrated \cite{ezhov01,ezhov05}. The measured
storage time in a test measurement was $(882 \pm 16)$ s, with no
depolarization being observed at this level of accuracy.

Another method that was suggested is to combine gravitational and
magnetic forces for spatial confinement (so-called spin trap)
\citep{zimmer00b}.


\medskip
{\it b. Neutron $\beta$ asymmetry parameter}
\medskip

Up to now the value of the mixing ratio $\lambda$ was usually
extracted from the $\beta$ asymmetry parameter $A$. However, as
the presently available results for $\lambda$ are not in very good
agreement (Fig.~\ref{fig:lambda-n}) new and more precise
determinations of the $A$ parameter are required.

\begin{figure}[!htb]
\begin{center}
\includegraphics[height = 6 cm, width = 8 cm]{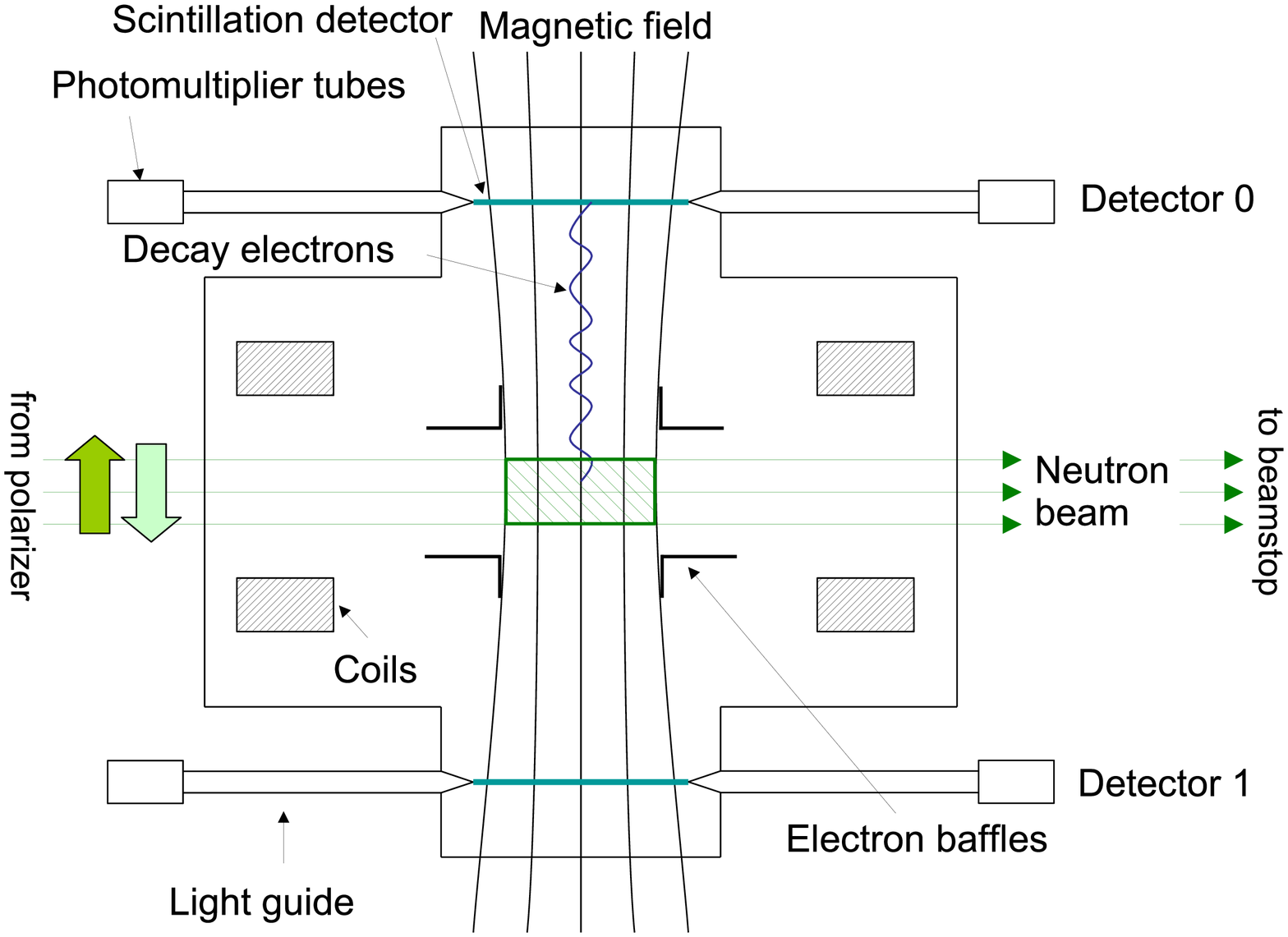}
\caption[PERKEO-II]{Schematic view of the PERKEO-II spectrometer.
The neutron beam passes through the apparatus. Electrons from
neutron decays in the center of the chamber are focused by a
strong magnetic field on two scintillator detectors. Adapted from
Reich $\emph{et al.}$ \citeyearpar{reich00}.
\label{fig:PERKEO-II}}
\end{center}
\end{figure}

The Heidelberg group has installed a ballistic super-mirror neutron
guide \citep{hase02} at the ILL.
%
This delivers the presently best and most intense polarized
neutron beam in the world, providing an increase of about a factor
of 6 in the cold neutron flux, corresponding to $2 \times 10^6$
neutron decays per second and per meter of beam length. By using
crossed super-mirror polarizers \cite{petoukhov03} a neutron
polarization larger than 99.5 \% is obtained over the full
cross-section of the neutron beam ($6 \times 20$ cm$^{2}$). The
neutron polarization is determined with a new polarimeter which is
based on spin-dependent neutron absorption in polarized $^{3}$He
and which yields a precision of about 0.1 \%
\citep{heil98,zimmer99}. Following these upgrades a new
high-precision determination of the $A$-parameter is being
prepared with the PERKEO-II set-up (Fig.~\ref{fig:PERKEO-II})
\citep{reich00}. The magnitude of the main correction is expected
to be reduced from 1.1\% to less than 0.5\% with an error of
0.1\%, which would lead to precision of 0.1\% or better on
$\lambda$ \citep{abele03}.

At LANSCE (Los Alamos) the UCNA collaboration has made progress
toward measuring the electron asymmetry parameter $A$ with
neutrons from a spallation-driven solid deuterium UCN source
\citep{carr00}. The use of UCN in combination with a
superconducting solenoidal spectrometer that ensures $4\pi$
coverage for the decay electrons, and a wire chamber/scintillator
combination as electron detector will greatly suppress the
backscattering of electrons at the surface of the detector
\cite{young01}. The precision aimed for is at the level of 0.3\% .

A group at PNPI (St.Petersburg, Russia) is preparing a new set-up
to measure the $A$ coefficient using cold neutrons and the axial
magnetic field in the shape of a ``bottle'' provided by a
superconducting magnet system \cite{serebrov05b}. Such
configuration permits to extract the decay electrons inside a
small solid angle with high accuracy. Background will be
suppressed by the use of electron-proton coincidences. An accuracy
at the level of a few $10^{-3}$ is being pursued.

A simultaneous measurement of the coefficients $A$ and $B$,
eliminating the need to determine the neutron polarization with
high precision, was carried out by \citet{mostovoi01}, yielding a
precision of 0.4\% on $\lambda$.

The abBA collaboration \cite{wilburn01,bowman04} prepares a
detector that would be able to measure the correlations $a$, $b$,
$A$, and $B$ with a precision of approximately $10^{-4}$, using a
pulsed neutron beam at the SNS in Oak Ridge. The experiment uses
an electromagnetic spectrometer combined with two large-area
segmented Si detectors to detect the decay proton and electron in
coincidence, with $4\pi$ acceptance for both particles. Measuring
four correlation coefficients with the same apparatus enables a
redundant determination of $\lambda$, with multiple cross checks
on systematic effects.

\begin{figure}[!htb]
\begin{center}
\includegraphics[height = 6 cm, width = 8 cm]{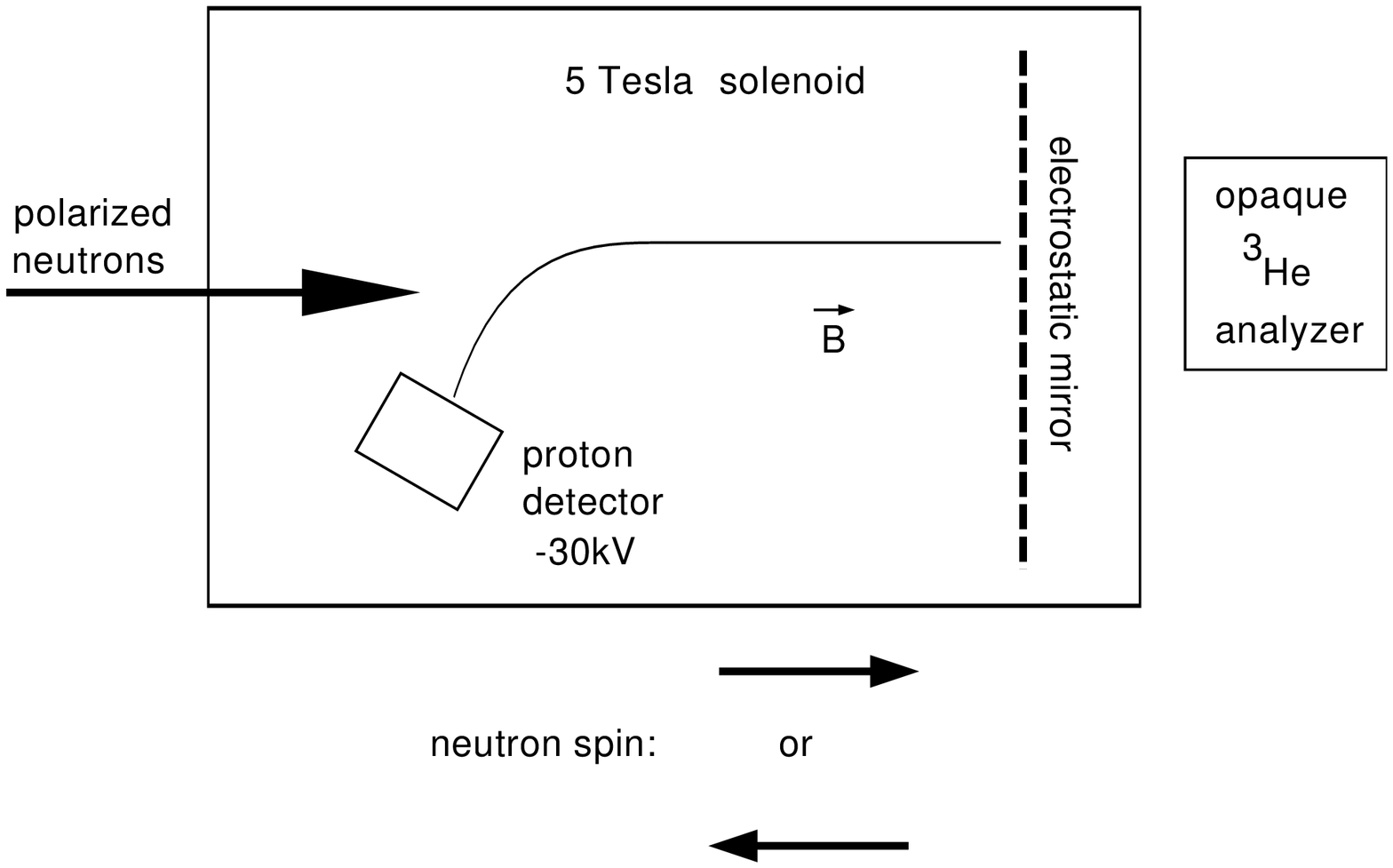}
\caption[C-NIST]{Schematic of the proposed set-up to measure the
spin-proton asymmetry coefficient $C$ at NIST. Details are given
in the text. From Dewey \citeyearpar{dewey01}. \label{fig:C-NIST}}
\end{center}
\end{figure}

Finally, at NIST an experiment is being set up to measure the
so-called spin-proton asymmetry parameter, $C$, in polarized
neutron decay \citep{dewey01}. This is proportional to $A+B$ and
is related to $\lambda$ via \cite{gluck96}
\begin{equation}
C \propto \lambda / (1 + 3 \lambda^2).
\end{equation}
In the proposed experiment (Fig.~\ref{fig:C-NIST}) longitudinally
polarized neutrons will be guided into a 5 T solenoid and the
decay protons, reflected by an electrostatic mirror, will then be
counted with a silicon detector. The number of decay protons
emitted parallel versus anti-parallel to the neutron polarization
yields the proton asymmetry $C$. Polarized $^{3}$He neutron spin
filters will be used for high accuracy neutron polarimetry. A
0.5\% measurement of $\lambda$ is envisaged with this method.

It is to be noted that since experimental precisions below 1\% are
now possible for $A$ (\citealp{abele}), the inclusion of
recoil-order effects and radiative corrections
\cite{holstein72,holstein74,holstein76,gluck97,gluck98b,gardner01,garcia01}
in the interpretation of the experimental data has to be
considered.

\medskip
{\it c. Beta-neutrino correlation in neutron decay}
\label{sec:beta-nu-neutron}
\medskip

A measurement of the beta-neutrino angular correlation coefficient
$a$ in neutron decay has a similar sensitivity to $\lambda$ as the
beta asymmetry parameter $A$. However, measurements of the
correlation coefficient $a$ are more difficult than measurements
of the $A$ parameter since the low energy ({\em viz.} $<$ 751 eV)
recoil protons have to be detected. Only a few precision
measurements of this coefficient have been carried out
\citep{grigoriev68,stratowa78,byrne02}. The two most precise
measurements, which yielded $a = -0.1017(51)$ \citep{stratowa78},
and $a = -0.1054(55)$ \citep{byrne02}, achieved a similar
precision, corresponding to $\lambda = -1.259(15)$ and $\lambda =
-1.271(18)$. Comparing these values with the present best result
from a measurement of the asymmetry parameter $A$, {\em i.e.}
$\lambda = -1.2739(19)$ \citep{abele02} (see also
Fig.~\ref{fig:lambda-n}), it is clear that the precision in
beta-neutrino correlation measurements has to be improved by
almost an order of magnitude in order to be competitive with
measurements of the $A$ parameter.
%
%
%
Note also that in view of the fact that the consistency of the
results for the asymmetry parameter $A$ is not very satisfactory
(Fig.~\ref{fig:lambda-n} and \citealp{garcia01}), it is important
that measurements leading to an improved precision for $a$ be
pursued.

In the most recent measurement \citep{byrne02}, $a$ was deduced
from the shape of the integrated energy spectrum of the recoil
protons from the $\beta$ decay of unpolarized neutrons.


In an experiment that is being prepared at NIST (``aCORN'')
\citep{dewey01,wietfeldt05} a new approach will be pursued. It
relies on a coincidence measurement between the decay electron and
the recoil proton and on the construction of an asymmetry that
directly yields $a$ without requiring precise proton spectroscopy.
The electron energy and the time-of-flight between electron and
proton detection will be measured. A new spectrometer was designed
for this. A statistical precision of less than 1 \% on $a$ is
anticipated, while it is planned to control all expected
systematic effects at the level of 0.5\% or less.

\begin{figure}[!htb]
\begin{center}
\includegraphics[height = 6 cm, width = 9 cm]{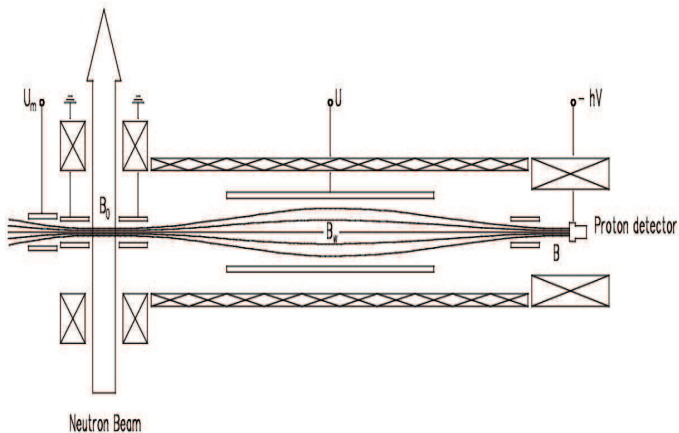}
\caption[aspect]{Schematic of the  $a$spect spectrometer. Details
are given in the text. From Zimmer $\emph{et al.}$
\citeyearpar{zimmer00a}. \label{fig:aspect}}
\end{center}
\end{figure}

Another method \citep{zimmer00a} is based on a magnetic
spectrometer with electrostatic retardation potentials. This
spectrometer, called $a$spect, is currently being developed at
Mainz and will be set up at the ILL. The main idea is to increase
the precision by completely separating the source of decay protons
from the spectroscopy part of the apparatus. The set-up is shown
schematically in Fig.~\ref{fig:aspect}. The neutron beam passes
through a region with a strong and rather homogeneous magnetic
field $B_0$. Decay protons which have an initial momentum
component towards the proton detector spiral along the field lines
and reach a region with a weak magnetic field $B_w$. In an
adiabatic motion, most of their initial kinetic energy
perpendicular to the field is transformed into longitudinal
kinetic energy, the exact fraction depending on the ratio $B_w /
B_0$. In the weak field region an electrostatic potential $U$ is
applied for the energy selection of the arriving protons. Protons
with total kinetic energy $T$ can overcome this potential barrier
only if their longitudinal energy is larger than $U$. A second
region with strong magnetic field $B \simeq B_0$ is used for
magnetic focusing of the protons onto the detector. Protons which
had enough energy to overcome the potential barrier are
post-accelerated in this region to a final energy of $\sim 30$~keV
in order to obtain a measurable signal. Counting the number of
protons as a function of the retardation potential $U$ permits to
measure the proton recoil energy spectrum which can be fitted to
obtain the beta-neutrino correlation coefficient $a$. This
experiment aims at a statistical error of about $2.5 \times
10^{-3}$ and a systematic error of about $1-2 \times 10^{-3}$
\citep{zimmer00a}.

Finally, a new measurement of $a$ is also planned at the UCN
source at Los Alamos \citep{young02} and at the SNS at Oak Ridge,
as discussed above \citep{bowman04}.

\medskip
{\it d. Rare neutron decay} \label{sec:radiative-decay}
\medskip

Efforts are ongoing to observe for the first time the radiative
decay mode of the free neutron. Whereas this decay branch is well
investigated already for other particles no efforts were done as
yet for the neutron. Recent theoretical calculations
(\citealp{gaponov96}) estimate this contribution to be about 1.5\%
of the total neutron $\beta$ decay probability and about 0.1\% for
the experimentally rather easily accessible energy region between
35 keV and 100 keV (above 100 keV the probability becomes
negligible). Recently, an experiment at the ILL-Grenoble has
yielded an upper limit of $6.9 \times 10^{-3}$ (90\% C.L.) for
this energy region~\cite{beck02}. In this experiment the radiative
decay mode is singled out by triple electron-proton-gamma
coincidences, with electron-proton coincidences signaling a normal
neutron $\beta$ decay. To reduce correlated background events from
bremsstrahlung emitted by the electron traveling through the
detector, a sectioned electron-gamma detector is used with the six
segments of the CsI gamma detector being placed at $35^{\circ}$
with respect to the axis of the plastic scintillator electron
detector. The experiment was recently moved to the new FRM-II
reactor in M{\"u}nchen. In a first phase a precision of about 10\%
on the branching ratio is aimed at.

At NIST a neutron radiative decay experiment is being set up
too~\cite{dewey01,fisher04}. This experiment will use the existing
apparatus for the lifetime measurement mentioned above, which can
provide substantial background reduction by using an
electron-proton coincidence trigger.

Note that if the radiative decay mode of the neutron can be
established, new correlations and polarization features in neutron
decay may be studied, including additionally the momentum or the
polarization of the radiative photon.

\subsection{Exotic interactions}
\label{sec:exotic}


In addition to the observed $V$ and $A$ type interactions the
general $\beta$ decay Hamiltonian includes also scalar ($S$) and
tensor ($T$) interactions, Eq.~(\ref{eqn:Hgeneral}). At the tree
level scalar type interactions in the $d \rightarrow u e^{-}
\overline{\nu}_e $ decay can arise from the exchange of Higgs
bosons and spin-zero or spin-one leptoquarks. In supersymmetric
models with R-parity violation it can be due to the exchange of
sleptons \citep{herczeg01}. They can appear also in so-called
composite models in the form of contact interactions
\citep{herczeg01, cornet97, zeppenfeld99}. Tensor type
interactions can arise from the exchange of spin-zero leptoquarks
and as contact interactions in composite models \citep{herczeg01}.

Constraints on $S$- and $T$-couplings in $\beta$ decay are usually
obtained either from the Fierz interference term $b$ or from the
$\beta$-$\nu$ correlation coefficient $a$.

The Fierz interference term $b$ depends linearly on the coupling
constants. In the standard model with only $V$ and $A$ couplings,
$b$ = 0. A measurement of $b$ yields a narrow unlimited band as
constraint in the $C_i$ versus $C_i^{'}$ ($i$= $S$ or $T$)
parameter plane. In addition, $b$ is identically zero if the
exotic couplings are purely right-handed ($C_i = -C_i^{'}$). Since
the Fierz interference term does not depend on any particular spin
or momentum vector it is an integral part of most measurements in
$\beta$ decay. It can easily be shown that in most correlation
measurements the actual quantity that is determined experimentally
is not $X$ but
\begin{equation}
\widetilde{X} = \frac{X} { 1 + \langle b^\prime \rangle }
\end{equation}
with $X = a, A, B, D, R$, etc., $b^\prime \equiv (m / E_e) b$ and
where $\langle \, \rangle$ stands for the weighted average over
the observed part of the $\beta$ spectrum.

The $\beta$-$\nu$ correlation coefficient $a$ depends
quadratically on the exotic couplings. A higher experimental
precision is thus needed in this case in order to get the same
absolute constraints on the couplings compared to measurements of
the Fierz interference term. However, a measurement of $a$
constrains a closed region in the parameter plane and is
independent of the helicity properties of the different
interaction types. Note that for a Fermi transition one has $a_F$
= +1 for a pure $V$-interaction and $a_F = -1$ for a pure
$S$-interaction, while for a Gamow-Teller transition $a_{GT} =
-1/3$ for a pure $A$-interaction and $a_{GT}$ = +1/3 for a pure
$T$-interaction.

Recently, a comprehensive analysis of experimental data for the
neutron lifetime and the correlation coefficients $a, A$ and $B$
in neutron decay was carried out \citep*{mostovoi00}. The analysis
assumed right-handed couplings for scalar and tensor interactions
and yielded (68\% C.L.) $|C^{(\prime)}_S/C_V| < 0.11$ and
$|C^{(\prime)}_T/C_A| < 0.08$.
%
%
%
Under the same assumptions the present analysis of the data set
including results from both neutron and nuclear $\beta$ decay
experiments, yields (95.5\% C.L.)
(Sec.~\ref{sec:fit-realcouplings}, case 4)
$|C_S/C_V| < 0.07$ and $|C_T/C_A| < 0.08$,
while the most general fit of neutron and nuclear $\beta$ decay
data (Sec.~\ref{sec:fit-conclusion}, case 6) yields (95.5\% C.L.)
$|C^{(\prime)}_S/C_V| < 0.07$ and $|C^{(\prime)}_T/C_A| < 0.09$.
%
%
%

Thus, 40 years after it was established that the weak interaction
is dominated by $V$ and $A$ currents \citep{allen59}, scalar and
tensor interactions are ruled out only to the level of about 5 to
10\% of the $V$- and $A$-interactions. The present constraints
still allow to accommodate sizable contributions of scalar and
tensor interactions without affecting our conclusions on the
phenomenology of semi-leptonic weak processes.


\subsubsection{Fierz interference term}
\label{sec:Fierz-scalar-tensor}

Strong limits on exotic couplings were recently obtained from the
Fierz interference term extracted from the $\mathcal{F}t$-value of
the superallowed $0^+ \rightarrow 0^+ $ transitions and from the
so-called polarization asymmetry correlation.

Assuming a non-zero Fierz interference coefficient $b$, the
$\mathcal{F}t$-value for the superallowed $0^+ \rightarrow 0^+ $
transitions is written as

\begin{equation}
\mathcal{F} t = \frac{K}{2 G^2_F V^2_{ud} (1+\Delta^V_R) }
\frac{1}{(1 + \langle b^\prime_F \rangle)}
\end{equation}
where $b^\prime_F$ is the Fermi part of the Fierz interference
term defined in Eq.~(\ref{eqn:bFierzapprox})\footnote{Note that
the factor $\gamma m / E_e$ which appears explicitly in a similar
expression in~\citet{towner03} has been included here in the
definition of the Fierz interference term $b^\prime$.}. The latest
analysis \citep{hardy05a} has yielded $(C_S + C_S^\prime)/C_V =
-0.0001(26)$ (assuming maximal parity violation for the vector
interaction), corresponding to $-0.0044 < (C_S + C_S^\prime)/C_V <
0.0044$ (90\% C.L.).

Strong limits for tensor couplings were previously obtained
\citep*{boothroyd84} from a measurement of the $b$ coefficient in
the decay of $^{22}$Na. However, these limits can be questioned
because of the large log$ft$-value for this beta transition such
that effects of higher order matrix elements can be important.

More recently, limits for tensor couplings were obtained from the
Fierz interference term in a so-called polarization asymmetry
correlation experiment where the longitudinal polarization of
positrons emitted by polarized $^{107}$In nuclei (log$ft$ = 5.6)
was measured (see Sec.~\ref{sec:rhc-longpol}), yielding $-0.034 <
(C_T + C_T^\prime)/C_A < 0.005$ (90\% C.L.)
\citep{severijns00,camps97}.



\subsubsection{Beta-neutrino correlation}
\label{sec:betanu-exotic}

Since neutrinos are very hard to detect, the $\beta$-$\nu$
correlation in semi-leptonic processes is usually investigated by
observing the $\beta$ particle and/or the recoiling nucleus,
taking into account the kinematics of the decay.

\medskip
{\it a. Indirect measurements of the recoiling nucleus}
\medskip

\citet{macfarlane71} and later \citet{clifford83,clifford89}
showed that the $\beta$-$\nu$ correlation can be obtained from the
kinematic broadening of $\beta$ delayed $\alpha$ particles. More
recently, several experiments were carried out to determine the
$\beta$-$\nu$ correlation from the Doppler shift of gamma rays
following the $\beta$ decay to an excited state of the daughter
nucleus. For $^{18}$Ne this yielded $a = 1.06 \pm 0.10$
\citep{egorov97}. The precision was limited by a systematic error
related to the effects of the slowing down of $^{18}$Ne in the
beryllium-oxide target. A similar measurement with $^{14}$O did
not yield a final result for $a$ due to unexpected problems
related to molecular binding effects \citep{vorobel03}.

\begin{figure}[!htb]
\begin{center}
\includegraphics[height = 5 cm, width = 6cm]{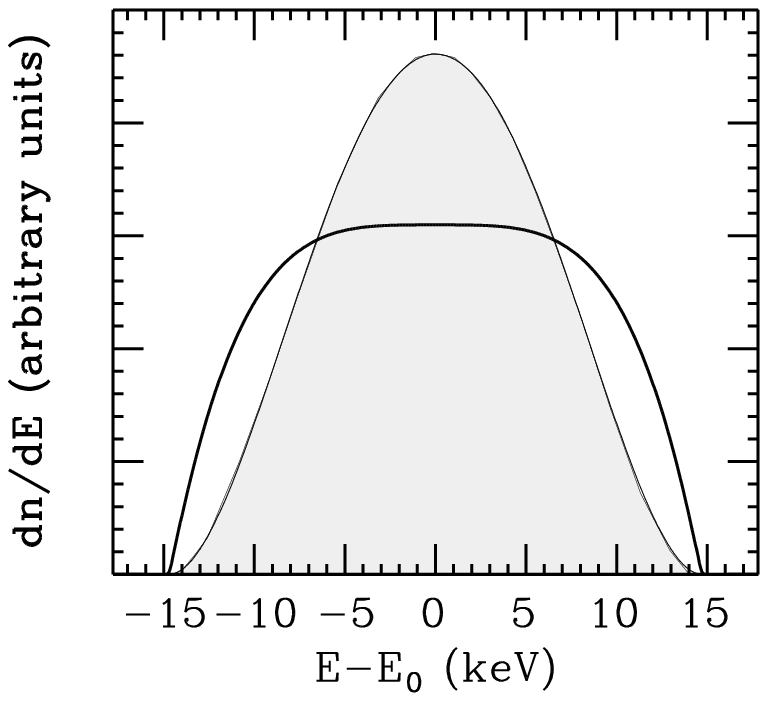}
\includegraphics[height = 6 cm, width = 5cm]{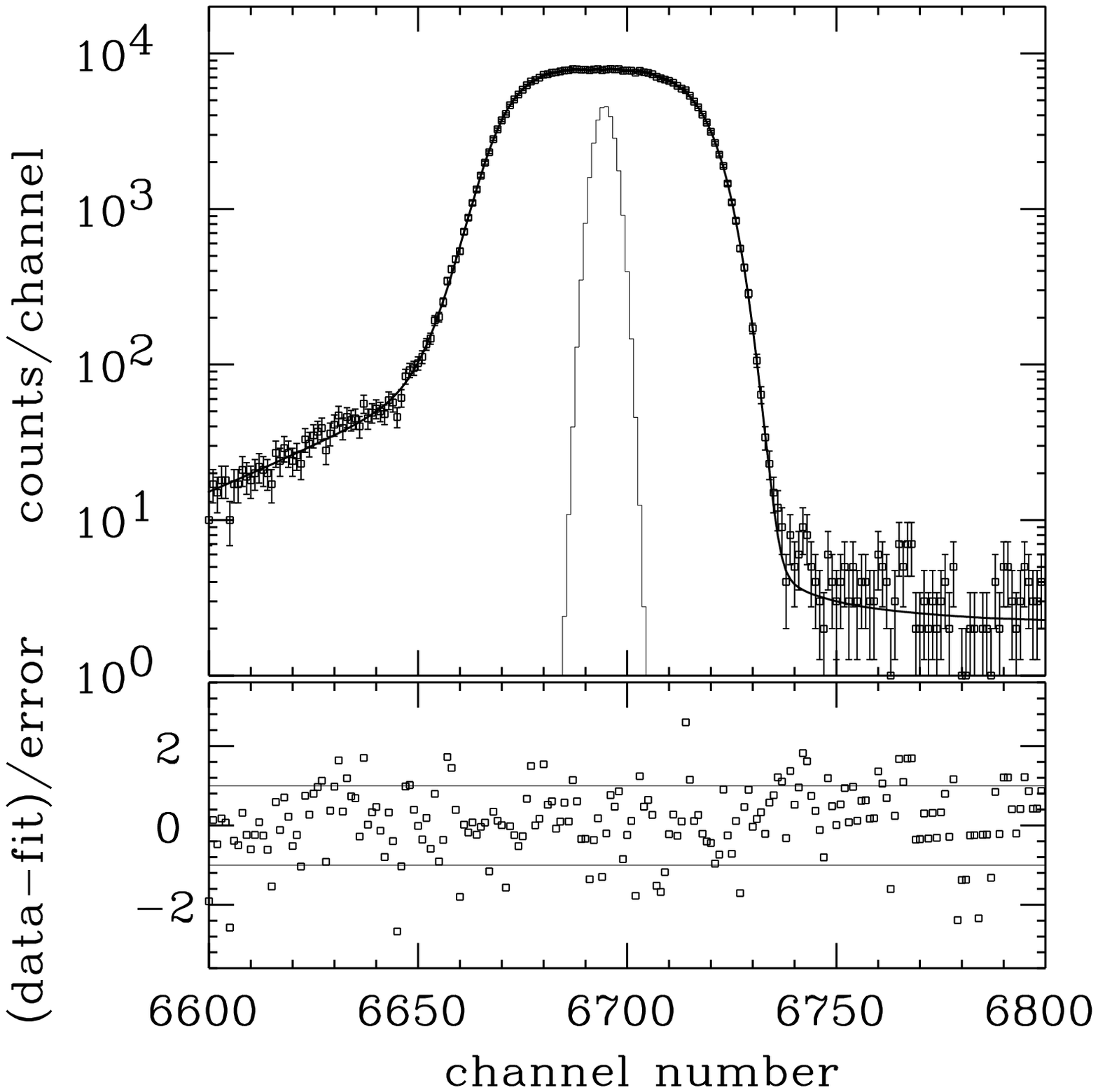}
\caption[32ArISOLDE]{Top: Shapes of the $\beta$ delayed proton
group from $^{32}Ar$ $0^+ \rightarrow 0^+$ decay for $a=+1, b=0$
(pure V interaction; flat curve) and $a=-1, b=0$ (pure S
interaction; 'Gaussian'-like curve). Bottom: Fit (upper panel) and
residuals (lower panel) of the proton peak (0.500 keV/channel).
The narrow pulser peak in the upper panel shows the electronic
resolution. From Adelberger $\emph{et al.}$
\citeyearpar{adelberger99}. \label{fig:32ArISOLDE}}
\end{center}
\end{figure}

\citet{schardt93} measured the kinematic broadening of $\beta$
delayed protons in the pure Fermi decay of $^{32}$Ar and the mixed
decay of $^{33}$Ar. These measurements were repeated at
ISOLDE-CERN with improved precision \citep{adelberger99,garcia00}.
The result for $^{32}$Ar is compared with theoretical expectations
for pure $S$- and $V$-interactions in Fig.~\ref{fig:32ArISOLDE}.
Fitting the shape of this delayed proton group yielded $\tilde{a}
= 0.9989 \pm 0.0052_{stat} \pm 0.0039_{syst}$, improving the
limits on a possible scalar contribution. The systematic error is
mainly due to the adopted error on the mass of $^{32}$Ar that was
obtained from a fit of the Isobaric Multiplet Mass Equation. A
direct mass measurement of $^{32}$Ar was meanwhile performed at
ISOLDE \citep{blaum03}. The reanalysis of the above mentioned
experiment, taking into account the measured mass of $^{32}$Ar is
in progress \citep{garcia03}.

\medskip
{\it b. Direct measurements of the recoil}
\medskip

The advent of ion and atom traps in nuclear physics has led to a
new series of measurements of the $\beta$-$\nu$ correlation $a$
and the $\beta$ emission asymmetry parameter $A$
\citep{sprouse97,kluge02,behr03}. These tools enable $\beta$
particles and recoil ions from $\beta$ decay to be detected with
minimal disturbance from scattering and slowing down effects.

\begin{figure}[!htb]
\begin{center}
\includegraphics[width=\columnwidth]{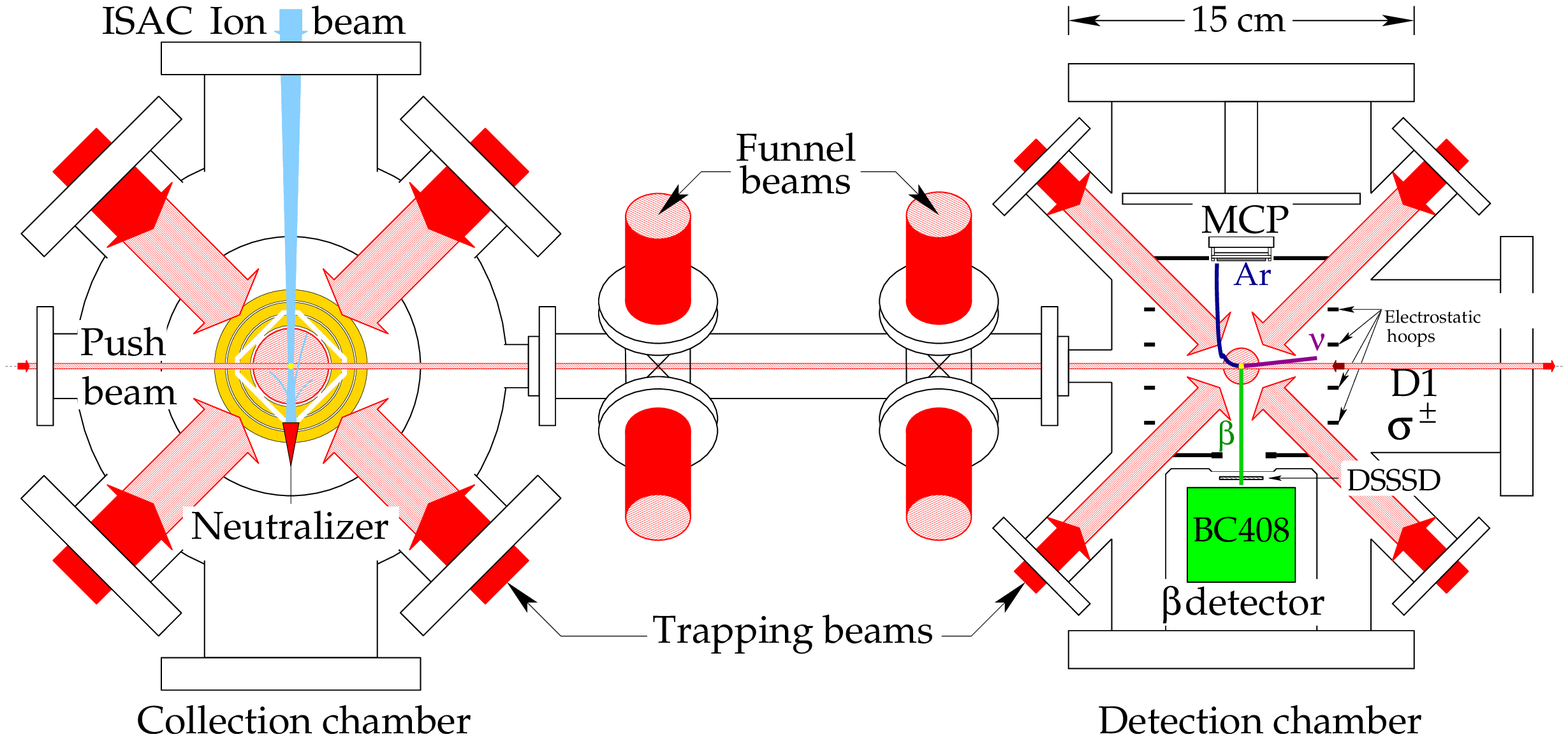}
\includegraphics[width=\columnwidth]{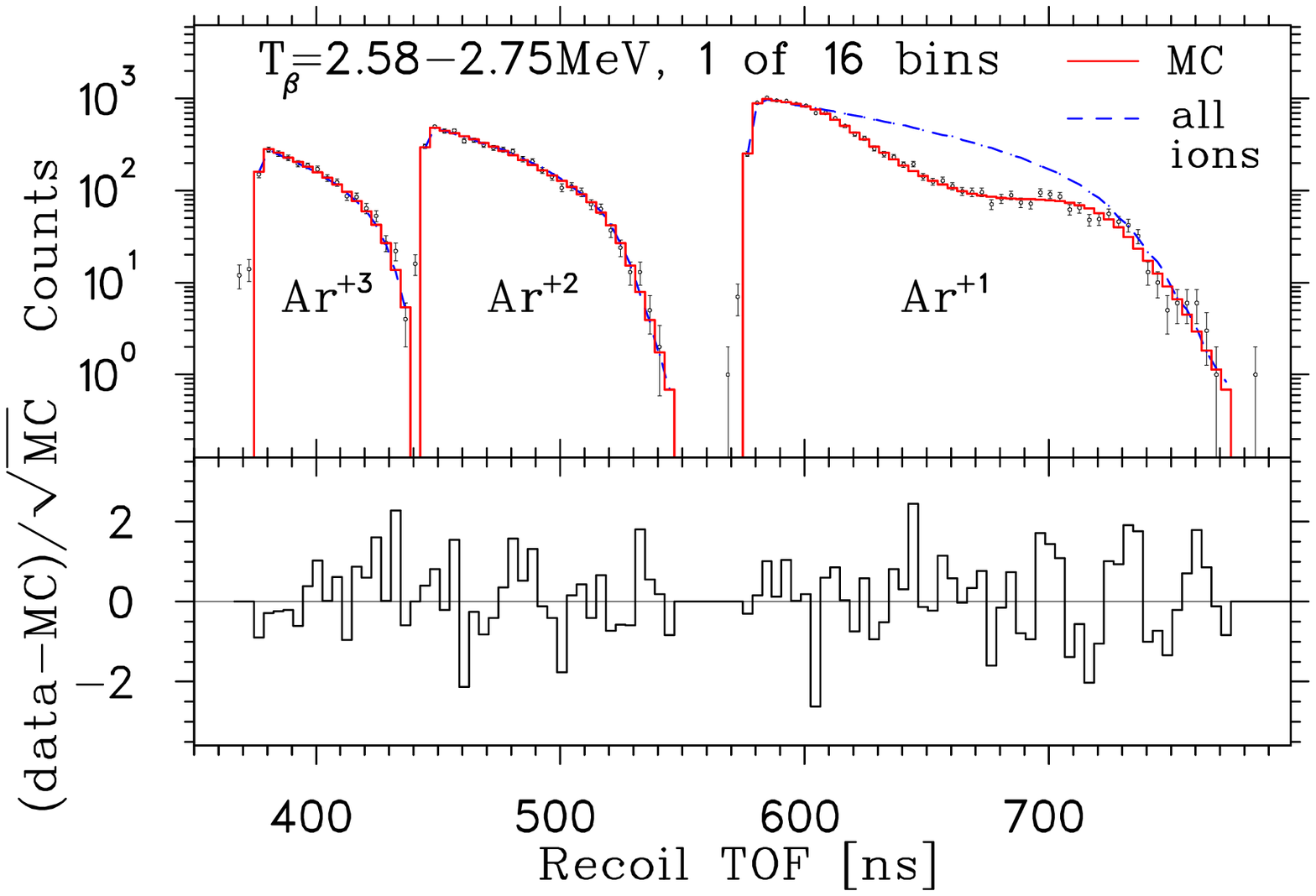}
\caption[TRINAT]{Top: Top view of the TRINAT two-MOT apparatus.
The ion beam is implanted in a neutralizer Zr foil in the trap at
the left. Atoms that leave the foil after heating are trapped with
six laser beams in three mutually perpendicular directions. At
regular intervals the trapped atom cloud is transferred to the
measurement trap at the right where the decay $\beta$ particles
and recoil ions are observed. The second MOT chamber is 15 cm in
diameter. Bottom: Time-of-flight of Ar recoils from $^{38}$K
decay. Ar charge states are separated by a 800 V$/$cm field. From
Gorelov {\em et al.}
\citeyearpar{gorelov00,gorelov05}.\label{fig:TRINAT}}
\end{center}
\end{figure}

The first successful application of an atom trap in a correlation
measurement in nuclear $\beta$ decay was the TRINAT experiment at
TRIUMF \citep{gorelov00,gorelov05} which uses two Magneto Optical
Traps (MOT) (Fig.~\ref{fig:TRINAT}) and is set up at the ISAC
isotope separator. The possible presence of a scalar interaction
was probed by investigating the $\beta$-$\nu$ correlation in the
pure Fermi decay of $^{38m}$K. The $^{38m}$K ions were implanted
in a Zr foil that was periodically heated in order to release the
atoms which were then trapped in a first MOT. To avoid the large
background from untrapped atoms, the trapped atoms were
transferred to a second MOT by a laser push beam and 2-D
magneto-optical funnels. A telescope detector for the $\beta$
particles and a Z-stack of three micro channel plates to detect
the recoil ions were installed in this second MOT. The recoil ion
energy was determined by its time-of-flight with the $\beta$
particle providing the start signal. Typical recoil time-of-flight
spectra are shown in Fig.~\ref{fig:TRINAT}.
The result $\tilde{a}$=0.9981$\pm$0.0030$^{+0.0032}_{-0.0037}$
\citep{gorelov05} is in agreement with the standard model.

At Berkeley a MOT was used to study the $\beta$-$\nu$ correlation
in the mixed decay of the mirror nucleus $^{21}$Na
\cite{scielzo03a,scielzo04}. As this transition is mainly (67\%)
of Fermi character, this experiment was predominantly sensitive to
scalar currents.
A $^{21}$Na atomic beam was produced with a proton beam from the
LBL 88" cyclotron. The $^{21}$Na atoms were loaded into a MOT
using a Zeeman slower. The correlation coefficient $a$ was
obtained from the time-of-flight spectrum of the recoiling
$^{21}$Ne ions from $^{21}$Na $\beta$ decays in the trap. The
result, $a = 0.5243(92)$, differs by about 3$\sigma$ from the
value of 0.558(3) calculated within the standard model using the
experimental $ft$-value \cite{naviliat91}. This deviation could be
caused by a systematic dependence of the result on the ion-trap
population~\cite{scielzo04}. Another possible explanation for the
discrepancy is the reliability of the $ft$-value. In particular,
there is a several percent branch to an excited daughter state to
be considered \citep{firestone96}. Several measurements of this
branching ratio have been carried out but the consistency of the
results is not satisfactory. It is planned to remeasure this
branching ratio with better precision both at TRIUMF
\citep{scielzo03b} and at the KVI-Groningen \citep{achouri04}.


Measurements of the $\beta$-$\nu$ correlation with radioactive
isotopes ($^{19}$Ne, $^{20}$Na and $^{21}$Na) stored in a MOT atom
trap are also planned at the new TRI$\mu$P facility at the
KVI-Groningen \cite{berg03a,berg03b}. Here radioactive isotopes
are produced in inverse-kinematics from fragmentation reactions
initiated with heavy ions accelerated in the superconducting
cyclotron AGOR. Reaction products are separated from the primary
beam in a dual-mode recoil and fragment separator. Beams of
isotopes of interest will be transformed into a low-energy,
high-quality, bunched beam and, after neutralization, stored in a
MOT for measurement.


%
Experiments to measure the $\beta$-$\nu$ correlation using
electromagnetic traps are being prepared too, one at GANIL
\citep{delahaye02,ban05a} and the other at ISOLDE-CERN
\citep*{beck03a,beck03b}.

\begin{figure}[!htb]
\includegraphics[width = 8cm]{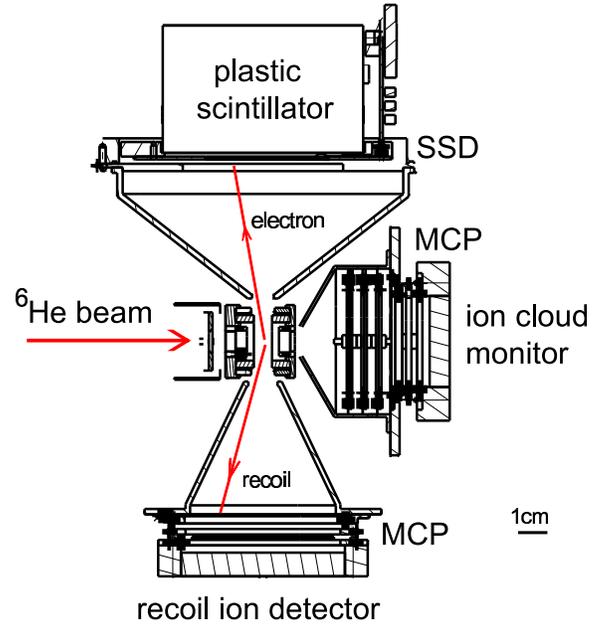}
\caption{Schematic lay-out of the LPC Caen transparent Paul trap
set-up with the beta and recoil detectors, and the ion cloud
monitor. The system is mounted on the low energy beam line of the
SPIRAL facility at GANIL (see text for details).}
\label{fig:LPCTrap-setup}
\end{figure}

The first experiment aims at determining the $\beta$-$\nu$ angular
correlation in the decay of $^{6}$He. This is a pure GT transition
and is thus sensitive to tensor couplings. The goal is to improve
the old experiment of \citet{johnson63} which determined $a_{GT}$
with a relative precision of about 1\%. The $^{6}$He nuclei are
produced by heavy ion reactions in the target/ion source system of
the SPIRAL facility at GANIL. The $^{6}$He$^+$ ions are extracted
from the source with energies in the range 10-35~keV. The
radioactive ion beam is then cooled and bunched in order to
increase the injection efficiency of the ions into a Paul trap.
The cooling and bunching is performed by a Radio Frequency
Quadrupole using the buffer gas cooling technique \cite{darius04}.
The cooling of $^{4}$He$^+$ ions using H$_2$ as buffer gas has
very recently been demonstrated \citep{ban04}, yielding
transmissions of up to 10\% which are enough to trap sufficient
ions into the Paul trap. The trap (Fig.~\ref{fig:LPCTrap-setup})
has been designed with an open geometry to reduce the scattering
of electrons on the electrodes while enabling the detection of the
decay products. The quadrupole trapping field is generated by four
concentric ring electrodes \citep{ban05a} mounted around the beam
axis. The $\beta$-$\nu$ correlation will be deduced from time of
flight measurements between the $\beta$ particles and the recoil
ions. The $\beta$ particles are detected by a telescope consisting
of a 300~$\mu$m silicon strip detector (SSD) and a thick plastic
scintillator while the recoiling ions are counted with a position
sensitive micro-channel plate. An additional ion detector is
mounted along the beam line to monitor the phase space of the
trapped ions cloud. The set-up has been commissioned and the proof
of principle has recently been demonstrated \cite{mery05}.

\begin{figure}[!htb]
\begin{center}
\includegraphics[width=\columnwidth] {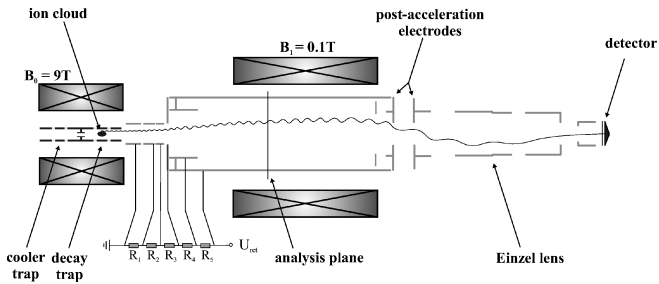}
\caption[WITCH-Trap]{Spectrometer section of the WITCH set-up.
Details are given in the text. From Beck {\em et al.}
\citeyearpar{beck03a}. \label{fig:WITCH-trap}}
\end{center}
\end{figure}

The second set-up (``WITCH'') is based on a magnetic spectrometer
with electrostatic retardation potentials and was installed at
ISOLDE-CERN. \citep{beck03a,beck03b} (Fig.~\ref{fig:WITCH-trap}).
A pulsed radioactive beam coming from the REXTRAP Penning trap at
ISOLDE is slowed down in a pulsed drift tube and caught in a first
(cooler) Penning trap situated in a 9 T magnetic field. The cooled
ions are transferred to a second (decay) Penning trap where they
are stored for some time. Recoil ions from decays in this second
trap spiral adiabatically from the high magnetic field to a low
magnetic field region (0.1 T) where a retardation potential is
applied. While the ions travel from the high to the low field
region most of their energy is converted into axial energy which
is then probed by the retardation potential. This is the same
principle which has been discussed already for the $a$spect
experiment (Sec.~\ref{sec:unitarity_neutron}) and that is used
also for the direct neutrino mass measurements
(Sec.~\ref{sec:numass}). Recoil ions with longitudinal energy
large enough to overcome the retardation potential are
re-accelerated and electrostatically focused onto a micro-channel
plate detector. The recoil energy spectrum, the shape of which
depends on the $\beta$-$\nu$ correlation coefficient $a$, is then
measured by scanning the retardation potential. The
WITCH-experiment will first focus on $^{35}$Ar. Since the
Gamow-Teller component in the mirror $\beta$ decay of $^{35}$Ar is
small (7\%), this will thus mainly probe the existence of scalar
weak currents. Eventually, also pure $0^+$-$0^+$ transitions
($^{26m}$Ar) and pure Gamow-Teller transitions ($^{122m}$In) will
be measured. The aim is to reach a precision on $a$ of about 0.5
\% or better.

\begin{figure}[!htb]
\begin{center}
\includegraphics{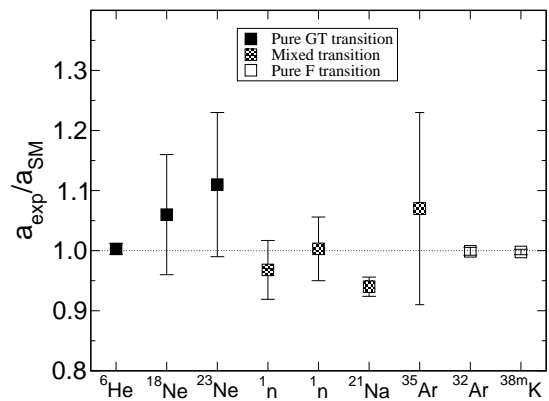}
\caption[a_best]{Results of the early $\beta$-$\nu$ correlation
measurements of \citet{allen59} for $^{23}$Ne and $^{35}$Ar,
compared to more recent measurements. Only results with a
precision better than 10\% are included while, in addition, for a
given isotope only the most precise result is shown. The about
$3\sigma$ deviation from the standard model for $^{21}$Na could be
caused by a systematic dependence of the result on the ion-trap
population, or by a problem with the value of the branching ratio
that was used to calculate the $ft$-value~\cite{scielzo04} (see
text). ($^{6}$He:~\citealp{johnson63};
$^{18}$Ne:~\citealp{egorov97}; $^{23}$Ne:~\citealp{allen59};
n:~\citealp{stratowa78} and ~\citealp{byrne02};
$^{21}$Na:~\citealp{scielzo04}; $^{35}$Ar:~\citealp{allen59};
$^{32}$Ar:~\citealp{adelberger99}; $^{38m}$K:~\citealp{gorelov05})
\label{fig:a_total}}
\end{center}
\end{figure}

Figure~\ref{fig:a_total} shows the results of $\beta$-$\nu$
correlation measurements with a precision better than 10\%
available to date.

\subsubsection{Beta-asymmetry parameter}
\label{sec:beta-asymmetry-exotic}

\begin{figure}[!htb]
\begin{center}
\includegraphics{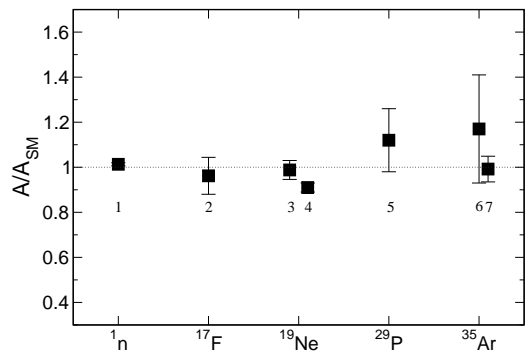}
\caption[a_best]{Results of the $\beta$ asymmetry parameter
measurements for the mixed transitions of the T=1/2 mirror nuclei.
For the neutron the weighted average value \cite{eidelman04} is
shown. The standard model values were calculated from the
experimental $ft$-values \cite{naviliat91}. Note that the second
result for $^{19}$Ne \cite{schreiber83} was never published.
More details can be found in Table~\ref{tab:A-mirror}.
(1:~\citealp{eidelman04}; 2:~\citealp{severijns89a};
3:~\citealp{calaprice75}; 4:~\citealp{schreiber83};
5:~\citealp{masson90}; 6:~\citealp{garnett88};
7:~\citealp{converse93}) \label{fig:mirror_A}}
\end{center}
\end{figure}

The asymmetry parameter $A$ in neutron decay is not very sensitive
to the presence of either real scalar or tensor currents (see
Eq.~(\ref{eqn:Aparapprox_exot})). Also for nuclear decays $A$ is
not sensitive to scalar currents as it vanishes for a pure Fermi
transition. Over the years a number of measurements of $A$ for T =
1/2 mirror transitions were carried out, mainly as a test of CVC
(Table~\ref{tab:A-mirror} and Fig.~\ref{fig:mirror_A}). Precision
measurements of $A$ for pure Gamow-Teller transitions, on the
other hand, permit the existence of a tensor component in the weak
interaction to be probed. Only a limited number of measurements of
this type were carried out till now (Table~\ref{tab:nuclear}) and
several of these results have a poor precision \cite{vanneste86}.
For $^{60}$Co two rather precise results were reported ({\em i.e}
$A = -1.01(2)$ \cite{chirovsky80} and $A = 0.972(34)$
\cite{hung76}), but as log$ft$ = 7.5 for this transition the
effect of recoil effects like weak magnetism may be important in
this case. The possibilities of Gamow-Teller transitions to search
for exotic weak interactions has thus not extensively been
explored yet. \vspace{5mm}

\begin{table}
\begin{center}
\begin{ruledtabular}
\begin{tabular}{c|c|c|c|c}
 Isotope & $A/A_{SM}$ & $A$ & $A_{SM}$ & $ft$-value \\
    &   &   &   &  (s) \\
\hline
 $^1n$ & $0.9995(95)$ & $-0.1173(13)^1$ & $-0.11736(12)$ & $1052.7(10)$ \\
 $^{17}F$ & $0.962(82)$ & $0.960(82)^2$ & $0.99715(16)$ &
 $2314.0(69)$ \\
 $^{19}Ne$ & $0.988(42)$ & $-0.0391(14)^3$ & $-0.03957(80)$ & $1725.1(44)$ \\
 $^{19}Ne$ & $0.910(20)$ & $-0.0363(8)^4$ & $ -0.03991(16)$ & $1726.8(4)$ \\
 $^{29}P$ & $1.12(14)$ & $0.681(86)^5$ & $0.6061(44)$ & $4869(17)$ \\
$^{35}Ar$ & $1.14(23)$ & $0.49(10)^6$ & $0.4303(84)$ & $5718(14)$ \\
 $^{35}Ar$ & $0.992(57)$ & $0.427(23)^7$ & $0.4303(84)$ & $5718(14)$ \\
\end{tabular}
\caption{Results of the $\beta$ asymmetry parameter measurements
for the mixed $\beta$ transitions of the T=1/2 mirror nuclei.
$^{1}$: \citet{eidelman04}; $^{2}$: \citet{severijns89a}; $^{3}$:
\citet{calaprice75}; $^{4}$: \citet{schreiber83}; $^{5}$:
\citet{masson90}; $^{6}$: \citet{garnett88}; $^{7}$:
\citet{converse93}. $ft$-values are from \cite{naviliat91}.}
\label{tab:A-mirror}
\end{ruledtabular}
\end{center}
\end{table}
At present, several new efforts in this domain are ongoing. At the
Los Alamos National Laboratory a MOT-based experiment is being
carried out \citep{vieira00,crane01}. $^{82}$Rb ions from an
isotope separator are transformed into atoms, using a Zr catcher
foil, and trapped into a first MOT. The trapped atoms are then
transferred with a laser push beam to a second MOT where they are
re-trapped and polarized by optical pumping. Applying a rotating
bias field with which the nuclear spin vector is aligned then
permits to measure with a single detector the $\beta$ asymmetry
parameter $A$ as a function of the $\beta$ particle energy and the
angle between the $\beta$ particle and the nuclear spin vector. A
precision at the 1\% accuracy level is aimed at \cite{hausmann04}.


At Leuven a new set-up to polarize nuclei using the method of low
temperature nuclear orientation \citep{postma86,vanneste86} has
recently become operational \citep{kraev05}. The set-up includes a
17 T superconducting magnet and a Si pin-diode particle detector
operating at a temperature of about 10 K. The nuclei are embedded
in a non-magnetic host lattice in order to avoid uncertainties
related to the lattice position of the nuclei when hyperfine
fields in magnetic host lattices are used to polarize nuclei. Here
too a precision below 1\% is envisaged.

Finally, another type of $\beta$ asymmetry measurement to search
for tensor currents is being prepared
\cite{severijns05a,severijns05b} at the ISOLDE facility, using the
NICOLE low temperature nuclear orientation set-up. It is a
relative measurement of the $\beta$ asymmetry parameter for two
isotopes of a single element, one decaying via a $\beta^+$
transition the other via a $\beta^-$ transition. Such relative
measurements have a two times higher sensitivity to tensor
currents compared to a single absolute measurement and, in
addition, are less affected by systematic effects.

\subsubsection{Limits from other fields}
\label{sec:other fields-exotic}

It was recently shown \citep{campbell05} that limits on induced
pseudo-scalar interactions, which are strongly constrained by data
on $\pi^\pm \rightarrow l^\pm\nu_l$ decay, imply limits on the
underlying fundamental scalar interactions that are, in certain
cases, up to an order of magnitude stronger than limits on scalar
interactions from direct $\beta$ decay searches. However, if the
new physics responsible for the effective scalar interactions
arises at the electroweak scale from the explicit exchange of new
scalars, limits from direct $\beta$ decay searches are comparable
to those from $\pi^\pm \rightarrow l^\pm\nu_l$ decay. Depending on
the underlying assumptions, the indirect limits from this decay
can even be weaker than the $\beta$ decay limits, leaving the
interest in searches for new scalar interactions in $\beta$ decay
experiments undiminished.

Limits on scalar and tensor couplings are also obtained from
$K_{e3}$ and $K_{\mu3}$ decays \citep{eidelman04}, and from the
purely leptonic decay of the muon \citep{herczeg95a, fetscher95,
fetscher98, kuno01}. It is to be noted that $K$-decay, muon decay
and $\beta$ decay yield complementary information.

Recently, the PIBETA Collaboration has found a strong deficit of
the branching ratio of the radiative decay of positive pions at
rest in the high-$E_\gamma$/low-$E_e$ kinematic region
\cite{frlez04}. The same anomaly was observed before in another
experiment, although with less statistical significance
\cite{bolotov90}. This deficit could be caused by an inadequacy of
the present $V - A$ description of the radiative pion decay, along
with the radiative corrections, or by a small admixture of new
tensor interactions which may arise due to exchange of new spin
one chiral bosons which interact anomalously with matter
\cite{frlez04,chizhov04}. This result clearly calls for further
theoretical and experimental work.

Finally, \citet{ito05} derived order of magnitude constraints
$|(C_S - C^\prime_S)/C_V| <$ 10$^{-3}$ and $|(C_T -
C^\prime_T)/C_A| <$ 10$^{-2}$ from the upper limit on the neutrino
mass. These results are complementary to those from measurements
of $b$, which are sensitive to ($C_S + C^\prime_S$) and ($C_T +
C^\prime_T$) and measurements of $a$, which are sensitive to
$|C_S|^2 + |C^\prime_S|^2$ and $|C_T|^2 + |C^\prime_T|^2$.


\subsection{Parity violation}
\label{sec:parity}

Whereas the violation of parity in the weak interaction was
discovered almost 50 years ago \citep{wu57}, its origin is still
today not understood. The most popular models explaining the
seemingly maximal violation of parity in the weak interaction are
so-called left-right symmetric models
(Sec.~\ref{sec:righthanded}).

%
%

Constraints on right-handed currents from $\beta$ decay come from
longitudinal positron polarization experiments with unpolarized
nuclei \citep{vanklinken83,wichers87,carnoy90}, measurements of
the longitudinal polarization of positrons emitted by polarized
nuclei \citep{severijns93,allet96,camps97,severijns98,thomas01},
experiments in neutron decay
\citep{deutsch99,abele,kuznetsov95,serebrov98} and the
$\mathcal{F}$t values of the superallowed $0{^+}\rightarrow 0{^+}$
transitions \citep{hardy05a}.

There is strong interest in more precise tests of maximal parity
violation in nuclear $\beta$ decay as these would provide new
constraints on models with exotic fermions, with leptoquark
exchange or with contact interactions \citep{herczeg01}.

\subsubsection{$\mathcal{F}$t value of superallowed Fermi
transitions} \label{sec:zeta}

The average $\mathcal{F}$t value for the superallowed
$0{^+}\rightarrow 0{^+}$ pure Fermi transitions provides a
stringent constraint on the mixing angle $\zeta$ between the left-
and right-handed $W$ gauge bosons. In a model where right-handed
currents are assumed, one can write the $\mathcal{F}$t value as

\begin{equation}
\mathcal{F} t = \frac{K} {2 G^2_F V^2_{ud} (1 - 2\zeta) (1 +
\Delta^V_R) }
\end{equation}
Using the previously cited value $\mathcal{F}$t = $3073.5(12)$~s
\citep{hardy05a} and the values for $|V_{us}|$ and $|V_{ub}|$
recommended by the Particle Data Group \citep{eidelman04}, and
requiring that $V_{ud}^2$ satisfies unitarity, \citet{hardy05a}
found $\zeta = 0.0018(7)$. This value deviates by about
$2.5\sigma$ from zero, the value that corresponds to maximal
parity violation. However, when the above mentioned weighted
average of the new values for $V_{us}$ obtained from measurements
in $K$ decays is used ($viz.$ $|V_{us}|=0.2251(21)$;
Sec.~\ref{sec:status-unitarity}), one has $\zeta$ = 0.0006(7). The
mixing angle for the left- and right-handed gauge bosons is thus
clearly limited to the milliradian region: $-0.0006 \leq \zeta
\leq 0.0018$ (90 \% C.L.). This is currently the strongest limit
on $\zeta$ from $\beta$ decay.

\subsubsection{Beta-asymmetry parameter}
\label{sec:rhc-beta-asym}

Parity violation in the weak interaction was first observed by
measuring the asymmetry parameter $A$ in the ($5^+ \rightarrow
4^+$) $\beta^-$ decay of polarized $^{60}$Co nuclei ~\citep{wu57}.
Twenty years later this experiment was repeated with a more
advanced set-up where the nuclear polarization could be rotated
using two crossed magnetic coils, yielding $A = -1.01(2)$
~\citep{chirovsky80,chirovsky84}.
%
%
%
Using Eq.~(\ref{eqn:Aparapprox_rhcGT}) the above result yields a
lower limit of 245 GeV/c$^2$ (90\% C.L.) for the mass of a vector
boson coupling to right-handed particles (assuming $\zeta = 0$).
However, this result should be taken with some caution as long as
the effect of recoil order corrections, like weak magnetism, has
not been evaluated for this transition.

The problem with recoil corrections can be avoided by measuring
the $\beta$ asymmetry parameter $A$ in the $\beta$ decay between
analog states of $T=1/2$ mirror nuclei. Due to the superallowed
character of these mirror $\beta$ transitions nuclear structure
dependent corrections are very small. A survey of the sensitivity
of such experiments for the mixed transitions of $T=1/2$ mirror
nuclei from $^{11}$C up to $^{43}$Ti has indicated
\citep{naviliat91} that for most transitions the required
experimental precision is of the order of 0.5\% in order to be
sensitive to a right-handed boson mass of 300 GeV/$c^2$ (90\%
C.L.). This requires a very precise determination of the degree of
nuclear polarization. A way to overcome this difficulty of
absolute measurements is to carry out relative measurements,
comparing the asymmetry parameter for the mixed mirror $\beta$
transition to that of a pure Gamow-Teller transition from the same
isotope. This is possible for several mirror nuclei, such as
$^{21}$Na, $^{23}$Mg, $^{29}$P and $^{35}$Ar, and was demonstrated
in the case of $^{29}$P ~\citep{masson90} and $^{35}$Ar
\citep{converse93}. Since all $\beta$ transitions have the same
sensitivity to $\delta^2$ ($\delta = (m_1/m_2)^2$, with $m_1$
($m_2$) the mass of the $W_1$ ($W_2$) boson; see
Sec.\ref{sec:righthanded}) independent of their Fermi/Gamow-Teller
character (see Eq.~(\ref{eqn:AparmirGT_rhc})), the sensitivity to
$\delta^2$ is lost in such relative measurements.
%

%
%

In order to obtain a competitive limit for the mass of a $W_R$
boson, a precision of at least 0.5\% is needed, both for pure
Gamow-Teller transitions and for the $T=1/2$ mirror $\beta$
transitions.



\subsubsection{Longitudinal polarization} \label{sec:rhc-longpol}

The early measurements of the longitudinal polarization, $P_L$, of
beta particles from unpolarized nuclei in pure Fermi and pure
Gamow-Teller transitions have reached a precision of a few percent
\citep{vanklinken78,vanklinken83}. However, the uncertainties in
the recoil order corrections in the decays of $^{32}$P and
$^{60}$Co used in those measurements, hamper the extraction of
reliable conclusions on weak interaction properties. The only
measurement free of this problem is that for the mixed transition
in the decay of tritium which yielded $P_L
=~-1.005(26)$~\citep{koks76}. However, it has been pointed out
previously \citep{deutsch95} that some additional concern
regarding the accuracy of Mott scattering polarimetry for very low
electron energies \citep{fletcher86} may make the error estimate
of this measurement optimistic.

Sub-percent sensitivity was reached only in relative measurements
\citep{wichers87,skalsey89,carnoy90,carnoy91} where the ratio of
the longitudinal polarization in pure Fermi and pure Gamow-Teller
transitions was determined. This ratio is sensitive to the product
$\delta \zeta$
\begin{equation}
P^F_L / P^{GT}_L \simeq 1 + 8 \delta \zeta \; .
\end{equation}
All measurements performed so far have been carried out in
$\beta^+$ transitions. In the first experiment ~\citep{wichers87}
the positron polarization was determined using Bhabha scattering.
Later experiments ~\citep{skalsey89,carnoy90,carnoy91} used the
method of time-resolved spectroscopy of positronium hyperfine
states. This technique makes use of the magnetic field dependence
of both the lifetime and the population of the singlet and the $m
= 0$ triplet positronium states
\citep{dick63,vanhouse84,carnoy91}. The weighted average result of
all these experiments yielded $-4.0 < \delta \zeta \times 10^4 <
7.0$ (90\% C.L.) \citep{carnoy90}. Note that, although these
limits are very stringent, they are not sensitive to the mass of a
possible $W_2$-boson with right-handed couplings in the limit
$\zeta = 0$ (Sec.~\ref{sec:zeta} and Fig.~\ref{fig:rhc-results}).

\subsubsection{Polarization asymmetry correlation}
\label{sec:rhc-polasycorr}

Limits on right-handed currents complementary to those presented
above and also more stringent, can be obtained from the
measurement of the longitudinal polarization of $\beta$ particles
emitted by polarized nuclei, the so-called polarization-asymmetry
correlation \citep{quin89}. This observable determines the
parameter $(\delta + \zeta)^2$ which is sensitive to $\delta$ and
hence to the mass of a right-handed $W_2$ boson even when $\zeta =
0$. Four measurements of this correlation were carried out, all
using the method of time-resolved spectroscopy of positronium
hyperfine states to determine the longitudinal polarization of the
decay positrons. The experimental quantity that was addressed was
either the ratio of positron polarizations $P^-$ and $P^+$ for
positrons emitted in two opposite directions with respect to the
nuclear spin direction, or the ratio of the polarization of
positrons emitted opposite to the nuclear spin direction, $P^-$,
and emitted by unpolarized nuclei, $P^0$. In the first case
\begin{equation}
P^- / P^+ = R_0 \left[ 1 - \frac{8 \beta^2 \bfvec{\beta} \cdot
\bfvec{J} A} {\beta^4 - \left(\bfvec{\beta} \cdot \bfvec{J} A
\right)^2} (\delta + \zeta)^2 \right],
\end{equation}
\noindent where
\begin{equation}
R_0 = \left[\frac{\beta^2 - \bfvec{\beta} \cdot \bfvec{J}
A}{\beta^2 + \bfvec{\beta} \cdot \bfvec{J} A} \right]
\left[\frac{1 + \bfvec{\beta} \cdot \bfvec{J} A}{1 - \bfvec{\beta}
\cdot \bfvec{J} A} \right],
\end{equation}
\noindent is the standard model expectation value for $P^- / P^+$,
$\beta = v/c$ and $\bfvec{\beta}.\bfvec{J}A$ the experimental
$\beta$ asymmetry. In the second case
\begin{equation}
P^- / P^0 = R_0 \left[ 1 - \frac{4 \bfvec{\beta} \cdot \bfvec{J}
A} {\beta^2 - \left(\bfvec{\beta} \cdot \bfvec{J} A \right)}
(\delta + \zeta)^2 \right],
\end{equation}

\noindent with
\begin{equation}
R_0 = \frac{\beta^2 - \bfvec{\beta} \cdot \bfvec{J} A}{\beta^2 (1
- \bfvec{\beta} \cdot \bfvec{J} A) } \, .
\end{equation}
As can be seen, interesting candidates for this type of
experiments are nuclei for which a large degree of nuclear
polarization can be obtained and which decay via a pure
Gamow-Teller transition of the type $J \rightarrow J-1$, {\em
i.e.} with a maximal asymmetry parameter $A=1$. The sensitivity of
the $T=1/2$ mirror $\beta$ transitions has also been considered
\citep{govaerts95}. Due to the relative character of this type of
measurements a number of systematic effects are reduced
significantly or even eliminated.

\begin{figure}[!htb]
\begin{center}
\includegraphics[height = 9 cm, width = 8cm]{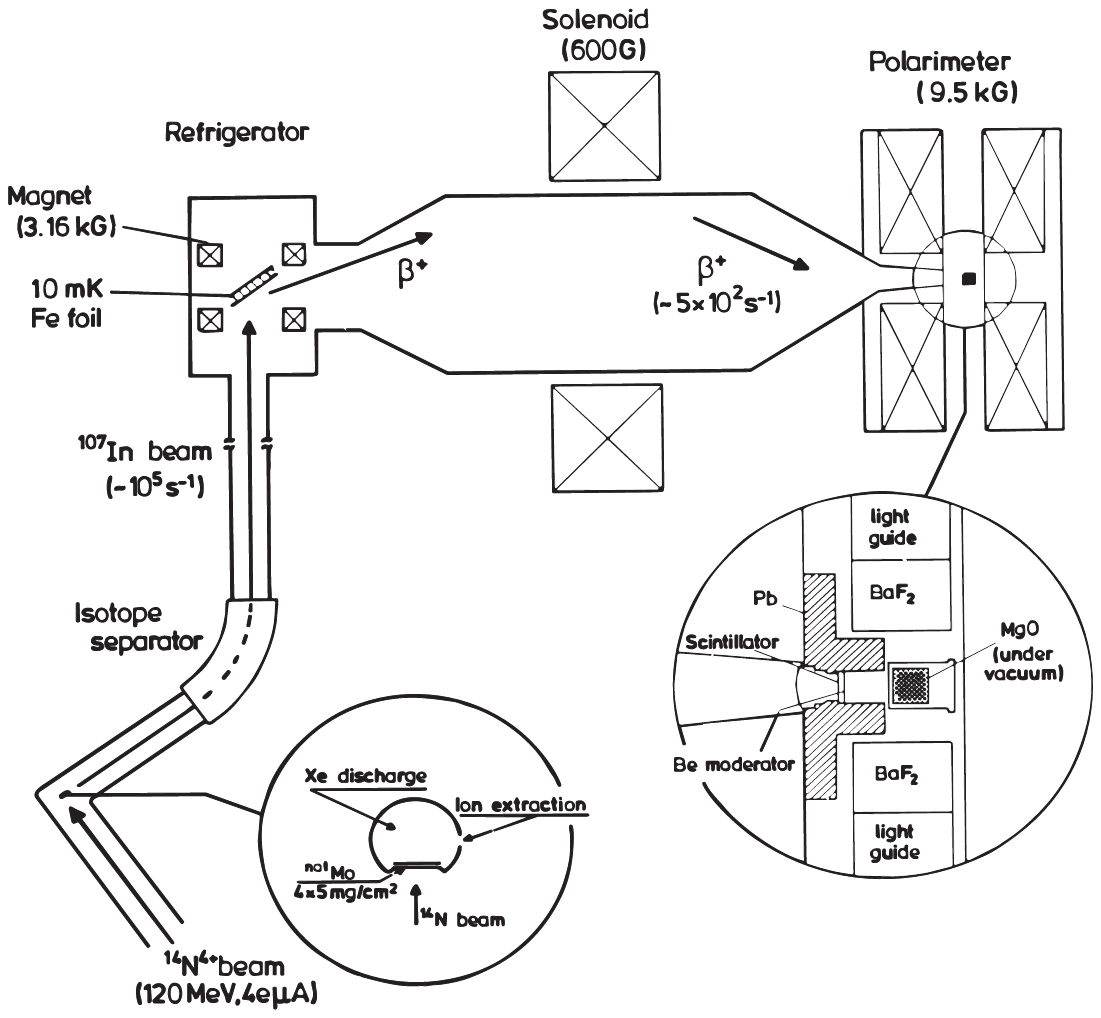}
\caption[107In set-up]{Experimental set-up to measure the
longitudinal polarization of positrons emitted in the decay of
polarized $^{107}$In nuclei. The radioactive ions delivered by the
isotope separator are implanted and oriented in an iron foil at a
temperature of 10 mK inside a dilution refrigerator. The positrons
emitted in the decay of the polarized nuclei are energy-selected
with a spectrometer and then slowed down and stopped in a MgO
pellet. One plastic and two BaF$_2$ scintillator detectors observe
the decay of the positronium that is formed in the MgO, from which
the longitudinal polarization of the positrons is then obtained.
From Severijns {\em et al.} \citeyearpar{severijns93}.
\label{fig:107In}}
\end{center}
\end{figure}

The first measurement of this type \citep{severijns93}, at the
LISOL isotope separator coupled to the CYCLONE cyclotron in
Louvain-la-Neuve, used the isotope $^{107}$In ($t_{1/2} = 32.4 $
m) (Fig.~\ref{fig:107In}), which was polarized with the method of
low temperature nuclear orientation
\citep{vandeplassche81,postma86}. This technique combines
temperatures in the millikelvin region obtained in a $^3$He-$^4$He
dilution refrigerator, with the large magnetic hyperfine fields,
ranging from a few Tesla to several hundreds of Tesla, that
impurity nuclei feel in a ferromagnetic host lattice. The second
measurement, carried out at the Paul Scherrer Institute, used
$^{12}$N that was produced and polarized in the $^{12}C(\vec{p},
n_0)^{12}\vec{N}$ polarization transfer reaction initiated by a
70\% polarized proton beam \citep{allet96}. With each isotope two
measurements were performed, the second one always after
considerable improvements of the experimental set-up. For
$^{107}$In an experimental $\beta$ asymmetry $\bfvec{\beta} \cdot
\bfvec{J} A$ $\simeq 0.50$ was obtained, corresponding to a
nuclear polarization of $\sim 65\%$. The final result was $(\delta
+ \zeta)^2 = 0.0021(17)$ \citep{camps97, severijns98}. With
$^{12}N$ an experimental $\beta$ asymmetry $\bfvec{\beta} \cdot
\bfvec{J} A$ $\simeq 0.13$ was obtained, corresponding to a
nuclear polarization of $\sim 15 \%$. This experiment yielded
$(\delta + \zeta)^2 = -0.0004(32)$ ~\citep{thomas01}. If
interpreted in manifest left-right symmetric models, these results
correspond to a lower limit for the mass of a $W_2$ vector boson
with right-handed couplings of 303 GeV/c$^2$ and 310 GeV/c$^2$
(90\% C.L.). These are the most sensitive tests of parity
violation in nuclear beta decay to date. The lower limit from the
combined result of both experiments is 320 GeV/c$^2$ (90\% C.L.).

A similar experiment was carried out with the mirror nucleus
$^{21}$Na, produced with a deuteron beam from the University of
Wisconsin tandem electrostatic accelerator. The $^{21}$Na was
polarized with circularly polarized light from a
copper-vapor-laser-dye-laser system tuned to the sodium $D_{1}$
line. The result $(\delta + \zeta)^2 = -0.037(70)$
\citep{schewe97} is in agreement with the measurements on
$^{107}$In and $^{12}$N.

Finally, a measurement of the polarization asymmetry correlation
was also carried out with $^{118}$Sb at the ISOLDE-CERN isotope
separator facility. This experiment  \citep{vereecke01} has
explored the limits of sensitivity of this type of experiments
when the technique of low temperature nuclear orientation is used
to polarize the nuclei. The analysis of the data showed that this
method is limited by the present knowledge of the spin rotation of
positrons when being scattered in the iron host foil in which the
nuclei have to be implanted in order to be polarized. As the
nuclear polarization obtained in polarization transfer reactions
is rather small ~\citep{miller91}, significant progress can not be
expected from this method either. An interesting option in this
respect, circumventing the above mentioned difficulties, would be
to couple a $\beta$ polarimeter to an ion or atom trap containing
a nuclear polarized sample.



\subsubsection{Neutron decay}
\label{sec:rhc-neutron}

The correlation coefficients in neutron decay that are most
sensitive to parity violation are the $\beta$ asymmetry parameter
$A$ and the neutrino asymmetry parameter $B$.

\begin{figure}[!htb]
\begin{center}
\includegraphics[height = 9 cm, width = 8cm]{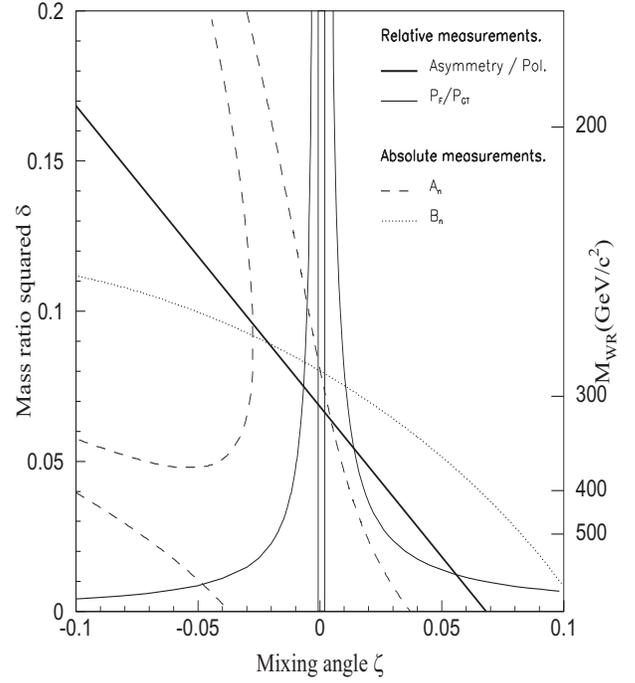}
\caption[rhc-results]{Constraints (90\% C.L.) on the right-handed
current parameters $\delta$ and $\zeta$ from the nuclear $\beta$
decay experiments discussed here.
The allowed regions are those containing $(\delta, \zeta) =
(0,0)$. The narrow vertical band around $\zeta$ = 0 is the region
allowed by unitarity and the $\mathcal{F}$t value for the
superallowed Fermi transitions (Sec.~\ref{sec:zeta}). Adapted from
Thomas {\em et al.} \citeyearpar{thomas01}.
\label{fig:rhc-results}}
\end{center}
\end{figure}

In Fig.~\ref{fig:rhc-results} the limits for the right-handed
current parameters $\delta$ and $\zeta$ (in manifest left-right
symmetric models) from the $A$ and $B$ parameters in neutron decay
\cite{eidelman04} are compared to the limits from superallowed
$\beta$-decay, the $P_F/P_{GT}$ measurements and the
polarization-asymmetry correlation measurements discussed in the
previous paragraphs (see also \citet{abele}).

The available values for the neutrino asymmetry parameter $B$ are
very consistent with each other (see Table~\ref{tab:neutron}). The
precision was significantly increased in the two most recent
measurements, which were both performed with cold polarized
neutrons at the WWR-M reactor of the Petersburg Nuclear Physics
Institute (PNPI) \citep{kuznetsov95, serebrov98}.
\begin{figure}[!htb]
\begin{center}
\includegraphics[height = 6 cm, width = 8cm]{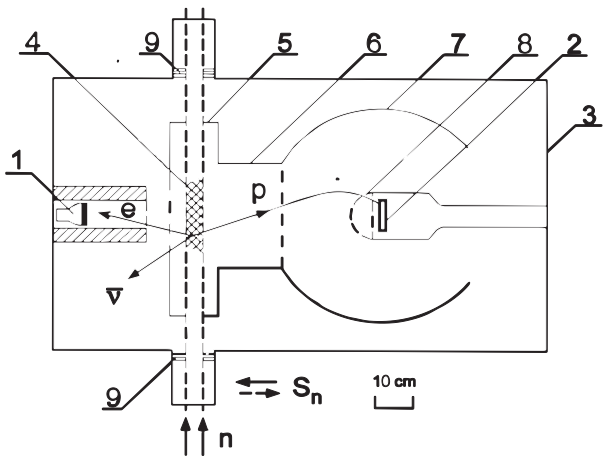}
\caption[B-parameter]{Experimental apparatus for measuring the
$B$-parameter in neutron decay. (1) Electron detector, (2) proton
detector, (3) vacuum chamber, (4) decay region, (5) cylindrical
electrode, (6) TOF chamber, (7) spherical electrode, (8) spherical
grid, and (9) LiF diaphragm. From Kuznetsov {\em et al.}
\citeyearpar{kuznetsov95}. \label{fig:B-parameter}}
\end{center}
\end{figure}
The set-up that was used is shown schematically in
Fig.~\ref{fig:B-parameter}. The momentum and angle of escape of
the undetected antineutrino were deduced from the coincident
detection of the decay electron and recoil proton, and the
subsequent measurement of their momenta. The electrons were
detected with a plastic scintillator. The protons were accelerated
and focused by an electric field onto the proton detector which
consisted of an assembly of two microchannel plates. This
permitted to determine the time of flight of each proton with an
accuracy of 10 ns. The weighted averaged result of the two
measurements, $B$ = 0.9821(40) yields a lower limit of 280
GeV/c$^2$ (90 \% C.L.) for the mass of a $W_2$ boson with
right-handed couplings (Fig.~\ref{fig:rhc-results}).

Recently a measurement of the neutrino asymmetry parameter $B$ was
also performed with the PERKEO II set-up
(Fig.~\ref{fig:PERKEO-II}) at the ILL \citep{kreuz03}. The
electron and the proton from the neutron decay were guided by a 1
T magnetic field towards two combined electron and proton
detectors. In order to better control systematics a detection
system which is able to detect both particles in both detectors
was chosen. The electrons are detected by plastic scintillators in
combination with photomultipliers. The protons are accelerated by
a negative potential towards a thin carbon foil where they create
secondary electrons which can then be detected in the electron
detector. Since the electric potential is considerably lower than
the electron energies observable in the experiment (threshold 60
$keV$), the electrons pass the foils unhindered. This method
reduces systematics and increases the sensitivity. The analysis is
still ongoing.



\subsubsection{Comparison with other fields}
\label{sec:other fields-parity}

A recent new determination of the Michel parameter $\rho$ in muon
decay by the TWIST Collaboration at TRIUMF \cite{musser04} has
yielded an upper limit $|\zeta| <$0.030 (90\% C.L.) on the $W_L -
W_R$ mixing angle. A measurement of the muon decay parameter
$\delta$ by the same collaboration \cite{gaponenko05} yielded a
lower limit on the $W_R$ mass of 420 GeV/c$^2$ (90 \% C.L.). Both
experiments improved slightly on the earlier results reported by
\citet{jodidio86,jodidio88}. The longitudinal polarization of
positrons emitted from polarized muons has been remeasured at PSI
\citet{morelle02}. The result is expected to provide a new value
of the Michel parameter $\xi"$ with significantly higher precision
but the anticipated limit on $W_R$ is, however, not expected to be
improved.

\begin{figure}[!htb]
\begin{center}
\includegraphics[width=9.3cm]{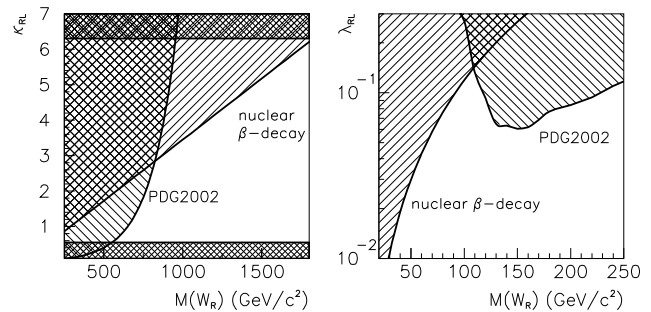}
\vspace{-45mm} \caption{Exclusion plots on parameters of
generalized left-right
        symmetric extensions of the standard model. The parameter
        $\kappa_{RL} = g_R/g_L$ characterizes the intrinsic gauge coupling
        of the right-handed sector relative to the left-handed one, while
        $\lambda_{RL} = |V^R_{ud}|/|V^L_{ud}|$ denotes the relative
        coupling strength of first generation quarks to an hypothetical
        right-handed gauge boson with mass $M(W_R)$. The hatched areas are
        excluded either by direct searches at colliders (PDG2002) or by
        precision experiments in nuclear $\beta$ decay. The horizontal
        bands in the left panel are bounds from theory. The contours in
        the left panel assume $|V^R_{ud}| = |V^L_{ud}|$; those in the
        right panel assume $g_R = g_L$. Adapted from \citet{thomas01}. \label{fig:nuclear-collider}}
\end{center}
\end{figure}

In manifest left-right symmetric models the limits from $\beta$
decay (Fig.~\ref{fig:rhc-results}) and from muon decay for the
mass of a $W_R$ boson with right-handed couplings are weaker than
the lower limit of 715 GeV/$c^2$ from a simultaneous fit to the
charged and neutral sectors \cite{czakon99} and the lower limit of
786 GeV/$c^2$ for the mass of a heavy $W^\prime$ boson from $p
\overline{p}$ collisions at Fermilab \citep{affolder01}. However,
results from $\beta$ decay, from muon decay and from collider
experiments are complementary when interpreted in more general
left-right symmetric extensions of the standard model, such as
models with different gauge coupling constants or different CKM
matrices in the left- and right-handed sectors, etc. This is
illustrated in Fig.~\ref{fig:nuclear-collider}. The
complementarity between experiments at low and at high energies is
also discussed by \citet{langacker98} and \citet{herczeg01}. Note
also that experiments in $\beta$ decay and in muon decay are
sensitive to the helicity of a heavy $W$ gauge boson, while $p
\overline{p}$ collisions are not.
%
%
%

Generally the weak interaction is ignored in atomic physics,
because it is much weaker than the electromagnetic interaction.
However, the valence electrons of an atom can experience the weak
interaction. Indeed, the neutral vector boson, $Z^0$, can be
exchanged between a nucleon and a valence electron, provided the
electron wave function has a non-zero amplitude at the nucleus
since the exchange is effectively a point-like interaction. This
means that parity-violation can be observed in atomic transitions.
Precise measurements of this atomic parity non-conservation
provide an important low-energy test of the electroweak standard
model, complementary to nuclear physics and particle physics
experiments. Reviews were published by \citet{bouchiat97} and
\citet{haxton01}.

\subsection{Time reversal violation}
\label{sec:timereversal}

At present there are two unambiguous pieces of evidence for time
reversal violation (T-violation) and CP-violation, i.e. the decay
of neutral $K$- and $B$-mesons
~\citep{christenson64,fanti99,alavi-harati00,browder03}, and the
excess of baryonic matter over antimatter in the Universe
\citep*{riotto99}. However, the CP-violation that is observed in
$K$- and $B$-meson decays, and which can be incorporated in the
standard model via the quark mixing mechanism, is too weak to
explain the excess of baryons over antibaryons. Cosmology
therefore provides a hint for the existence of an unknown source
of T-violation that is not included in the standard model.

The standard model predictions of T-violation, originating from
the quark mixing scheme, for systems built up of $u$ and $d$
quarks are by 7 to 10 orders of magnitude lower than the
experimental accuracies presently available \cite{herczeg97}.
Thus, because standard model contributions to T-violating electric
dipole moments and T-violating correlations in decay or scattering
processes are so strongly suppressed, any sign for the presence of
T-violation in these observables or processes would be a signature
of a new source of T-violation. New T-violating phenomena may be
generated by several mechanisms \cite{herczeg01} like the exchange
of multiplets of Higgs bosons, leptoquarks, right-handed bosons,
etc. These exotic particles may generate scalar or tensor variants
of the weak interaction or a phase different from 0 or $\pi$
between the vector and axial-vector coupling constants. It is a
general assumption that T-violation may originate from a tiny
admixture of such new exotic interaction terms. Weak decays
provide a favorable testing ground in a search for such new feeble
forces \citep{boehm95,herczeg95b}.

Direct searches for time reversal violation via correlation
experiments in $\beta$ decay require the measurement of terms
including an odd number of spin and/or momentum vectors. The $D$
triple correlation $\bfvec{J} \cdot (\bfvec{p_e} \times
\bfvec{p_\nu})$ is sensitive to P-even, T-odd interactions with
vector and axial-vector currents. To determine this correlation,
the momenta of the $\beta$ particle and neutrino emitted in
mutually perpendicular directions in a plane perpendicular to the
nuclear spin axis is to be determined. It also requires the use of
mixed transitions (see Eq.~(\ref{eqn:dcorr})). As an example, for
the neutron the standard model prediction for the magnitude of
this correlation coefficient, based on the observed CP-violation,
is $D$ $< 10^{-12}$ \cite{herczeg97}. Any value above the final
state effect level, which is typically $D_{FSI}$ $\approx
10^{-5}$, would thus indicate new physics. For leptoquark models
this experimental range is not excluded by measurements of other
observables, like electric dipole moments \citep{herczeg01} (see
also Sec.~\ref{sec:other fields-TRV}).

The other important correlation with respect to searches for
T-violation in $\beta$ decay is the $R$ triple correlation
$\bfvec{\sigma} \cdot (\bfvec{J} \times \bfvec{p_e})$,
Eq.~(\ref{eqn:rcorr}), which probes P-violating components of
T-violating scalar and tensor interactions. To determine the
$R$-correlation the transverse polarization of $\beta$ particles
emitted in a plane perpendicular to the polarized nuclear spin
axis is to be determined.

Measurements of the $D$- and $R$-triple correlations are very
difficult as they require the use of polarized nuclei and at the
same time the determination of either the neutrino momentum
through the recoil ion ($D$-correlation), or of the transverse
polarization of the $\beta$ particle ($R$-correlation).

\subsubsection{D-correlation}

The $D$-correlation was measured in neutron decay and for
$^{19}$Ne. Early measurements in neutron decay have yielded $D_n =
-0.0011(17)$ \citep{steinberg74}, $D_n = -0.0027(50)$
\citep{erozolimsky74} and $D_n = 0.0022(30)$
\citep{erozolimsky78}. Two new and more precise measurements of
$D_n$ have recently been carried out, one at NIST \citep{lising00}
and the other at the ILL \citep{soldner04}.

\begin{figure}[!htb]
\begin{center}
\includegraphics[height = 6.5 cm, width = 5 cm]{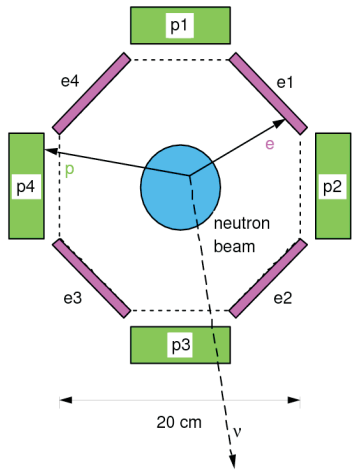}
\caption[emiT]{Principle of the {\em emi}T experiment to test time
reversal violation using an octagonal array of four each proton
(P) and electron (E) detectors. Adapted from Lising {\em et al.}
\citeyearpar{lising00}. \label{fig:emiT}}
\end{center}
\end{figure}

In the $emiT$ experiment at NIST a beam of cold neutrons is
polarized and collimated before it passes through a detection
chamber with four electron and four proton detectors in an
octagonal arrangement (Fig.~\ref{fig:emiT}). The octagonal
geometry places electron and proton detectors at relative angles
of $45^{\circ}$ and $135^{\circ}$. Coincidences are counted
between detectors at relative angles of $135^{\circ}$. While the
cross product $\bf{p}_e \times \bf{p}_\nu$ is largest at
$90^{\circ}$, the preference for larger electron-proton angles in
the decay makes placement of the detectors at $135^{\circ}$ the
best choice to achieve greater symmetry, greater acceptance and
greater sensitivity to $D$ \citep{lising00}, compared to previous
experiments which detected coincidences at $90^{\circ}$. Another
important improvement is the larger polarization, which is 96(2)\%
compared to about 70\% previously.
The initial run with this new set-up produced a statistically
limited result of $D_n = [-0.6 \pm 1.2(stat) \pm 0.5(syst)] \times
10^{-3}$ \citep{lising00}. A second run, with an improved set-up
\cite{mumm04} and aiming at a sensitivity of about 2 $\times
10^{-4}$ or better, was in the mean time completed.

The TRINE experiment at the ILL, detects the neutron decay
electrons by four plastic scintillators in coincidence with
multiwire proportional chambers. Four PIN diodes with thin
entrance windows are used for detecting the protons. These are
accelerated onto the PIN diodes in a focusing electrostatic field
provided by a high voltage electrode.
The neutron beam polarization was 97.4(26)\%. The main advantage
of TRINE with respect to other experiments is the suppression of
systematic effects that is obtained by using the spatial
resolution of the wire chambers and the high segmentation with 12
detector planes. In addition, thanks to the large signal to
background ratio, the statistics of the neutron beam can be used
completely. This resulted in $D_n = [-2.8 \pm 6.4(stat) \pm
3.0(syst)] \times 10^{-4}$ \citep{soldner04}. A new measurement
with TRINE with improved statistics and systematics was in the
mean time carried out \cite{plonka04}. Together with the new data
from $emi$T the world average for $D$ in neutron decay will soon
reach a precision in the very interesting range of $10^{-4}$.

%

The most precise measurements of the $D$-correlation in the decay
of the mirror nucleus $^{19}$Ne have yielded $D = -0.0005(10)$
\citep{baltrusaitis77} and $D = 0.0004(8)$ \citep{hallin84}. These
experiments have reached the limit imposed by final state effects,
which is at the $10^{-4}$ level~\citep{calaprice85}. The combined
result, $D = 0.0001(6)$ \citep{calaprice85} is at present the most
precise limit on a T-violating angular correlation in a weak decay
process. These measurements were also the first ever to test
T-invariance in any weak process at a level below the
characteristic $K$-decay CP violation of 2.3 $\times 10^{-3}$
\citep{christenson64} without any evidence for a violation of
T-invariance.

No sign for T-violation in the {\em V-A} weak interaction was thus
observed in nuclear beta decay so far.


%

\subsubsection{R-correlation}

Only two $R$-correlation measurements were carried out in nuclear
$\beta$ decay. The first with $^{19}$Ne \citep{schneider83} and
the second with $^{8}$Li \citep{sromicki96,huber03}.

The $^{19}$Ne \citep{schneider83} was polarized to essentially
100\% by deflection of an atomic beam in a ``Stern-Gerlach''
magnet. The polarized beam was captured in a holding cell which
assured a spin holding time that was long compared to the decay
lifetime. The transverse spin component of the $\beta$ particles
emitted perpendicularly to the nuclear spin polarization was
analyzed with four identical large acceptance Mott scattering
polarimeters with detector telescopes. A non-zero value of the
$R$-triple correlation coefficient would lead to a left-right
asymmetry in the scattering of the $\beta$ particles by a gold
Mott analyzing foil. The polarimeter analyzing power was a few
percent. The final result of this measurement was $R$($^{19}$Ne) =
0.079(53), the error being limited only by statistics.

\begin{figure}[!htb]
\begin{center}
\includegraphics[height = 6 cm, width = 8 cm]{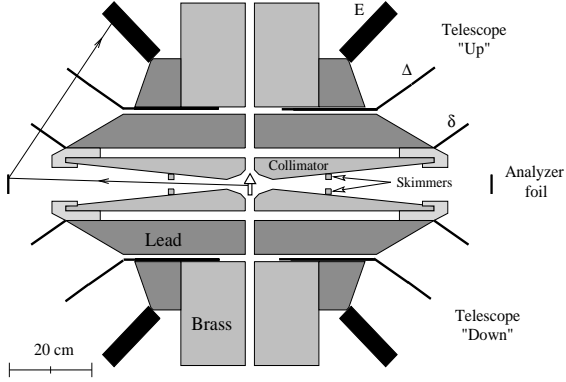}
\caption[8Li]{Vertical cross section through the Mott polarimeter
used in the $^8$Li $R$-correlation experiment. The direction of
incidence of the polarized deuteron beam is perpendicular to the
figure. The central arrow indicates the direction of the $^8$Li
spin in the target. A trajectory of an electron scattered on the
lead analyzer foil is also shown. The labels $\delta$, $\Delta$
and $E$ refer to two delta-E detectors and the energy detector.
More details are given in the text. From Huber {\em et al.}
\citeyearpar{huber03}. \label{fig:8Li}}
\end{center}
\end{figure}

A high-precision measurement of the $R$-parameter was carried out
in the 1990's for the Gamow-Teller decay of $^8$Li at the Paul
Scherrer Institute \citep{sromicki96,huber03}. The set-up for this
experiment (Fig.~\ref{fig:8Li}) has continuously been improved and
upgraded to finally reach a precision of $2 \times 10^{-3}$.
Polarized $^8$Li nuclei were produced by a vector-polarized
deuteron beam on an enriched $^7$Li metal target. This was cooled
to liquid helium temperature in order to achieve a long
polarization relaxation time ($t \geq 20$s), an order of magnitude
longer than the mean lifetime for $^8$Li ($\tau = 1.21$ s). The
transverse polarization of the $^8$Li decay electrons was deduced
from the measured asymmetry in Mott scattering at backward angles
using a lead foil as analyzer. To obtain a large solid angle the
detectors were arranged in a cylindrical geometry around the
$^8$Li polarization axis. In fact, the set-up was made of four
separate azimuthal segments, each containing an upper and a lower
telescope, thus providing four independent measurements of the
electron polarization. Each telescope consisted of two thin
transmission scintillators followed by a thick stopping
scintillator. Much attention was paid to the passive shielding of
the detectors against background radiation produced in the target
area. The weighted average result of six runs is $R$($^8$Li) =
0.0009(22) \citep{huber03}. This has been corrected for the
effects of the final state interaction which can mimic the genuine
time reversal violation in the $R$-correlation and which was
calculated to be $R_{FSI}=0.7(1) \times 10^{-3}$. This has
improved by about an order of magnitude the bounds for T-violating
tensor couplings, yielding $-0.008 < Im(C_T+C^\prime_T) / C_A <
0.014$ (90\% C.L.).

\begin{figure}[!htb]
\begin{center}
\includegraphics[height = 6 cm, width = 8 cm]{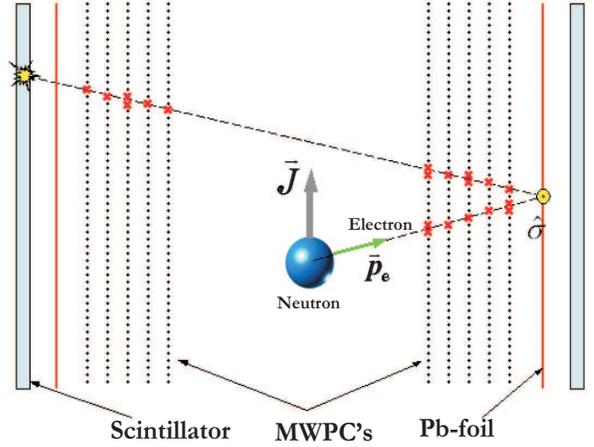}
\caption[R-n]{Principle of the R-correlation experiment in neutron
decay determining the amplitude of $\bfvec{\sigma} \cdot
(\bfvec{J} \times \bfvec{p_e})$. The labels $E$ and $V$ refer to
the energy detector and the veto detector. More details are given
in the text. Adapted from Bodek {\em et al.}
\citeyearpar{bodek03}. \label{fig:R-n}}
\end{center}
\end{figure}

The above mentioned result from $^{19}$Ne \citep{schneider83} is
in principle sensitive to both T-violating scalar as well as
tensor couplings but rather weak limits were obtained due to the
limited experimental precision. A high precision test for the
presence of a T-violating scalar component thus still remains to
be done. Therefore, the $R$-correlation in neutron decay is now
investigated at the polarized cold neutron facility FUNSPIN
\citep{bodek00,zejma05} at the spallation source SINQ at the Paul
Scherrer Institute. This experiment \citep{bodek03} aims at a
0.5\% measurement by determining the transverse polarization of
electrons emitted in polarized neutron decay using large angle
Mott scattering. Electrons emitted from polarized neutrons and
scattered from a Pb analyzer foil are tracked by a system of two
multiwire gas chambers and stopped in a plastic scintillator
(Fig.~\ref{fig:R-n}). True events, where the electron emitted in
the neutron decay was scattered from the analyzing foil, can thus
be selected by the reconstruction of the scattering vertex and the
electron energy information, thereby significantly reducing the
background. From the electron tracks, the scattering angle and the
Mott scattering asymmetry can be determined. The apparatus permits
a simultaneous determination of the time reversal invariant $N$
correlation parameter, Eq.~(\ref{eqn:ncorr}) at the 5\% (relative)
level. Because $N$ is proportional to $A$ ($viz.$ $N_{SM} =
-(\gamma m_e/E_e) A_{SM} \simeq 0.119 m_e/E_e$) and $A$ has been
measured to the 1\% level, determining $N$ provides a calibration
of the apparatus. Final state effects contribute only at the level
of 0.001, which is beyond the expected accuracy on the
$R$-coefficient.

Finally, it is to be noted that several other correlations in
neutron and nuclear $\beta$ decay are also sensitive to
T-violating couplings, either through final-state interaction
effects such as the $\beta$ asymmetry parameter $A$ and the
longitudinal beta particle polarization $G$, or through a
quadratic dependence on the norm of the coupling constants, as is
the case for the $\beta$-$\nu$ correlation coefficient $a$. In
this way, experimental limits on T-violating couplings, although
being somewhat less stringent, were obtained from the longitudinal
positron polarization in the decay of several light nuclei
\citep{carnoy91}, from the positron-neutrino correlation in the
$0^+ \rightarrow 0^+$ decay of $^{32}$Ar \citep{adelberger99} and
from the polarization asymmetry correlation for positrons in the
decay of polarized $^{107}$In \citep{severijns98}.

\subsubsection{Comparison with other fields} \label{sec:other fields-TRV}

T-violating and P-conserving ``$D$-type'' correlations as well as
T-violating and P-violating ``$R$-type'' correlations were also
investigated in the decay of the muon and in decays of kaons and
hyperons. In both cases the neutron and nuclear $\beta$ decays
discussed above yielded the most precise results. Other decays are
usually about a factor of 5 to 10 or even less precise. For the
$D$-type correlation a precision of a few times 10$^{-3}$ was
recently obtained in $K^+$-decay \citep{abe04}. $R$-type triple
correlation experiments in the decays of polarized $\Lambda_0$
particles have yielded results that are often one to two orders of
magnitude less precise than the $R$-correlation experiment with
$^8$Li \cite{sromicki96}. Recently, a precision of about $8 \times
10^{-3}$ was obtained for the $R$-type transverse positron
polarization in muon decay \citep{danneberg05}.

Another low energy search for time reversal violation is provided
by measurements of permanent electric dipole moments (EDM). Since
EDM's violate both parity and time reversal, the simultaneous
presence of even small amounts of violation of these two discrete
symmetries by the fundamental forces would result in small but
finite particle EDM's. Experiments searching for particle EDM's
have started in the 1950's. They played a crucial role in
eliminating theories put forward to explain the observation of
CP-violation in the $K^0$ system, because they usually predicted
too large dipole moments. The standard model, however, predicts
electric dipole moments that are well below the present
experimental sensitivity. EDM experiments are therefore an ideal
probe to search for new physics beyond the standard model. Over
the years, the accuracy of EDM experiments has improved by 7 to 8
orders of magnitude such that the sizes of EDM's predicted by {\em
e.g.} supersymmetric, left-right symmetric or multi-Higgs models
now lie within the detectable range \cite{pendlebury00}.

Of the best existing EDM measurements, the current limit on the
neutron EDM is 6.3~$\times 10^{-26}$~e$\cdot$cm (90\% C.L.)
\citep{harris99}, that on the electron is 1.6~$\times
10^{-27}$~e$\cdot$cm (90\% C.L.) \citep{regan02} and that on the
Hg atom is 2.1$\times$~10$^{-28}$~e$\cdot$cm (95\% C.L.)
\citep{romalis01}. New experiments in neutron decay, aiming at a
sensitivity limit of $10^{-28} \emph{e}$ cm, are being prepared
and/or planned at several facilities ($e.g.$ \cite{atchison05}).
For the electron EDM a significant increase in precision is
expected from the use of heavy atoms (Ra) or heavy polar molecules
(YbF) which have large enhancement factors, enabling in principle
to reach a sensitivity in the $10^{-30} \emph{e}$ cm range in the
next decade. Further, new EDM measurements for the muon and the
deuteron are planned too \citep{silenko03,aoki04}.

It is to be noted that even though the search for a neutron
electric dipole moment restricts the parameter space for many
extensions to the standard model \citep{ellis89}, the $D$-triple
correlation is more sensitive to CP violation induced by
leptoquarks which appear naturally in Grand Unified Theories
\citep{herczeg01}. Although determinations of the $R$-parameter
provide less stringent bounds than what is obtained from
experiments searching for atomic and molecular electric dipole
moments, the theoretical uncertainties associated with the last
ones could be large \citep{herczeg01} and therefore direct limits
on imaginary scalar and tensor couplings from $R$-correlation
measurements would still be useful.

\subsection{Neutrino mass}

\subsubsection{Neutrino oscillations}

Another sector of the standard model that is tested in $\beta$
decay is the one of neutrino masses \citep{mckeown04}. Whereas the
standard model assumes neutrinos to be massless, clear evidence
for non-zero neutrino masses were recently found in several types
of oscillation experiments. The origin of this was the long
standing solar neutrino problem, the large deficit that was
observed for the number of detected neutrinos coming from the sun
with respect to the amount that was expected on the basis of the
Standard Solar Model
(\citet{cleveland98,hampel99,abdurashitov99,fukuda94,fukuda98a}).

Neutrino oscillations imply that a neutrino from one specific
flavor, say a muon neutrino $\nu_{\mu}$, transforms into another
flavor eigenstate, {\em i.e.} an electron neutrino $\nu_{e}$ or a
tau neutrino $\nu_{\tau}$, while traveling from the source to the
detector. The existence of neutrino oscillations requires {\it
(i)\,} a non-trivial mixing between the three weak interaction
eigenstates ($\nu_{e}$, $\nu_{\mu}$, $\nu_{\tau}$) and the
corresponding neutrino mass states ($\nu_{1}$, $\nu_{2}$,
$\nu_{3}$) and {\it (ii)\,} that these masses are not degenerate,
{\em viz.} that the mass eigenvalues ($m_1, m_2, m_3$) differ from
each other. Consequently, the experimental evidence of neutrino
oscillations proves that at least some neutrinos have non-zero
masses. In addition, the existence of neutrino oscillations
implies the breakdown of lepton family number conservation.

Evidence for neutrino oscillations was first obtained in
measurements observing atmospheric neutrinos which are produced as
decay products in hadronic showers caused by collisions of cosmic
rays with nuclei in the upper atmosphere. A strong deficit of muon
neutrinos was reported by several experiments
\citep{fukuda94,becker-szendy92,allison97,fukuda98a}. Clear
evidence for this deficit being caused by neutrino oscillations
was obtained from the zenith angle dependence of the flux ratio
\citep{fukuda94,fukuda98b} which showed a clear deficit for the
upward-going muon neutrinos, which have to travel through the
earth before reaching the detector, in contrast to the
downward-going muon neutrinos. The atmospheric neutrino data is
consistent with neutrino oscillation from muon neutrinos to tau
neutrinos \citep{fukuda98b}.

Clear evidence has also been obtained for oscillations of solar
neutrinos \citep{ahmad01,ahmad02a,ahmad02b} and reactor neutrinos
\citep{eguchi03,ahn03}, while a number of new experiments to study
the nature of neutrino oscillations are in progress or planned as
well \citep{mckeown04}.

\subsubsection{Absolute neutrino mass determinations}
\label{sec:numass}

Neutrino oscillation experiments, whether they are observing
solar, atmospheric or reactor neutrinos, are only sensitive to the
differences of the squared masses of neutrinos, $\Delta m^{2}_{ij}
= |m_i^2 - m_j^2|$, and can therefore not determine the absolute
mass values. This absolute mass scale can be deduced from two
types of experiments: the search for neutrinoless double $\beta$
decay, a process that is forbidden in the standard model, and
direct kinematic neutrino mass experiments. Both approaches are
measuring different parameters and are complementary to each
other.

\medskip
{\it a. Direct searches}
\medskip

Experiments investigating the kinematics of weak decays by
measuring the charged decay products to determine absolute
neutrino masses have been performed for all three neutrino
flavors. The measurement of pion decays into muons and $\nu_\mu$
at PSI and the investigation of $\tau$-decays into five pions and
$\nu_\tau$ at LEP have yielded the upper limits $m(\nu_\mu)<$ 190
keV/$c^2$ (90\% C.L.) \citep{eidelman04} and $m(\nu_\tau)<$ 18.2
MeV/$c^2$ (95\% C.L.) \citep{barate98}. Experiments investigating
the mass of the electron neutrino by analyzing $\beta$ decays with
emission of electrons or positrons have reached a sensitivity in
the eV/$c^2$ mass range. The most sensitive direct searches for
the mass of the electron neutrino are based on the investigation
of the electron spectrum of tritium $\beta$ decay
\begin{equation}
^{3}{\rm H} \rightarrow ^{3}{\rm He}^{+} + e^{-} +
\overline{\nu}_{e} \, .
\end{equation}

\begin{figure}[!htb]
\begin{center}
\includegraphics[height=8cm,width=6cm,angle=90]{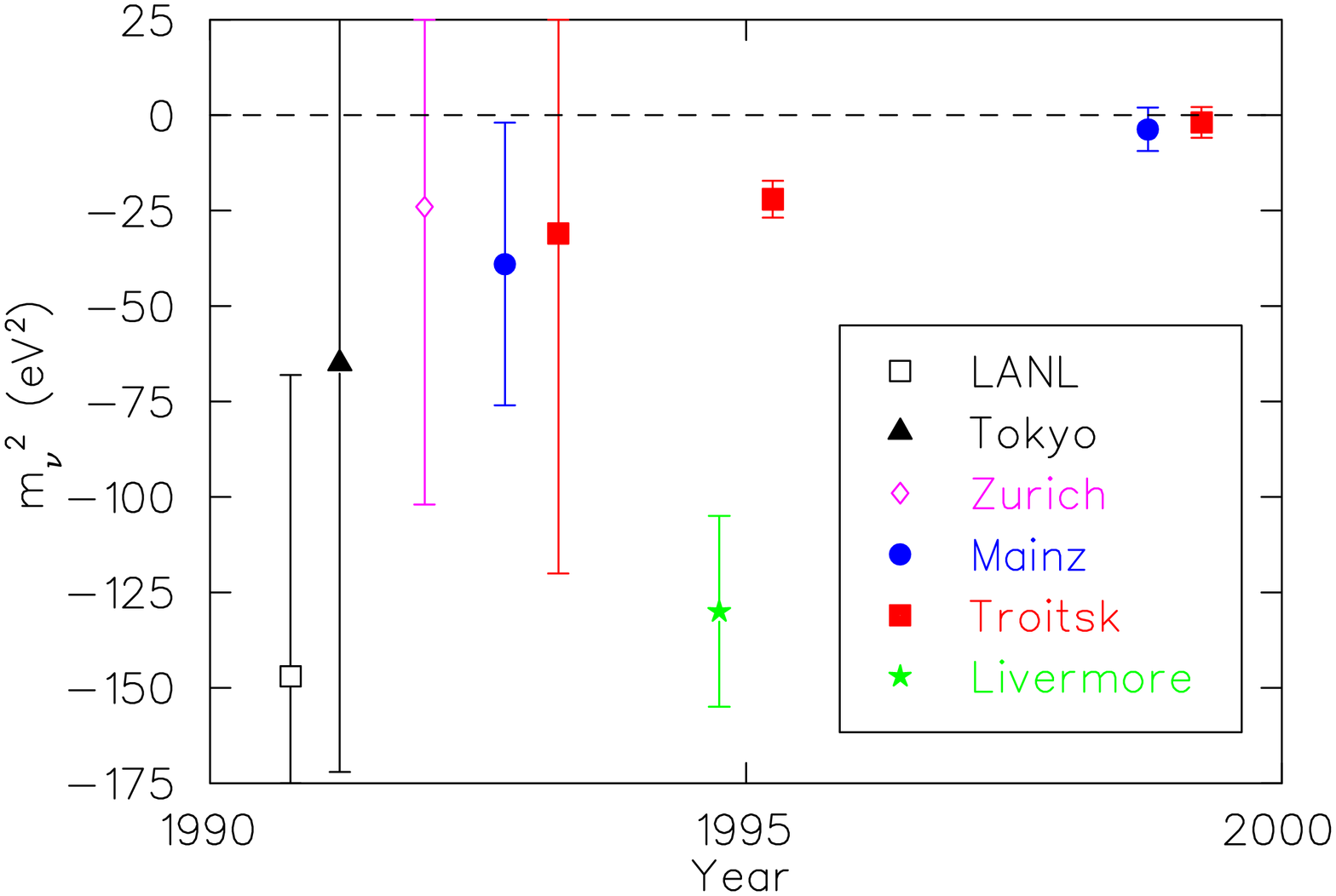}
\caption[tritium-Mainz]{Results of measurements of
$m^2_{\overline{\nu}_e}$ from tritium $\beta$ decay experiments
since 1990. From \citealp{mckeown04}. \label{fig:nu-e-mass}}
\end{center}
\end{figure}

A non-zero neutrino mass would slightly change the shape of the
$\beta$ electron energy spectrum at the upper end. However, the
interesting part of the $\beta$ spectrum is only one part in a
billion! A long series of experiments with tritium were carried
out over the last 20 years. During this period the error bar on
$m^{2}_{\nu}$ has decreased by almost two orders of magnitude
(Fig.~\ref{fig:nu-e-mass}).

\begin{figure}[!htb]
\begin{center}
\includegraphics[height = 6 cm, width = 8 cm]{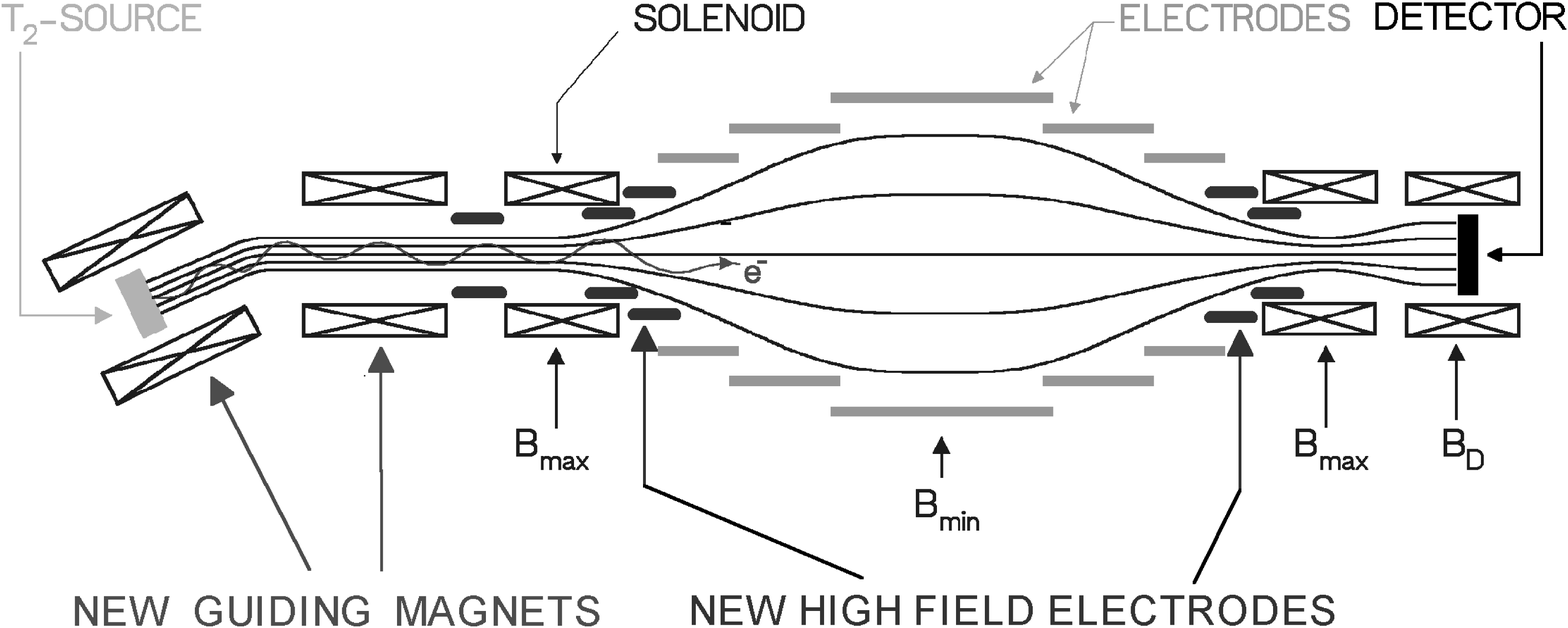}
\caption[tritium-Mainz]{Schematic drawing of the MAC-E filter used
at Mainz for tritium $\beta$ spectroscopy. From Bonn {\em et al.}
\citeyearpar{bonn99}. \label{fig:tritium-Mainz}}
\end{center}
\end{figure}

The problem of negative values for $m^{2}_{\nu}$ of the early
1990's \cite{eidelman04} (Fig.~\ref{fig:nu-e-mass}) has
disappeared due to a better understanding of systematics and
improvements in the experimental set-ups. The highest sensitivity
was reached in experiments at Troitsk and Mainz which used a new
type of spectrometers~\citep{lobashev99,weinheimer99}, so-called
MAC-E-Filters for Magnetic Adiabatic Collimation followed by an
Electrostatic Filter (Fig.~\ref{fig:tritium-Mainz}). This type of
spectrometer combines high luminosity and low background with a
high energy resolution. The $\beta$ electrons from the tritium
source placed in the first of a series of superconducting
solenoids are guided in a cyclotron motion around the magnetic
field lines into the forward hemisphere, resulting in a solid
angle acceptance of nearly 2$\pi$. Due to the very slow decrease
of the magnetic field, by nearly four orders of magnitude, between
the first solenoid and the center of the spectrometer, most of the
electrons cyclotron energy is transformed into longitudinal
motion. The isotropic distribution of the $\beta$ electrons at the
source is thus transformed into a broad beam of electrons flying
almost parallel to the magnetic field lines. This parallel
electron beam runs against an electrostatic potential created by a
set of cylindrical electrodes. Electrons with sufficient energy to
pass the electrostatic barrier are re-accelerated and focused onto
the detector. All other electrons are reflected. The spectrometer
thus acts as an integrating high-energy pass filter, the energy
resolution of which is only determined by the ratio between the
minimum and the maximum magnetic field in the spectrometer:
$\Delta E / E = B_{min} / B_{max}$. By scanning the electrostatic
retarding potential the $\beta$ spectrum can be measured in an
integrating mode.

For the Troitsk experiment \citep{lobashev03} the fit of the
$\beta$ spectrum yielded the limit $m_\nu c^2 \leq 2.05$ eV (95\%
C.L.), whereas the Mainz experiment \citep{kraus03,kraus05}
obtained $m_\nu c^2 \leq 2.3$ eV (95\% C.L.). Both experiments
have reached their intrinsic sensitivity limit. A new project,
called KATRIN \citep{osipowicz01,weinheimer02}, was recently
started at the Forschungszentrum Karlsruhe with the aim of
improving the sensitivity to about $m(\nu_{e}) = 0.35 eV/c^{2}$
(90\% C.L.) assuming three years of measuring time. The main
components of this system comprise two tritium sources, two
electrostatic MAC-E filter electron spectrometers, and a segmented
solid state detector. The overall length of the set-up is about 70
m and the energy resolution is 1 eV.
%



\medskip
{\it b. Neutrinoless double $\beta$ decay}
\medskip

An alternative and very sensitive means to search for non-zero
neutrino masses is neutrinoless double $\beta$ decay ($0 \nu \beta
\beta$). Physically this means that two $\beta$ decays occur at
the same moment in the same nucleus. This is only observable if
the ground state of the $\beta$ decay daughter of a nucleus has a
higher energy than the parent state, such that the parent nucleus
must immediately decay to its grand-daughter isotope. The normal
double $\beta$ decay with emission of two electron (anti)neutrinos
($2 \nu \beta \beta$) was observed more than 15 years ago
\citep{elliot87}. This is, however, a process allowed by the
standard model, albeit with an extremely low probability. For
neutrinoless double $\beta$ decay to occur the neutrino emitted by
one $\beta$ decaying nucleon inside the nucleus  ($n \rightarrow p
+ e^{-} + \bar{\nu}_e$) has to be absorbed by the second nucleon
undergoing inverse $\beta$ decay $\nu_{e} + n \rightarrow p +
e^{-}$. For this to be possible, the neutrino has to have a mass
and to be its own antiparticle, that is it has to be a Majorana
particle. The process of $0 \nu \beta \beta$ violates lepton
number.

The search for neutrinoless double beta decay gives new
information on the nature of the neutrino \cite{bahcall04}. It is
the only feasible experimental technique that could establish
whether the neutrino is a Majorana particle. If this is indeed the
case, these experiments are also sensitive to the mass of the
neutrino. Current mass limits from neutrinoless double beta decay
are about one order of magnitude more stringent than limits from
direct measurements and proposed or suggested experiments will
further improve on this \cite{bahcall04}. It is to be noted though
that neutrinoless double $\beta$ decay is sensitive to an
effective neutrino mass $m_{ee}(\nu)$, which is a coherent sum of
all neutrino mass states $\nu_{i}$ contributing to the electron
neutrino $\nu_{e}$ according to their mixing described by the
neutrino mixing matrix elements $U_{ei}$, and to their Majorana
phases $e^{i \phi_{i}}$:
\begin{equation}
m_{ee}(\nu) = \left| \sum e^{i \phi_{i}} U_{ei}^{2} m(\nu_{i})
\right|
\end{equation}

\noindent As the Majorana phases $e^{i \phi_{i}}$ are unknown and
the mixing matrix elements $U_{ei}$ are in general complex,
cancellations can occur and $m_{ee}(\nu)$ can become zero or very
small even when the mass values $m(\nu_{i})$ are non-zero.

No clear indication for neutrinoless double $\beta$ decay has been
obtained so far, although one experiment has reported a possible
indication for this decay mode in $^{76}$Ge \citep{klapdor01} (see
also~\citealp{aalseth02}). Recent reviews of this field can be
found in \citet{elliot02,zdesenko02}.

\subsection{Tests of CVC and searches for second class currents}
\label{sec:CVC_SCC}

The presently best tested consequence of CVC
(Sec.~\ref{sec:higherorder}) is the requirement that $g_V(q^2=0)$
should be constant irrespective of the nucleus considered. Many
careful measurements of decay parameters in superallowed $0^+
\rightarrow 0^+$ transitions between $T = 1$ states
(\ref{sec:ftvalues}) have confirmed the CVC hypothesis at the $3
\times 10^{-4}$ precision level \citep{hardy05a}. However, tests
of the so-called strong form of CVC, relating the weak magnetism
form factor uniquely to the corresponding electromagnetic form
factor (obtained from the transition rate for the analogous
$\gamma$ decay) are still far from reaching a similar precision.
Experimental tests comprise {\em e.g.} $\beta$ spectrum shape
measurements, measurements of correlations in $\beta$ decay as
well as $\beta-\alpha$ and $\beta-\gamma$ correlation
measurements. Reviews of this work can be found in
\citet{grenacs85,towner95}. Whereas the CVC hypothesis was
confirmed only to an accuracy of $\pm 25\%$ in $\beta-\alpha$ and
$\beta-\gamma$ correlation measurements, it was confirmed to an
accuracy of about 5\% in a measurement of the $^{20}$F $\beta$
spectrum shape \citep{elmbt87} and in $\beta$ correlation
measurements \citep{towner95}.

It is to be noted that in various $\beta$ correlation experiments
the interference between the weak magnetism form factor $f_M$ and
the dominant axial form factor $g_A$, which introduces a deviation
from the behavior of the correlation as determined by the $g_A$
form factor alone, appears always in conjunction with an
interference term arising from an eventual second class
contribution from the induced tensor form factor $f_T$
(Sec.~\ref{sec:higherorder}). Correlation experiments therefore
test CVC only if one assumes the absence of second class currents.
Conversely, experiments designed to observe second class currents
have to rely on the CVC predicted value for $f_M$ or on its direct
measurement. In correlation experiments on the $\mu^- + ^{12}C
\rightarrow ^{12}B + \nu_{\mu}$ muon capture transition, the
prediction of CVC was verified with a precision of about 6\%
\citep{grenacs85,possoz77}.

Originally, a very sensitive method to search for second class
currents seemed to be the comparison of $ft$-values in mirror
transitions which, in the impulse approximation, can be written as
\citep{towner95}

\begin{equation}
\frac{ft^+}{ft^-} \propto 1 - \frac{4}{3} \frac{W^+_0 + W^-_0}{2M}
\frac{g_T}{g_A} + \delta_{nucl}
\end{equation}

\noindent with $W^{+}_{0}$ and $W^{-}_{0}$ the respective endpoint
energies and $\delta_{nucl}$ a nuclear structure correction that
arises from the fact that the wave functions of the two mirror
nuclei are not exact isospin eigenstates. However, the average
reliability of the calculated values for $\delta_{nucl}$ hamper
the extraction of any information on possible second class
currents, even at the present level of many-body techniques
\citep{smirnova03}.

An overall analysis \citep{wilkinson00} of relevant experimental
data in the $\beta$ decay of complex nuclei has yielded a limit on
the second class tensor coupling constant of $|f_T/f_M| < 0.1$
(90\% C.L.). The most precise results in this respect were
obtained in recent precision measurements of the $\beta$ angular
distributions from the aligned mirror pair nuclei $^{12}$B and
$^{12}$N \citep{minamisono98, minamisono03}. This conclusion is
supported by evidence from particle studies although there the
limits obtained are less stringent \citep{wilkinson00}.

An alternative approach to test CVC and search for second class
currents might be precision correlation measurements in neutron
decay. As pointed out by \citet{gardner01}, if both the $\beta$
asymmetry parameter $A$ and the $\beta-\nu$ correlation $a$ could
be determined with a precision of 1\% or better, this would permit
to test the CVC hypothesis and to search independently for
second-class currents.

Finally, the scalar form factor was extracted with high precision
from the $\mathcal{F}$t value of the superallowed
$0{^+}\rightarrow 0{^+}$ Fermi transitions, yielding $f_S/g_V$ =
-0.00005(130) \citep{hardy05a}. This form factor should vanish
both because of CVC and because it is a second class term.

%
%
%
%
%
\section{SUMMARY AND CONCLUSION}
\label{sec:conclusion}

The current experimental status of the weak interaction in nuclear
and neutron $\beta$-decay and their potential to search for
physics beyond the standard model were discussed. A description of
the different formalisms that are used in the literature was given
as well as approximate expressions for a number of correlation
coefficients. In addition, overall fits of selected data have
provided new values and limits for the coupling constants that
describe $\beta$-decay processes.

Experiments in nuclear $\beta$ decay have significantly
contributed in the past to the determination of basic aspects of
the weak interaction. They continue to be a powerful tool to test
the underlying symmetries, to determine the structure in more
detail and to search for physics beyond the standard model.

Progress in the development of a number of new and advanced
experimental techniques, often combined with improved isotope
yields, resulted in a series of new precision experiments in
nuclear $\beta$-decay. These provided important new tests of
parity violation and time reversal invariance, new constraints on
scalar and tensor contributions, new experimental as well as
theoretical results related to the $V_{ud}$ element of the CKM
quark mixing matrix, and more stringent direct limits on the
neutrino mass.

The development of more intense cold as well as ultra-cold neutron
beams and of improved techniques for neutron polarization,
polarimetry and detection led to a significant increase in
precision for lifetime measurements and for different correlations
in neutron beta decay. Recent correlation experiments have mainly
concentrated on improving the determination of the ratio between
the axial-vector and the vector form factors, $g_{A}/g_{V}$, for
an alternative determination of $V_{ud}$ in order to test the
unitarity of the CKM matrix. Stringent P-violation tests and tests
of T-invariance were carried out as well.

An important problem during the last decade has been the possible
violation of unitarity of the CKM matrix. The value of $V_{ud}$
deduced from the $\mathcal{F}$t-values of superallowed pure Fermi
transitions has resulted for a long time in a 2 to 2.5$\sigma$
deviation from unitarity when combined with the adopted value for
the $V_{us}$ matrix element. Nevertheless, the
$\mathcal{F}$t-values are consistent at the $3 \times 10^{-4}$
level confirming the CVC hypothesis. A similar deviation from
unitarity was reported in neutron decay, where nuclear structure
effects are absent. The result combines the neutron lifetime with
the $\beta$-asymmetry parameter. This has triggered new
determinations of $V_{us}$ in kaon decay as well as a new analysis
of existing hyperon beta decay data. All results obtained are
consistent, leading to a new value for $V_{us}$ which resolves the
long standing unitarity problem. Additional experiments to confirm
this new value are ongoing and planned while the form factor
$f_+(0)$, which takes into account $SU(3)$ symmetry breaking, is
addressed again. Assuming the observed shift in the value of
$V_{us}$ is genuine, the unitarity condition is validated for the
first row of the CKM matrix and, in addition, the nuclear
corrections are put on a solid basis. The very precise average
$\mathcal{F}$t-value for the superallowed transitions can now be
used to test the understanding of isospin effects in nuclei at an
unprecedented level of precision and to carry out detailed studies
of the $pn$-interaction near the $N=Z$ line. In turn, the control
of nuclear structure effects will permit further tests of the
symmetries of the standard model and new searches for physics
beyond. Accurate determinations of $\mathcal{F} t$-values for
superallowed transitions should therefore be pursued whenever
possible. Such determinations require the measurements of the
half-life, the corresponding branching and the $Q_{EC}$-value of
the transitions. Mass measurements at the level of 10$^{-8}$ are
required to determine these $Q_{EC}$-values.

New measurements of the neutron lifetime and of the
$\beta$-asymmetry parameter led to a significant improvement in
the determination of the fundamental ratio $g_A/g_V$. With
currently ongoing developments further progress can be expected
and it is important that this type of measurements be continued.

Significant progress was also made over the past decade in the
search for possible scalar and tensor contributions to the weak
interaction. New measurements of different correlations between
the spins and momenta of the particles involved in $\beta$-decay
resulted in improved limits on the couplings and masses of bosons
which could generate phenomenological scalar and tensor
interactions. Under identical assumptions these indirect limits
are often tighter than those obtained by direct searches in
collider experiments.
%
%
Important contributions in the search for scalar currents were
provided by the $\beta$-$\nu$ correlation experiments with
$^{32}$Ar and $^{38m}$K. In the search for tensor currents the
contribution of the polarization-asymmetry parameter measurement
with $^{107}$In is important.

A remarkable development in the field was the introduction of atom
and ion traps. These tools have enabled new types of $\beta$-$\nu$
correlation and $\beta$-asymmetry measurements, free of scattering
effects and with undisturbed nuclear recoils. Recently, first
results from an experiment with $^{38m}$K became available, while
several other experiments are currently ongoing. Using different
experimental methods these $\beta-\nu$ correlation measurements
reach the 0.5\% precision level. Future experiments could consider
to improve this to the 0.1\% level. This requires the production
of clean and high intensity beams, a detailed understanding of
systematic effects and the eventual inclusion of recoil order
effects in the data analysis.

New experimental methods which do not use traps are also being
developed. These new approaches avoid or at least significantly
reduce a number of systematic effects. The new experiments will
focus on relative measurements of the $\beta$ asymmetry parameter,
on $\beta$ asymmetry measurements in a 17 T external magnetic
field with polarized nuclei, and on the determination of the
$\beta$-$\nu$ correlation in neutron decay with a retardation
spectrometer. A precision of 0.5\% to 1\% is anticipated.

In the last decade significant progress was also made to test the
discrete symmetries of parity and time reversal. The first
measurements of the polarization-asymmetry correlation, carried
out with $^{107}$In and with $^{12}$N, yielded the most precise
tests of parity violation in nuclear $\beta$ decay to date. A
similar precision was obtained in a measurement of the neutrino
asymmetry parameter $B$ in neutron decay. There is strong interest
in more precise tests of parity violation in nuclear $\beta$
decays as these provide stringent constraints on several new
extensions of the standard model. Any measurement reaching the
level of 500\,GeV/c$^2$ for a possible $W$ boson with right-handed
couplings would be very valuable. New techniques should be
developed to improve both the yields of the isotopes of interest
for such measurements as well as the nuclear polarization.

Important progress in the search for deviations from maximal
parity violation can be expected from new measurements of the
neutron lifetime, and the $\beta$ asymmetry as well as the
neutrino asymmetry in neutron decay. This is due to recent
improvements in the techniques to polarize neutrons and to
accurately determine this polarization, and to the development of
techniques to keep the neutrons in the measurement volume for a
time of the order of their lifetime.


From the observed matter-antimatter asymmetry in the Universe it
appears that there should be a new mechanism of time reversal
violation in addition to the observed CP-violation that is
incorporated in the standard model. New T-violation searches
should therefore be pursued vigorously. For the triple
correlations there are several orders of magnitude between the
current experimental level of sensitivity and the manifestation of
a standard model CP-violating effect and there is therefore a wide
window available to search for new T-violating mechanisms. Any
system is good for such measurements, provided the final state
interactions are well under control.

The determination of the $R$ triple correlation in the decay of
$^{8}$Li has yielded very stringent limits on the presence of
T-violating tensor couplings. An ongoing experiment to measure for
the first time the $R$ correlation in neutron decay will search
for a T-violating scalar component. Two new measurements of the
$D$ triple correlation in neutron decay provided new tests of time
reversal invariance in the vector and axial-vector parts of the
weak interaction. The present results will further be improved in
the second phase of both experiments which will potentially reach
the 10$^{-4}$ sensitivity level.

Tests for the presence of second class currents, such as in the
$A=12$ system, should be continued and improved. In addition,
better tests of the strong CVC, which was so far tested at the 5\%
level, should be pursued too.

Finally, the efforts to directly determine the electron neutrino
mass from a precision measurement of the tritium $\beta$ spectrum
shape near the endpoint have been pursued in recent years by two
very precise experiments with retardation spectrometers. Both
experiments have now reached their limits of sensitivity and have
yielded the most stringent direct upper limits on the mass of the
electron neutrino. Based on the experience gained with these
set-ups a new facility is now being developed which will be
sensitive to a neutrino mass at the 0.3\,eV level.

\medskip
In conclusion, the past two decades have witnessed significant
progress in the study of fundamental properties of the weak
interaction in both nuclear and neutron $\beta$-decay. Many
efforts to further improve the sensitivity for this type of
experiments are either ongoing or planned. In the years to come,
experiments at low energy will thus continue to contribute to the
study of weak interaction properties, providing information that
is complementary to experiments in muon decay, at colliders or in
underground neutrino laboratories.

%
%
%
%
%
\section{ACKNOWLEDGMENT}
\label{sec:acknow}

We are indebted to many colleagues for very useful discussions to
prepare this review and for providing information on the status
and progress of their activities. We warmly thank J.A.~Behr, B.
Blank, K. Blaum, K.~Bodek, J.~Bonn, J.~Byrne, J.~Deutsch,
M.S.~Dewey, V.~Egorov, Yu.V.~Gaponov, P.~Geltenbort, F.~Gl{\"u}ck,
J.C.~Hardy, P.~Herczeg, B.R.~Holstein, K.~Kirch, J.S.~Nico,
Ch.~Plonka, R.~Prieels, P.A.~Quin, A.~Serebrov, W.M.~Snow,
E.~Thomas, I.S.~Towner, D.J.~Vieira, H.~Wilschut, and O.~Zimmer.
This work was partly supported by a Franco-Belgian Integrated
Action Program {\em Tournesol} under contract 07014PB and by the
Fund for Scientific Research Flanders (FWO).
%
%
%
%
%
\appendix
\section{METRIC AND CONVENTIONS}
\label{sec:appendix:conv}

The analysis of the status of the {\em V-A} theory presented in
Sec.~\ref{sec:fit} and the discussion of the experimental tests in
Sec.~\ref{sec:experiments} refer to the experimental correlation
coefficients which are expressed in terms of the couplings $C_i$
and $C_i'$, defined in Eq.~(\ref{eqn:Hgeneral}). This equation is
quoted from \citet{jackson57a} so that we adopted here their
convention for the $\gamma$ matrices.

Eqs.~(\ref{eqn:Hhpf1}-\ref{eqn:Hhpf3}) are adapted from
\citet*{herczeg01} and Eqs.~(\ref{eqn:lepto1}-\ref{eqn:lepto4})
are adapted from \citet*{herczeg95a} who uses a representation of
the $\gamma$ matrices --the \citet*{bjorken63} convention-- which
differs from the one used by \citet{jackson57a}. In particular,
the signs of $\gamma_5$, which enters the projection operators,
are opposite in the two representations. All the signs of
$\gamma_5$ in the equations quoted from
\citet*{herczeg95a,herczeg01} have hence been changed. We refer
the interested reader to the original works for further details.

\section{CORRELATION COEFFICIENTS}
\label{sec:appendix:coeffs}

We list here the expressions of the correlation coefficients
calculated by \citet{jackson57b} for allowed transitions. They
contain the model- and nucleus-independent Coulomb corrections of
order $\alpha Z$. Numerical calculations \cite{vogel83} show that
the approximation used to get to these corrections are accurate at
the $10\%$ level. When higher precisions are needed the effect of
higher order effects like the induced weak currents, forbidden
matrix elements, radiative corrections, the finite size of the
nucleus, the electronic environments, etc., should be considered.

In the expressions below, the following notation is adopted: $M_F$
and $M_{GT}$ are the Fermi and Gamow-Teller matrix elements
respectively; $J$ and $J^\prime$ are the spins of the initial and
final nuclear states; the upper (lower) sign refers to $\beta ^-$
($\beta ^+$)-decay.

In addition,
\begin{equation}
\renewcommand{\arraystretch}{2}
\begin{array}{ll}
 \lambda_{ J^\prime J}= &
 \left\{
 \begin{array}{cl}
 \displaystyle 1 & J \to J^{\prime }=J-1 \\
 \displaystyle \frac{1}{J+1} & J \to J^{\prime }=J \\
  \displaystyle -\frac{J}{J+1} & J \to J^{\prime }=J+1 \\
 \end{array}
\right.
\end{array}
\end{equation}

\noindent $Z$ is the atomic number of the daughter nucleus,
$\alpha$ is the fine structure constant and $\gamma = \sqrt{(1 -
\alpha^2 Z^2)}$.

\begin{eqnarray}
\label{eqn:normalization} \xi & = &  \vert M_F \vert ^2 \left(
\vert C_S \vert ^2 + \vert C_V \vert ^2 + \vert C^\prime _S \vert
^2 + \vert
C^\prime _V \vert ^2 \right) \nonumber \\
& & + \vert M_{GT} \vert ^2 \left(  \vert C_T \vert ^2 + \vert C_A
\vert ^2 +  \vert C^\prime _T \vert ^2 +  \vert C^\prime _A \vert
^2 \right)
\end{eqnarray}

\begin{eqnarray}
\label{eqn:enangcorr} a \xi & = &  \vert M_F \vert ^2 \biggl[
-\vert C_S \vert ^2 + \vert
 C_V \vert ^2 - \vert C^\prime _S \vert ^2 + \vert C^\prime _V \vert
 ^2 \nonumber \\
 & & \mp 2 \frac{\alpha Z m}{p_e} Im \left( C_S C_V^* +
C^\prime _S C^{\prime *} _V \right) \biggr] \nonumber \\
 & & + \frac{\vert M_{GT} \vert ^2}{3} \biggl[ \vert C_T \vert ^2 -
 \vert C_A \vert ^2 +  \vert C^\prime _T \vert ^2 -  \vert C^\prime _A
 \vert ^2 \nonumber \\
 & & \pm  2 \frac{\alpha Z m}{p_e} Im \left(C_T C_A^* +
C^\prime _T C^{\prime *} _A \right) \biggr]
\end{eqnarray}

\begin{eqnarray}
\label{eqn:fierzinterference} b \xi & = & \pm 2 \gamma Re \biggl[
\vert M_F \vert ^2 \left(C_S C_V^* +
C_S^\prime C_V^{\prime *} \right) \nonumber \\
& & +  \vert M_{GT} \vert ^2 \left(C_T C_A^* + C_T^\prime
C_A^{\prime *} \right) \biggr]
\end{eqnarray}

\begin{eqnarray}
\label{eqn:betaasym} A \xi & = & \vert M_{GT} \vert ^2
\lambda_{J^\prime J} \biggl[ \pm 2 Re
 \left( C_T C_T^{\prime *} - C_A C_A^{\prime *} \right) \nonumber \\
 & & + 2 \frac{\alpha Z m}{p_e} Im \left( C_T C_A^{\prime *} +
 C^\prime _T C^* _A \right) \biggr] \nonumber \\
 & &  + \delta _{J^\prime J} M_F M_{GT} \sqrt{\frac{J}{J+1}} \nonumber \\
 & & \times \biggl[ 2 Re \left( C_S C_T^{\prime *} + C_S^\prime C_T^* -
C_V C_A^{\prime *} - C_V^\prime C_A^* \right) \nonumber \\
 & & \pm  2 \frac{\alpha Z m}{p_e} Im \left( C_S C_A^{\prime *}
 + C_S^\prime C_A^* \right. \nonumber \\
 & & \left. -C_V C_T^{\prime *} - C_V^\prime C_T^* \right) \biggr]
\end{eqnarray}

\begin{eqnarray}
\label{eqn:nuasym} B \xi & = & 2 Re \biggl\{ \vert M_{GT} \vert ^2
\lambda_{J^\prime J}
\nonumber \\
& & \times \left[ \frac{\gamma m}{E_e} \left( C_T C_A^{\prime *} +
C_T^\prime C_A^* \right) \pm \left( C_T C_T^{\prime *} +
C_A C_A^{\prime *} \right) \right] \nonumber \\
& & - \delta_{J^\prime J} M_F M_{GT} \sqrt{\frac{J}{J+1}}
\nonumber \\
& & \times \biggl[ \left( C_S C_T^{\prime *} + C_S^\prime C_T^* +
C_V C_A^{\prime *} + C_V^\prime C_A^* \right) \nonumber \\
& & \left. \left. \pm \frac{\gamma m}{E_e} \left( C_S C_A^{\prime
*}
+ C_S^\prime C_A^* \right. \right. \right. \nonumber \\
& & + C_V C_T^{\prime *} + C_V^\prime C_T^* \biggr) \biggr]
\biggr\}
\end{eqnarray}

\begin{eqnarray}
\label{eqn:dcorr} D \xi & = & \delta_{J^\prime J} M_F M_{GT}
\sqrt{\frac{J}{J+1}}
 \nonumber \\
 & & \biggl[ 2 Im \left( C_S C_T^* - C_V C_A^* + C_S^\prime C_T^{\prime *}
- C_V^\prime C_A^{\prime *} \right) \nonumber \\
 & & \left. \mp  2 \frac{\alpha Z m}{p_e} Re \left( C_S C_A^* - C_V
 C_T^* \right. \right. \nonumber \\
 & & \left. + C_S^\prime C_A^{\prime *} - C_V^\prime C_T^{\prime *}
\right) \biggr]
\end{eqnarray}

\begin{eqnarray}
\label{eqn:gcorr} G \xi & = &  \vert M_F \vert ^2 \biggl[ \pm 2 Re
\left( C_S
 C_S^{\prime *} - C_V C_V^{\prime *} \right) \nonumber \\
 & & \left. + 2 \frac{\alpha Z m}{p_e} Im \left( C_S C_V^{\prime *} +
C_S^\prime C_V^* \right) \right]  \nonumber \\
 & & +  \vert M_{GT} \vert ^2 \biggl[ \pm 2 Re \left( C_T C_T^{\prime
 *} - C_A C_A^{\prime *} \right) \nonumber \\
 & & + 2 \frac{\alpha Z m}{p_e} Im \left( C_T C_A^{\prime *} +
C_T^\prime C_A^* \right) \biggr]
\end{eqnarray}

\begin{eqnarray}
\label{eqn:ncorr} N \xi & = & 2 Re \biggl\{ \vert M_{GT} \vert ^2
\lambda_{J^\prime J}
 \left[ \frac{1}{2} \frac{\gamma m}{E_e} \right. \nonumber \\
 & & \times \left( \vert C_T \vert ^2 + \vert C_A \vert ^2 +
 \vert C_T^\prime \vert ^2 +  \vert C_A^\prime \vert ^2 \right)
 \nonumber \\
 & & \pm \left( C_T C_A^* + C_T^\prime C_A^{\prime *} \right) \biggr] \nonumber \\
 & & + \delta_{J^\prime J} M_F M_{GT} \sqrt{\frac{J}{J+1}}
 \nonumber \\
 & & \times \biggl[ \left( C_S C_A^* + C_V C_T^* +
C_S^\prime C_A^{\prime *} + C_V^\prime C_T^{\prime *} \right) \nonumber \\
 & & \pm \frac{\gamma m}{E_e} \left( C_S C_T^* + C_V C_A^* \right. \nonumber \\
 & & \left. + C_S^\prime C_T^{\prime *} +
C_V^\prime C_A^{\prime *} \right) \biggr] \biggr\}
\end{eqnarray}

\begin{eqnarray}
\label{eqn:rcorr}
R \xi & = & \vert M_{GT} \vert ^2 \lambda_{J^\prime J} \nonumber \\
 & & \biggl[ \pm 2 Im \left( C_T C_A^{\prime *} + C_T^\prime C_A^*
 \right) \nonumber \\
 & & - 2 \frac{\alpha Z m}{p_e} Re \left( C_T C_T^{\prime *} -
C_A C_A^{\prime *} \right) \biggr] \nonumber \\
 & & + \delta_{J^\prime J} M_F M_{GT} \sqrt{\frac{J}{J+1}} \nonumber
 \\
 & & \times \biggl[ 2 Im \left( C_S C_A^{\prime *} + C_S^\prime C_A^* -
C_V C_T^{\prime *} - C_V^\prime C_T^* \right) \nonumber \\
 & & \mp 2 \frac{\alpha Z m}{p_e} Re \left( C_S C_T^{\prime *} +
C_S^\prime C_T^*  \right. \nonumber \\
 & & \left. - C_V C_A^{\prime *} - C_V^\prime C_A^* \right) \biggr]
\end{eqnarray}

\medskip
\section{LIMITS AND APPROXIMATIONS}

We summarize here several useful limits and approximations of the
correlation coefficients presented in
Appendix~\ref{sec:appendix:coeffs}.

\subsection{Standard model expressions}
\label{sec:appendix:SM}

The standard model assumes only vector and axial-vector
interactions with maximal parity violation. In addition it is
expected that the effects due to CP (or T) violation are
negligible in the light quark sector at the present level of
precision. These assumptions result in the conditions $C^\prime_V
= C_V, C^\prime_A = C_A, C_S = C^\prime_S = C_T = C^\prime_T = 0$
and $Im(C^\prime_i) = Im(C_i) = 0$ for $i = V, A$. Neglecting
Coulomb as well as induced recoil effects one then obtains, for
the $\beta$-neutrino angular correlation coefficient

\begin{equation}
\label{eqn:acorr-SM} a_{SM} = \frac{1 - \rho^2/3}{1 + \rho^2}
\end{equation}

\noindent where $\rho$ is the mixing ratio

\begin{equation}
\rho = \frac{ C_A  M_{GT} }{ C_V  M_{F} },
\end{equation}

\noindent for the $\beta$-decay asymmetry

\begin{eqnarray}
\label{eqn:Acoeff-SM} A_{SM}  =  \frac{ \mp \lambda_{J^\prime J}
\rho^2 - 2 \delta_{J^\prime J} \sqrt{\frac{J}{J+1}} \rho } {1 +
\rho^2},
\end{eqnarray}

\noindent for the neutrino decay asymmetry

\begin{eqnarray}
\label{eqn:Bcoeff-SM} B_{SM}  =  \frac{ \pm \lambda_{J^\prime J}
\rho^2 - 2 \delta_{J^\prime J} \sqrt{\frac{J}{J+1}} \rho } {1 +
\rho^2},
\end{eqnarray}

\noindent for the beta-particle longitudinal polarization

\begin{equation}
\label{eqn:Gcoeff-SM} G_{SM} = \mp 1 ,
\end{equation}

\noindent and for the other coefficients

\begin{equation}
b_{SM} = D_{SM} = N_{SM} = R_{SM} = 0 .
\end{equation}

It is to be noted that the triple correlation coefficients, $N$
and $R$, are non-zero when Coulomb corrections are included.

\subsection{Approximations for searches of exotic couplings}
\label{sec:appendix:approx}

Approximate expressions of the coefficients given in
Appendix~\ref{sec:appendix:coeffs} can be obtained for $C_i \ll 1$
and $C'_i \ll 1$, with $i = S,T$, by lowest order developments in
terms of these exotic couplings. These expressions show more
explicitly the sensitivity of the correlation coefficients to
these couplings. In deriving the approximations one assumes
maximal parity violation and time reversal invariance for the $V$-
and $A$-interactions ({\em i.e.} $C^\prime _V = C_V$, $C^\prime _A
= C_A$ with both couplings real), except for the triple
correlation $D$ where the possibility for complex couplings has
been allowed. In the expressions below the case $\rho = 0$
corresponds to pure Fermi transitions and the limit $\rho
\rightarrow \infty$ corresponds to Gamow-Teller transitions. The
highest sensitivity to terms containing scalar and/or tensor
coupling constants is obtained for pure transitions.

Under these assumptions, the Fierz interference term $b^\prime
\equiv b m/E_e$ is approximated as

\begin{eqnarray}
\label{eqn:bFierzapprox} b^\prime & \simeq & \pm \frac{\gamma
m}{E_e} \frac{1}{ 1 + \rho ^2} \left[ Re \left( \frac{ C_S +
C^\prime _S }{ C_V } \right) \right. \nonumber \\
& & \left. + \rho ^2 Re \left( \frac{ C_T + C^\prime _T }{ C_A }
\right) \right]
\end{eqnarray}

\medskip
The beta-neutrino correlation coefficient can be written as

\begin{eqnarray}
\label{eqn:acorrapprox_exot}
a & \simeq & a_{SM} \nonumber \\
 & & - \frac{1}{\left( 1 + \rho ^2 \right)^2 } \left[ \left(
1 + \frac{1}{3} \rho ^2  \right) \frac{\vert C_S \vert ^2 + \vert
C^\prime _S \vert ^2}{ C_V ^2} \right. \nonumber \\
 & & \left. + \frac{1}{3} \rho ^2 \left( 1 -
\rho ^2 \right) \frac{\vert C_T \vert ^2 + \vert C^\prime _T \vert
^2}{ C_A ^2} \right] \nonumber \\
 & &  + \frac{\alpha Z m}{p_e} \frac{1}{ 1 + \rho ^2 } \left[ \mp Im
 \left( \frac{ C_S + C^\prime _S }{ C_V } \right) \right. \nonumber \\
 & & \left. \pm \frac{ \rho ^2
 }{ 3 } Im \left( \frac{ C_T + C^\prime _T }{ C_A } \right)
 \right].
\end{eqnarray}

\medskip
The beta asymmetry parameter becomes

\begin{eqnarray}
\label{eqn:Aparapprox_exot}
A & \simeq & A_{SM} \nonumber \\
 & & + \frac{ \alpha Z m } {p_e} \left[ \frac{ \lambda_{J^\prime J}
\rho^2 \pm \delta_{J^\prime J} \sqrt{\frac{J}{J+1}} \rho } {1 +
\rho^2} Im \left( \frac{ C_T + C^\prime _T }{ C_A } \right) \right. \nonumber \\
 & & \left. \pm \frac{\delta_{J^\prime J} \sqrt{\frac{J}{J+1}} \rho }
 {1 + \rho^2} Im \left( \frac{ C_S + C^\prime _S }{ C_V } \right) \right]
\end{eqnarray}

It is seen that the beta asymmetry parameter cancels for pure
Fermi transitions and that it is sensitive to the exotic couplings
only via the Coulomb correction term. For a pure Gamow-Teller
transition the quantity $\widetilde{A} \equiv A / (1 + b^\prime$)
becomes

\begin{eqnarray}
\label{eqn:AparapproxGT_tilde} \widetilde{A}_{GT} & = & \frac{
A_{GT} } {1 + b^\prime } \nonumber \\
 & & \simeq \lambda_{J^\prime J} \left[ \mp 1 +  \frac{ \alpha Z m }
{p_e} Im \left( \frac{ C_T + C^\prime _T }{ C_A } \right) \right. \nonumber \\
 & & + \left. \frac{\gamma m} {E_e} Re \left( \frac{ C_T + C^\prime _T }{ C_A }
\right) \right]
\end{eqnarray}

\noindent where $A_{GT}$ is obtained from
Eq.~(\ref{eqn:Aparapprox_exot}) for $\rho \rightarrow \infty$.

\medskip
The neutrino asymmetry parameter can be approximated as

\begin{eqnarray}
\label{eqn:Bparapprox} B & \simeq & B_{SM} \nonumber \\
 & & + \frac{ 1 }{1 + \rho^2} \biggl\{ \lambda_{J^\prime J} \rho^2
\frac{\gamma m} {E_e}
Re\left( \frac{ C_T + C^\prime _T }{ C_A } \right) \nonumber \\
& & \mp \delta_{J^\prime J} \rho \sqrt{\frac{J}{J+1}} \times
\frac{\gamma m}{E_e}
\left[ Re\left( \frac{ C_S + C^\prime _S }{ C_V } \right) \right. \nonumber \\
& & \left. + Re\left( \frac{ C^*_T + C^{*\prime}_T }{ C_A }
\right) \right] \biggr\}
\end{eqnarray}

\noindent For a pure Fermi transition $B = 0$. It is seen that $B$
and $\widetilde{B} \equiv B / (1+b^\prime) $ are insensitive to
time reversal violating interactions.

\medskip
The longitudinal beta polarization coefficient can be written as

\begin{eqnarray}
\label{eqn:Gcorrapprox} G & \simeq & G_{SM} + \frac{1}{ 1 +
\rho^2}
 \frac{\alpha Z m}{p_e} \left[ Im \left( \frac{ C_S + C^\prime
_S }{ C_V } \right) \right. \nonumber \\
 & & \left. + \rho^2  Im \left( \frac{ C_T + C^\prime _T
}{ C_A } \right) \right]
\end{eqnarray}

From the comparison of Eq.~(\ref{eqn:Gcorrapprox}) and
Eq.~(\ref{eqn:Aparapprox_exot}) it appears that $G$ probes the
same couplings as $A$ but with different sensitivities.

\medskip
The P-odd and T-odd triple correlation coefficient, $R$, which
probes the existence of time reversal violating scalar and/or
tensor components can be written as

\begin{eqnarray}
\label{eqn:Rcorrapprox} R & \simeq & \frac{1} { 1 + \rho^2 } \nonumber \\
& & \times \left[ \left( \pm \lambda_{J^\prime J} \rho^2 +
\delta_{J^\prime J} \sqrt{\frac{J}{J+1}} \rho \right) Im \left(
\frac{ C_T + C^\prime
_T }{ C_A } \right) \right. \nonumber \\
 & & + \delta_{J^\prime J} \sqrt{\frac{J}{J+1}} \rho Im \left(
 \frac{ C_S + C^\prime _S }{ C_V } \right) \nonumber \\
 & & + \left. \frac{ \alpha Z m } {p_e} \left( \lambda_{J^\prime
J} \rho^2 \pm 2 \delta_{J^\prime J} \sqrt{\frac{J}{J+1}} \rho
\right) \right]
\end{eqnarray}

\noindent This correlation cancels for a pure Fermi transition and
for a pure Gamow-Teller transition it reduces to

\begin{equation}
\label{eqn:RcorrapproxGT} R \simeq \lambda_{J^\prime J} \left[ \pm
Im \left( \frac{ C_T + C^\prime _T }{ C_A } \right) + \frac{
\alpha Z m } {p_e} \right]
\end{equation}

Finally, if we relax the assumption stated above that the $V$- and
$A$-couplings be both real, allowing for an imaginary phase
between them, then the P-even triple correlation coefficient, $D$,
becomes

\begin{eqnarray}
\label{eqn:Dcorrapprox} D & \simeq & \frac{- \rho} { 1 + \rho^2}
\biggl\{ \delta_{J^\prime J} \sqrt{\frac{J}{J+1}} \left[ 2
\frac{Im
\left(C_V C^*_A \right)} {C_V C^*_A} \right. \nonumber \\
 & & \left. + \frac{\alpha Z m}{p_e} Re
\left( \frac{ C_S + C^\prime _S }{ C_V }
 - \frac{ C_T + C^\prime
_T }{ C_A^* } \right)\right]\biggr\}
\end{eqnarray}

\noindent The sensitivity of this correlation to the terms between
brackets is non-zero only for mixed transitions, {\em i.e.} $\rho
\neq 0$.

\medskip
\subsection{Right-handed couplings}
\label{sec:appendix:rhc}

We restrict here to the simplest case of so-called manifest
left-right symmetric models, assuming no CP-violation in the
right-handed sector. The correlation coefficients can then be
expressed in terms of two parameters: $\delta = (m_1 / m_2)^2$ and
$\zeta$ (see Sec.~\ref{sec:righthanded} and
Sec.~\ref{sec:parity}). In developing the equations that follow,
all scalar and tensor couplings were assumed to be zero and time
reversal invariance was assumed to hold for the $V$- and
$A$-interactions.

\medskip
Assuming the presence of right-handed currents the beta-neutrino
correlation coefficient can be written as:

\begin{eqnarray}
\label{eqn:acorrapprox_rhc} a & \simeq & \frac{ \left[ 1 -
\frac{1}{3} \rho ^2 \right] \left[ 1 + \left( \delta + \zeta
\right)^2 \right] - 4 \delta \zeta} { \left[ 1 + \rho ^2 \right]
\left[ 1 + \left( \delta + \zeta \right)^2 \right] - 4 \delta
\zeta}
\end{eqnarray}

\noindent This coefficient looses its sensitivity to right-handed
currents in the limit of no mixing, $\zeta \rightarrow 0$.

\medskip
The beta-asymmetry parameter can here be written as

\begin{eqnarray}
\label{eqn:Aparexpan_rhc} A & \simeq &  A_{SM} \left( 1 +
\alpha_{\delta\delta} \delta^2 + \alpha_{\delta\zeta} \delta\zeta
+ \alpha_{\zeta\zeta} \zeta^2 \right)
\end{eqnarray}

\noindent with

\begin{equation}
\alpha_{\delta\delta} = -2 \; , \label{eq:var-delta-delta}
\end{equation}

\begin{equation}
\alpha_{\delta\zeta} = \frac{ -4 \lambda_{J^\prime J} \rho^3 \mp 4
\delta_{J^\prime J} \sqrt{\frac{J}{J+1}} \left( 1 - \rho^2 \right)
} { \left( \lambda_{J^\prime J} \rho \mp 2 \delta_{J^\prime J}
\sqrt{\frac{J}{J+1}} \right) \left( 1 + \rho^2 \right) } \; ,
\end{equation}

\noindent and

\begin{equation}
\alpha_{\zeta\zeta} = \frac{ -2 \lambda_{J^\prime J} \rho } {
\lambda_{J^\prime J} \rho \mp 2 \delta_{J^\prime J}
\sqrt{\frac{J}{J+1}} } \; .
\end{equation}

\noindent The sensitivity to $\delta^2$ in
Eq.~(\ref{eqn:Aparexpan_rhc}), driven by the factor
$\alpha_{\delta\delta} = -2$, and is then the same for all types
of transitions, whatever their Fermi/Gamow-Teller character. For a
pure Gamow-Teller transition one obtains

\begin{equation}
\label{eqn:Aparapprox_rhcGT} A_{GT} \simeq \mp \lambda_{J^\prime
J} \left[ 1 - 2 \left( \delta + \zeta \right)^2 \right] \; .
\end{equation}

\noindent The ratio between the asymmetry parameters of a mixed
and of a pure Gamow-Teller transition then becomes

\begin{equation}
\label{eqn:AparmirGT_rhc} \frac {A^{mix}} {A^{GT}} \simeq \frac
{A^{mix}_{SM}} {\lambda_{J^\prime J}^{GT}} \left[ 1 + \left(4 +
\alpha_{\delta\zeta} \right) \delta\zeta + \left(2 +
\alpha_{\zeta\zeta} \right) \zeta^2 \right].
\end{equation}

\noindent where $A^{mix}$ is given by
Eq.~(\ref{eqn:Aparexpan_rhc}). Here again, this ratio looses its
sensitivity to right-handed currents in the limit of no mixing,
$\zeta \rightarrow 0$.

\medskip
The neutrino asymmetry parameter can be written as

\begin{eqnarray}
\label{eqn:Bparapprox_rhc} B & \simeq & \biggl[ \pm
\lambda_{J^\prime J} \rho^2 \left( 1 - y^2 \right) - 2
\delta_{J^\prime J} \sqrt{\frac{J}{J+1}} \rho \left( 1 - x y
\right) \biggr] \nonumber \\
& & \times \biggl[ \left( 1 + x^2 \right) + \rho^2 \left( 1 + y^2
\right) \biggr]^{-1}
\end{eqnarray}

\noindent where $x = \delta - \zeta$ and $y = \delta + \zeta$.

\medskip
The longitudinal polarization of beta particles is given by

\begin{eqnarray}
\label{eqn:Gparapprox_rhc} G & \simeq & \mp \left[ 1 - \frac{ 2
\left( x^2 + \rho^2 y^2 \right) } { 1 + \rho^2 } \right].
\end{eqnarray}

One should stress here that the expressions in
Eqs.~(\ref{eqn:acorrapprox_rhc},
~\ref{eqn:Aparexpan_rhc},~\ref{eqn:AparmirGT_rhc},~\ref{eqn:Bparapprox_rhc}
and \ref{eqn:Gparapprox_rhc}) depend all from the mixing ratio
$\rho$ which has to be determined from an independent observable.
This last observable might also be sensitive to effects due to
right-handed currents so that the actual sensitivity of the
parameters listed above will change accordingly. Such an effect
has been studied more quantitatively by \citet{naviliat91}, for
the $\beta$-asymmetry parameter in mirror decays.

\subsection{Coefficients in neutron decay}
\label{app:corr_neutron}

The SM predictions for the correlation coefficients in the decay
of the neutron, neglecting Coulomb corrections as well as induced
recoil effects, are usually expressed in terms of a single
parameter, $\lambda = |\lambda| e^{-i\phi} = g_A/g_V = C_A/C_V$.
The assumptions are here identical to those stated in
Appendix~\ref{sec:appendix:SM}. In neutron decay we have $J = J' =
1/2$, $M_F=1$ and $M_{GT}=\sqrt{3}$, such that $\rho=C_A M_{GT} /
C_V M_F=\sqrt{3} \lambda$.

The SM expression in this particular case are then

\begin{eqnarray}
a_n & = & \frac{ 1 - \vert \lambda \vert ^2}{1 + 3 \vert \lambda
\vert ^2}
\label{eqn:aneutronTRV} \\
A_n & = & -2 \frac{\vert \lambda \vert ^2 + Re \lambda } {1 + 3
\vert \lambda \vert ^2}
\label{eqn:AneutronTRV} \\
G_n & = & -1
\label{eqn:GneutronTRV} \\
B_n & = & 2 \frac{\vert \lambda \vert ^2 - Re \lambda } {1 + 3
\vert \lambda \vert ^2}
\label{eqn:BneutronTRV} \\
D_n & = & 2 \frac{ Im \lambda } {1 + 3 \vert \lambda \vert ^2}
\label{eqn:DneutronTRV}
\end{eqnarray}

Under the assumptions above $b_n = N_n = R_n = 0$ and $Im \lambda
= 0$ such that $D_n = 0$. Again, it should be noted that the
triple correlation coefficients, $N_n$ and $R_n$, are non-zero
when Coulomb corrections are included. The parameter $\lambda$, or
its absolute value, can be determined from a measurement of either
$a_n$, $A_n$ or $B_n$.

%
%
%

Similarly, in the presence of exotic couplings or in the framework
of manifest left-right symmetric models, the expressions of the
correlation coefficients in neutron decay can be derived from
those given in Appendices~\ref{sec:appendix:coeffs},
\ref{sec:appendix:approx} and \ref{sec:appendix:rhc} by choosing
the proper sign for the $\beta^-$-decay and setting
$\lambda_{J^\prime J}=2/3$, $\delta_{J^\prime J}=1$, $\sqrt{J /
J+1} = 1/\sqrt{3}$, and $\rho=\sqrt{3} \lambda$.

Note that because for the neutron $Z=1$, the factor $\alpha Z m /
p_e$ is very small. Even in the middle of the electron energy
spectrum its value is only 0.0042, rendering terms proportional to
this factor hardly accessible at the present level of precision.
On the other hand, the factor $\gamma m / E_e$ equals 0.40 at the
end of the electron energy spectrum and 0.57 in the middle of the
spectrum giving a sensitivity to the terms proportional to this
factor similar to that obtained in nuclear transitions. A more
detailed discussion can be found in \citet{gluck95}.

%
%
\bibliographystyle{apsrmp}
%
%

%

%

%
\end{document}